\documentclass[journal]{IEEEtran}
\usepackage{multirow}
\usepackage{booktabs}
\usepackage{makecell}
\usepackage{cite}
\usepackage[numbers,sort&compress]{natbib}
\usepackage{amsmath}
\usepackage{array}
\usepackage{graphicx}
\usepackage{float}
\usepackage{fixltx2e}
\usepackage{amssymb}
\usepackage{pifont}
\usepackage[table,xcdraw]{xcolor}
\graphicspath{{./figures/}}
\usepackage{url}  
\usepackage[noend]{algpseudocode}
\usepackage{algorithmicx,algorithm}
\usepackage{epstopdf}
\usepackage{color}
\usepackage{bbding} 
\usepackage{pifont}
\usepackage{xcolor}
\usepackage{natbib}

\ifCLASSOPTIONcompsoc
  \usepackage[caption=false,font=normalsize,labelfont=sf,textfont=sf]{subfig}
\else
  \usepackage[caption=false,font=footnotesize]{subfig}
\fi

\begin{document}

\title{Seismic Data Interpolation via Denoising Diffusion Implicit Models with Coherence-corrected Resampling}

\author{Xiaoli~Wei, Chunxia~Zhang, \textit{Member}, \textit{IEEE}, Hongtao Wang, Chengli Tan, Deng Xiong, Baisong Jiang, Jiangshe~Zhang, Sang-Woon~Kim, \textit{Life Senior Member}, \textit{IEEE}
\thanks{Corresponding author: Chunxia Zhang. E-mail: cxzhang@mail.xjtu.edu.cn.}
\thanks{Xiaoli Wei, Chunxia Zhang, Hongtao Wang, Chengli Tan, Baisong Jiang, Jiangshe~Zhang are with the School of Mathematics and Statistics, Xi'an Jiaotong University, Xi'an, Shaanxi, 710049, China.}
\thanks{Deng Xiong is with the Geophysical Technology Research and Development Center, BGP, Zhuozhou, Hebei, 072751, China}
\thanks{Sang-Woon~Kim is with the Department of Computer Engineering, Myongji University, Yongin, 17058, South Korea.}
\thanks{This research was supported by the National Natural Science Foundation of China (No. 12371512), the Science \& Technology Research and Development Project of CNPC (Grant No. 2021ZG03), and the Key Acquisition, Processing and Interpretation Techniques for Seismic Data Processing of
CNPC (Grant No. 01-03-2023)}
\thanks{This work has been submitted to TGRS for possible publication. Copyright may be transferred without notice, after which this version may no longer be accessible.}
}

\markboth{
Journal of \LaTeX\ Class Files,~Vol.~14, No.~8, August~2015}%
{Shell \MakeLowercase{\textit{et al.}}: Bare Demo of IEEEtran.cls for IEEE Journals
}

\maketitle

\begin{abstract}
Accurate interpolation of seismic data is crucial for improving the quality of imaging and interpretation. In recent years, deep learning models such as U-Net and generative adversarial networks have been widely applied to seismic data interpolation. However, they often underperform when the training and test missing patterns do not match. To alleviate this issue, here we propose a novel framework that is built upon the multi-modal adaptable diffusion models. In the training phase, following the common wisdom, we use the denoising diffusion probabilistic model with a cosine noise schedule. This cosine global noise configuration improves the use of seismic data by reducing the involvement of excessive noise stages. In the inference phase, we introduce the denoising diffusion implicit model to reduce the number of sampling steps. Different from the conventional unconditional generation, we incorporate the known trace information into each reverse sampling step for achieving conditional interpolation. To enhance the coherence and continuity between the revealed traces and the missing traces, we further propose two strategies, including successive coherence correction and resampling. Coherence correction penalizes the mismatches in the revealed traces, while resampling conducts cyclic interpolation between adjacent reverse steps. Extensive experiments on synthetic and field seismic data validate our model's superiority and demonstrate its generalization capability to various missing patterns and different noise levels with just one training session. In addition, uncertainty quantification and ablation studies are also investigated.
\end{abstract}


\IEEEpeerreviewmaketitle

\section{Introduction}
\IEEEPARstart{S}{eismic} exploration collects seismic data to image the subsurface and interpret geological and fluid information. In practice, acquiring complete seismic data is always challenging due to complex surface or underground obstacles and certain economic considerations.
Consequently, the degradation of data integrity is typically observed in the form of random or consecutive missing seismic traces or very sparse source or receiver arrangements in land or marine data \cite{chen2019interpolation}. 
Seismic data interpolation helps to acquire an optimal sampling to improve data imaging and interpretation by infilling regular or consecutive missing data, reconstructing irregularly arranged data to a regular position, and interpolating sparsely acquired data to a much denser volume. It is an essential tool for tackling aliasing data problems, filling gaps to enhance data quality, minimizing uncertainties, and reducing certain economic costs in acquisition.

Seismic data interpolation has been extensively investigated over the past decades. Initially developed traditional methods often rely on the assumption of global or local linear events to convert the problem into an autoregressive framework \cite{liu2011seismic}. Especially, prediction-filter-based methods, combined with the $t\text{-}x$ and $f\text{-}x$ regularization \cite{li2017multidimensional}, \cite{chen20215d}, occupy the research mainstream in this direction. 
Wave-equation-based methods can extrapolate and interpolate wave field \cite{fomel2003seismic}, whereas they require additional information, e.g., wave velocity. 
Sparsity-based methods introduce various sparse transforms and sampling functions to interpolate missing data \cite{zwartjes2007fourier, herrmann2008non, gan2015dealiased, Ray2006, gao2010irregular, wang2014dreamlet}. 
The low-rank constraint models recover incomplete data by using singular value decomposition on block Hankel matrix \cite{oropeza2011simultaneous, ma2013three, chen2019five}.
Traditional and model-driven methods theoretically support interpolation and are widely used in industry. However, drawbacks such as manual parameter selection and high computational costs are significant, particularly for large, high-dimensional seismic data collected through advanced technologies.
 
With the rapid advancement of deep learning-based generative models, the research focus for seismic data interpolation has shifted toward data-driven methods, which mainly include two categories, i.e., end-to-end neural networks and generative adversarial networks (GAN). 
The preliminary methods in the first category of data-driven models contain the convolutional autoencoder (CAE) \cite{wang2020seismic}, \cite{chai2020deep}, U-Net \cite{mandelli2018seismic}, \cite{fang2021seismic}, and residual network (ResNets) \cite{wang2019deep}, etc. Liu \textit{et al.} \cite{liu2022seismic} introduce the invertible discrete wavelet transform for replacing the pooling operations in the traditional U-Net model, thereby avoiding the loss of detailed features caused by the downsampling scheme. Some researchers have worked on improving the long-range feature correlation via different attention modules \cite{lou2022irregularly}, \cite{song9893067}, which are critical to maintain the global content consistency, especially under the circumstance of consecutively missing seismic traces \cite{Yu9390348}. Furthermore, regularization terms are important in finding the optimal interpolation function, e.g., spectrum suppression \cite{Yuan9757159} and regeneration constraint \cite{aoqi10007856}. Some studies also focus on improving the seismic feature extraction ability of neural networks, including the adoption of UNet++ with a nested architecture \cite{wu2023seismic} and dynamically updating the valid convolution region \cite{pan2020partial}. However, a standalone neural network is usually insufficient to capture the vast range of dynamic energy in seismic data. 
To resolve this issue, the coarse-refine network \cite{park9826799} and the multi-stage active learning method \cite{He9469034} have been proposed, which exploit the strengths of every sub-network to make the interpolation process more efficient and well-performed. 
The second category of data-driven models, GAN-based methods, has also achieved impressive results in seismic data interpolation. Kaur \textit{et al.} \cite{kaur2019seismic} adopt the framework of CycleGAN to perform self-learning on the seismic features. 
The conditional generative adversarial network (CGAN) is introduced to interpolate the seismic data with consecutively missing traces \cite{oliveira2018interpolating}. Based on CGAN, the dual-branch interpolation method combining the time and frequency domains improves the smoothness and quality of the reconstructed seismic data \cite{Chang2020}. The large obstacle is a common trouble in seismic exploration, which leads to big gaps in the collected seismic data and impairs further data processing. The promising results of conditional Wasserstein generative adversarial networks with gradient penalty (WGAN-GP) have revealed the seismic feature generation capability \cite{wei2022big}, whose gradient penalty enhances the fidelity of reconstructed signals at large intervals by enforcing the Lipschitz constraint. The coarse-to-fine learning strategy driven by the joint of different losses strengthens the connection between different stages and enables relativistic average least-square generative adversarial network (RaLSGAN) to produce more accurate and realistic signal details \cite{wei2022hybrid}. 

UNet-based and GAN-based interpolation methods hold a leading position, serving as the backbone of many current deep learning frameworks. However, their missing pattern-specific training routine incurs significant computational expense and harms the generalization capability for unseen patterns. Field seismic data often exhibits various missing forms due to ground obstacles and geophone layout conditions. The aforementioned methods typically cater to a specific form of missing seismic data and require retraining to interpolate seismic data with different missing ratios or forms. Diffusion models are experiencing a research boom for their multi-task and multi-modal generalization capability \cite{ho2020denoising, nichol2021improved, dhariwal2021diffusion, kawar2022denoising}. Alongside our work, many diffusion-based interpolation models \cite{durall2023deep, deng2024seismic, wang2024seisfusion, wang2024reconstructing, liu2024generative} have been proposed and shown promising application advantages. However, their inference processes are generally computationally expensive due to the intensive sampling processing, and they usually handle one or two missing patterns, lacking diversity in the interpolation scenarios. 

In this paper, we propose a new seismic denoising diffusion implicit model with coherence-corrected resampling (SeisDDIMCR), showing that it only needs to be trained once to complete the reconstruction tasks of different missing rates or missing forms. It also exhibits superior interpolation effects compared to the existing deep learning methods. Our denoising diffusion model-based approach retains the strong power of generative neural networks since the backbone can be inherited from state-of-the-art generative architectures. The main contributions of this paper are summarized below:

\begin{itemize}
\item We successfully apply the diffusion model to reconstruct degraded seismic data. The training framework is built on DDPM \cite{ho2020denoising}, enabling unconditional seismic data generation. We adopt the cosine noise schedule to focus the sampling process on learning seismic signals at low noise levels, thereby reducing the excessive high-noise steps introduced by the linear noise schedule. 

\item To efficiently reduce the number of reverse sampling, our model's inference process utilizes denoising diffusion implicit models (DDIM) \cite{song2020denoising}. Based on this framework, to integrate the known trace information into the unconditional sampling process, we progressively insert revealed seismic data in the reverse steps, achieving conditional interpolation guided by the known traces. 

\item To address the inconsistency and discontinuity between revealed and missing traces, we introduce a coherence-corrected resampling strategy. Coherence correction optimizes the overall consistency of the distribution by penalizing interpolation errors in the known traces, successively. Then, iterative resampling improves continuity through cyclic guidance between adjacent steps. 
\item Our SeisDDIMCR is a missing pattern-free model since it requires only one training session on the complete data. It exhibits good generalization capabilities for various missing types and effectively interpolates seismic data, even when complex missing forms coexist. In addition, our diffusion-based model can provide uncertainty estimates for interpolation results.
\end{itemize}
The remainder of this paper is organized as follows. In Section \ref{Background}, we recap the background of DDPM \cite{ho2020denoising}. Section \ref{secMethodology} introduces our SeisDDIMCR method, including the training strategy, implicit conditional interpolation, coherence correction, and resampling. In Section \ref{secExperiments}, experiments with various missing interpolations are performed for both synthetic and field seismic data. The generalization of our method is demonstrated by comparing it with popular methods. Furthermore, to indicate the stronger advantages of our model in practical application scenarios, we conduct uncertainty quantification and model generalization validation. Section \ref{secabalation} presents some ablation studies. Finally, we make conclusions in Section \ref{secConclusion}.

\section{Background}\label{Background}
\begin{figure*}[!htbp]
	\centering
    \includegraphics[width=6.5in]{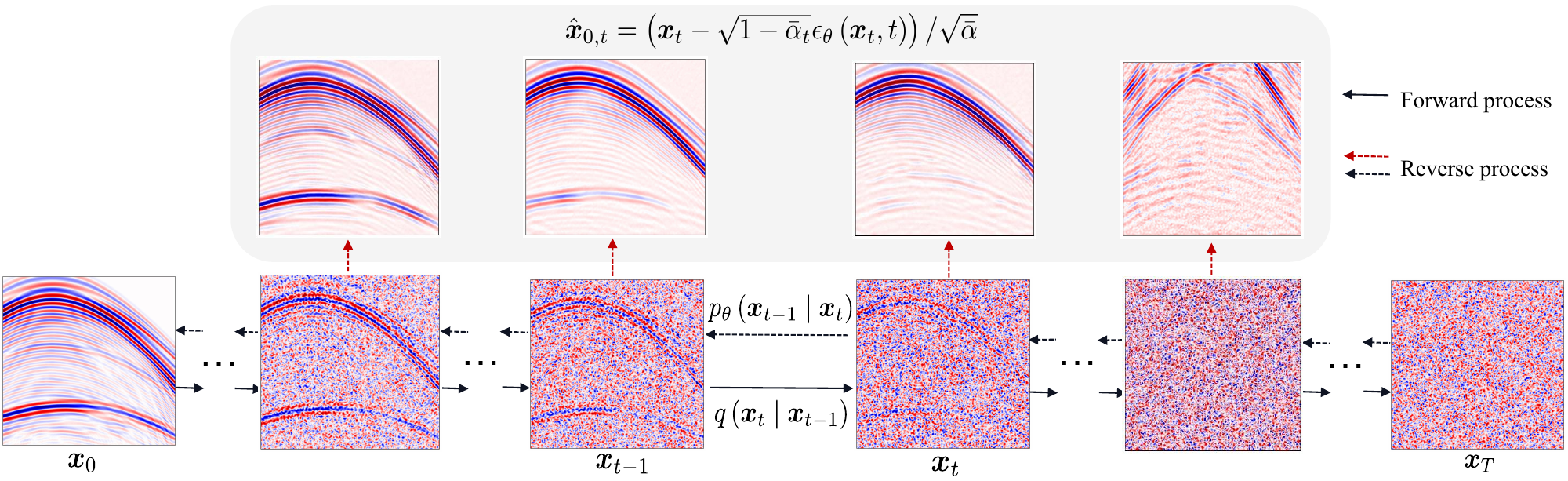}
    \vspace{-2mm}
	\caption{The pipeline of the seismic DDPM. It comprises two stages, i.e., the forward process and the reverse process. The forward process fixedly converts the complete seismic data $\boldsymbol{x}_0$ into a series of noise-added seismic data until $\boldsymbol{x}_T$ converges to an isotropic Gaussian noise, and the reverse process uses the neural network to learn the distribution parameters of each time step. Then $\boldsymbol{x}_0$ can be obtained by step-by-step iterative denoising. Especially, $\boldsymbol{x}_0$ can also be estimated at each reverse time step and denoted as $\hat{\boldsymbol{x}}_{0,t}$.}
	\label{fig:seiddpm}
	\vspace{-4mm}
\end{figure*}
 Combined with the background of seismic data interpolation, DDPM \cite{ho2020denoising} consists of two main processes, i.e., the training process for estimating the parameters of seismic DDPM and the sampling process for generating seismic data. Fig. \ref{fig:seiddpm} illustrates the detailed stream of the seismic DDPM. The forward process does not require training and directly converts $\boldsymbol{x}_0$ to the isotropic Gaussian noise. In the reverse process, the denoising model learns to predict the added noise for each time step. 

\subsection{Training Process}\label{Training_Diffusion}
Given the complete seismic data samples $\boldsymbol{x}_0 \sim q\left(\boldsymbol{x}_0\right)$, DDPM relies on the generative Markov chain process and the noise matching network to gradually learn the target distribution $p_\theta\left(\boldsymbol{x}_0\right)$. 
The forward diffusion process is a deterministic Markov chain starting from the initial input $\boldsymbol{x}_0$ and using a pre-specified noise schedule to gradually add Gaussian noise to perturb the data distribution. Given the latent variables $\boldsymbol{x}_1, \ldots, \boldsymbol{x}_T$ derived from the same sample space with $\boldsymbol{x}_0$, the diffusion process is defined as 
\begin{equation}
q\left(\boldsymbol{x}_{1: T} \mid \boldsymbol{x}_0\right):=\prod_{t=1}^T q\left(\boldsymbol{x}_t \mid \boldsymbol{x}_{t-1}\right),
\end{equation}
where 
\begin{equation}
\label{equ:q_t_t1}
q\left(\boldsymbol{x}_t \mid \boldsymbol{x}_{t-1}\right):=\mathcal{N}\left(\boldsymbol{x}_t ; \sqrt{1-\beta_t} \boldsymbol{x}_{t-1}, \beta_t \mathbf{I}\right).
\end{equation}
Here, $\beta_t\in(0,1)$ is a pre-designed increasing variance schedule of Gaussian noise. The closed form of sampling $\boldsymbol{x}_t$ given by Ho \textit{et al.} \cite{ho2020denoising} reveals the progressive changes during the middle time of the forward process. Letting $\alpha_t:=1-\beta_t$ and $\bar{\alpha}_t:=\prod_{s=1}^t \alpha_s$, it can be denoted as
\begin{equation}
\label{equ:q_t}
q\left(\boldsymbol{x}_t \mid \boldsymbol{x}_0\right)=\mathcal{N}\left(\boldsymbol{x}_t ; \sqrt{\bar{\alpha}_t} \boldsymbol{x}_0,\left(1-\bar{\alpha}_t\right) \mathbf{I}\right).
\end{equation}
As $t$ continues to increase, the final data distribution converges to a given prior distribution, i.e., a standard Gaussian for $\boldsymbol{x}_0$. Correspondingly, the reverse process will gradually denoise for each step of the forward process starting from $p\left(\boldsymbol{x}_T\right)=\mathcal{N}\left(\boldsymbol{x}_T ; \mathbf{0}, \mathbf{I}\right)$ under the Markov chain transition
\begin{equation}
p_\theta\left(\boldsymbol{x}_{0: T}\right):=p\left(\boldsymbol{x}_T\right) \prod_{t=1}^T p_\theta\left(\boldsymbol{x}_{t-1} \mid \boldsymbol{x}_t\right),
\end{equation}
where $p_\theta\left(\boldsymbol{x}_{t-1} \mid \boldsymbol{x}_t\right):=\mathcal{N}\left(\boldsymbol{x}_{t-1} ; \boldsymbol{\mu}_\theta\left(\boldsymbol{x}_t, t\right), \mathbf{\Sigma}_\theta\left(\boldsymbol{x}_t, t\right)\right)$ and the network parameter $\theta$ is shared across different reverse stages. This optimization problem of fitting the data distribution $q\left(\boldsymbol{x}_0\right)$ can be converted into the minimization of a variational lower bound (VLB) for the negative log-likelihood by introducing Jensen’s inequality
\begin{equation}
L_{\text{vlb}}:=\mathbb{E}_{q\left(\boldsymbol{x}_{0: T}\right)}\left[\log \frac{q\left(\boldsymbol{x}_{1: T} \mid \boldsymbol{x}_0\right)}{p_\theta\left(\boldsymbol{x}_{0: T}\right)}\right] \geq-\mathbb{E}_{q\left(\boldsymbol{x}_0\right)} \log p_\theta\left(\boldsymbol{x}_0\right).
\end{equation}
VLB is decomposed into the following KL-divergence form
between two Gaussian distributions by including the Markov property in the denoising diffusion model and the definition form of the forward process
\begin{equation}
\label{equ:vlb}
\begin{aligned}
L_{\text{vlb}}&=
\mathbb{E}_q[D_{\mathrm{KL}}\left(q\left(\boldsymbol{x}_T \mid \boldsymbol{x}_0\right) \| p\left(\boldsymbol{x}_T\right)\right)] -\mathbb{E}_q[\log p_\theta\left(\boldsymbol{x}_0 \mid \boldsymbol{x}_1\right)]\\
&+\mathbb{E}_q[\sum_{t=2}^T D_{\mathrm{KL}}\left(q\left(\boldsymbol{x}_{t-1} \mid \boldsymbol{x}_t, \boldsymbol{x}_0\right) \| p_\theta\left(\boldsymbol{x}_{t-1} 
\mid \boldsymbol{x}_t\right)\right)].
\end{aligned}
\end{equation}
According to Ho \textit{et al.} \cite{ho2020denoising}, the Gaussian distribution $q\left(\boldsymbol{x}_{t-1} \mid \boldsymbol{x}_t, \boldsymbol{x}_0\right)$ can be tractable as
\begin{equation}
q\left(\boldsymbol{x}_{t-1} \mid \boldsymbol{x}_t, \boldsymbol{x}_0\right)=\mathcal{N}\left(\boldsymbol{x}_{t-1} ; \tilde{\boldsymbol{\mu}}_t\left(\boldsymbol{x}_t, \boldsymbol{x}_0\right), \tilde{\beta}_t \mathbf{I}\right),
\end{equation}
where 
\begin{equation*}
\tilde{\boldsymbol{\mu}}_t\left(\boldsymbol{x}_t, \boldsymbol{x}_0\right):=\frac{\sqrt{\bar{\alpha}_{t-1}} \beta_t}{1-\bar{\alpha}_t} \boldsymbol{x}_0+\frac{\sqrt{\alpha_t}\left(1-\bar{\alpha}_{t-1}\right)}{1-\bar{\alpha}_t} \boldsymbol{x}_t
\end{equation*} 
and 
\begin{equation}
\tilde{\beta}_t:=\frac{1-\bar{\alpha}_{t-1}}{1-\bar{\alpha}_t} \beta_t.
\end{equation}
Finally, the popular loss used in DDPM is finally formulated as
\begin{equation}
\label{equ:l_simple}
\begin{aligned}
\footnotesize
L_\text{simple}=\mathbb{E}_{\boldsymbol{x}_0 \sim q\left(\boldsymbol{x}_0\right), \epsilon_t \sim \mathcal{N}(\mathbf{0}, \boldsymbol{I})}\left[\left\|\boldsymbol{\epsilon}_t-\epsilon_\theta\left(\sqrt{\bar{\alpha}_t} \boldsymbol{x}_0+\sqrt{1-\bar{\alpha}_t} \boldsymbol{\epsilon}_t, t\right)\right\|^2\right].
\end{aligned}
\end{equation}
Therefore, the network parameters are optimized by the mean squared error (MSE) loss between the Gaussian noise predicted by the network and the real noise for all time nodes of the reverse process except for $t=1$.   

\begin{figure*}[!t]
  \centering
  {\includegraphics[width=0.85\textwidth]{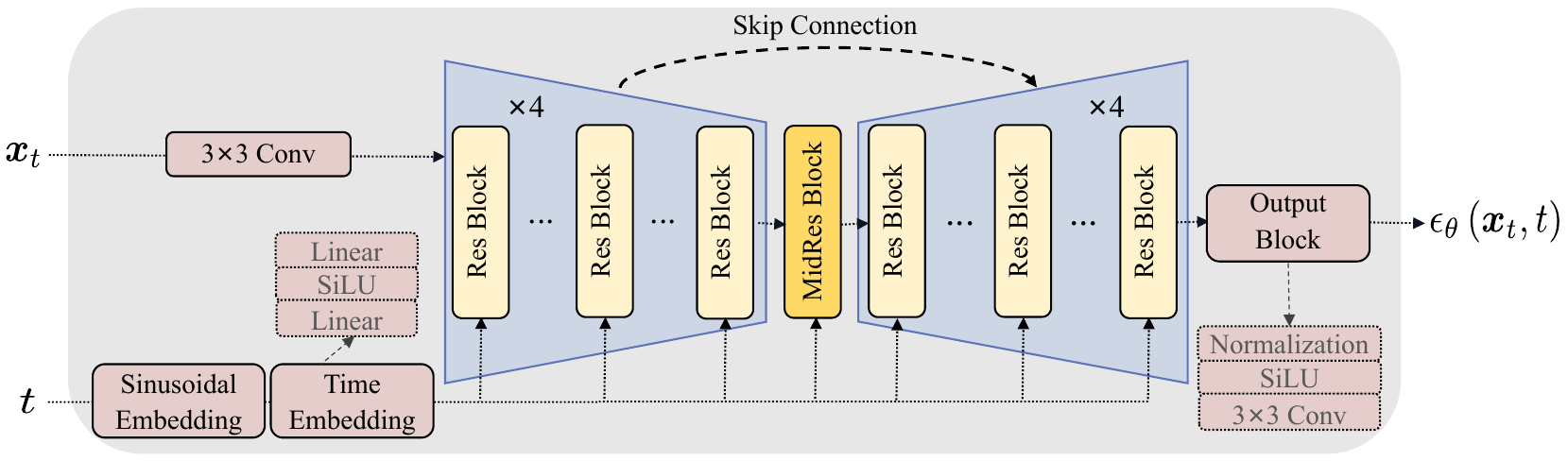}}
  \vspace{-2mm}
  \caption{The overall architecture of the noise matching network with a depth of 4. At each step, the network takes the noise sample $\boldsymbol{x}_{t}$ and its corresponding timestamp $t$ as input and produces the predicted noise $\epsilon_\theta\left(\boldsymbol{x}_t, t\right)$ as output. The main structure of the network is based on U-Net. The inputs $\boldsymbol{x}_{t}$ and $t$ are processed through convolution and time embedding, respectively, to adjust them to the same dimension. They are then fed together into the first layer of the network. The detailed structure of the time embedding and output block is displayed at the location indicated by the dashed arrow lines.}
  \label{fig:Noise Matching Network}
  \vspace{-4mm}
\end{figure*}

\subsection{Sampling Process}\label{Sampling_Process}
Once the training accomplished, sampling $\boldsymbol{x}_{t-1}$ from $p_\theta\left(\boldsymbol{x}_{t-1} \mid \boldsymbol{x}_t\right)$ can be conducted with the following iterative update formula 
\begin{equation}
\boldsymbol{x}_{t-1}=\frac{1}{\sqrt{\alpha_t}}\left(\boldsymbol{x}_t-\frac{1-\alpha_t}{\sqrt{1-\bar{\alpha}_t}} \boldsymbol{\epsilon}_\theta\left(\boldsymbol{x}_t, t\right)\right)+\sigma_t \boldsymbol{z},
\end{equation}
where $\boldsymbol{z} \sim \mathcal{N}(\mathbf{0}, \mathbf{I}) $ $(t>1)$ or $\boldsymbol{z} = \mathbf{0}$ $(t=1)$. 
In the reverse process, as shown in Fig. \ref{fig:seiddpm}, the estimated value of $\boldsymbol{x}_0$ can also be obtained at each time step according to 
\begin{equation}
\label{equ:pre_x0}
\hat{\boldsymbol{x}}_{0, t}=\sqrt{\frac{1}{\bar{\alpha}_t}}\boldsymbol{x}_t -\sqrt{\frac{1-\bar{\alpha}_t}{\bar{\alpha}_t}}\boldsymbol{\epsilon}_\theta\left(\boldsymbol{x}_t, t\right),
\end{equation}
even though it may not be satisfactory during mid-time stamps. 

\section{Methodology}\label{secMethodology}
Our SeisDDIMCR model includes two main stages: DDPM training and DDIM inference. Section \ref{Seismic_DDPM_Training} outlines our training strategy and network structure, and Section \ref{Seismic_DDIM_Inference} describes the detailed operations of the inference process.

\subsection{Seismic DDPM Training}\label{Seismic_DDPM_Training}
\subsubsection{Noise Matching Network}\label{NoiseMatching}
The overall architecture is displayed in Fig. \ref{fig:Noise Matching Network} using stacked residual blocks (Res Block) for the encoder and decoder of U-Net. $\boldsymbol{x}_{t}$ is used as the network input for the denoising learning process to obtain predicted noise $\epsilon_\theta\left(\boldsymbol{x}_t, t\right)$, and the accompanying timestamp $t$ is fed to each layer to embed time information by using the following Transformer sinusoidal time embedding (TE) \cite{vaswani2017attention}
\begin{equation}
\begin{aligned}
T E_{(t, 2 i)} & =\sin \left(t / 10000^{2 i /  d}\right) \\
T E_{(t, 2 i+1)} & =\cos \left(t / 10000^{2 i / d}\right),
\end{aligned}
\end{equation}
where $d$ stands for the dimension of embedding vectors, $t$ is the original time, and $i$ is the dimension. Figuratively speaking, it serves for $\boldsymbol{x}_{t}$ to inform each layer about the current step of reverse diffusion.

\begin{figure}[htbp]
  \centering
  \subfloat[The Res Block and MidRes Block (from left to right).]
  {\includegraphics[width=0.25\textwidth]{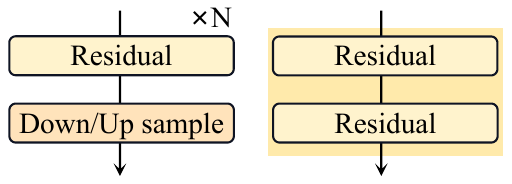}\label{fig:res attention block:subfig1}}
      \hspace{6mm}
  \subfloat[The residual module.]
  {\includegraphics[width=0.15\textwidth]{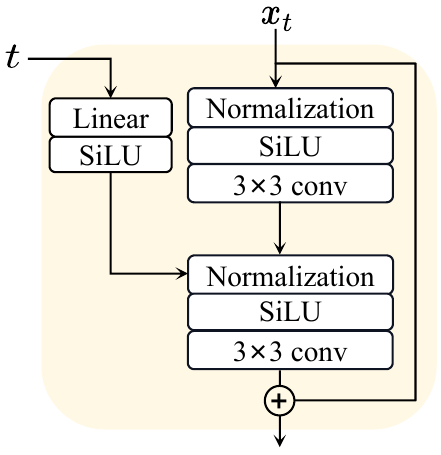}\label{fig:res block:subfig2}}
  \vspace{-1mm}
  \caption{Residual blocks.}
  \vspace{-2mm}
  \label{fig:res attention block}
\end{figure}

Fig. \ref{fig:res attention block:subfig1} displays the detailed components of the Res Block and MidRes Block from left to right, where $\mathrm{N}=2$ for the encoding process and $\mathrm{N}=3$ for the decoding process. Upsampling and downsampling are executed after Res Block, except for the bottom layer, for a total of four operations. As illustrated in Fig. \ref{fig:res block:subfig2}, the residual module is implemented to incorporate temporal information. 

\subsubsection{Cosine Noise Schedule}\label{CosineNoise}
\begin{figure}[htbp]
  \centering
  \subfloat[The changing trend of $\alpha_t$ versus the step of diffusion.]
  {\includegraphics[width=0.235\textwidth]{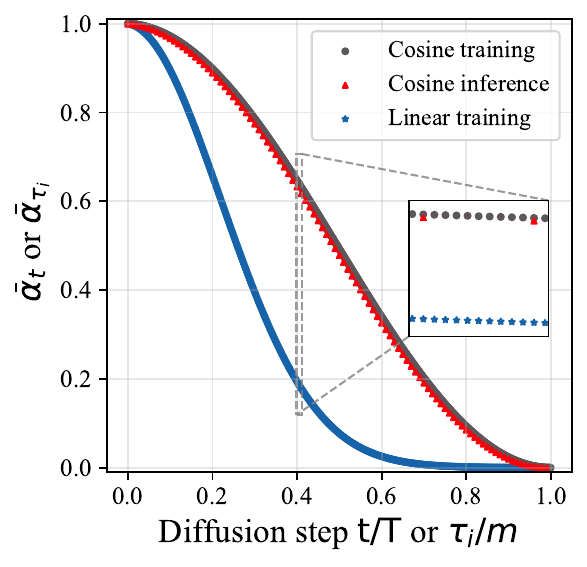}\label{fig:alpha_subfig1}}
      \hspace{1mm}
  \subfloat[The changing trend of $\beta_t$ versus the step of diffusion.]
  {\includegraphics[width=0.235\textwidth]{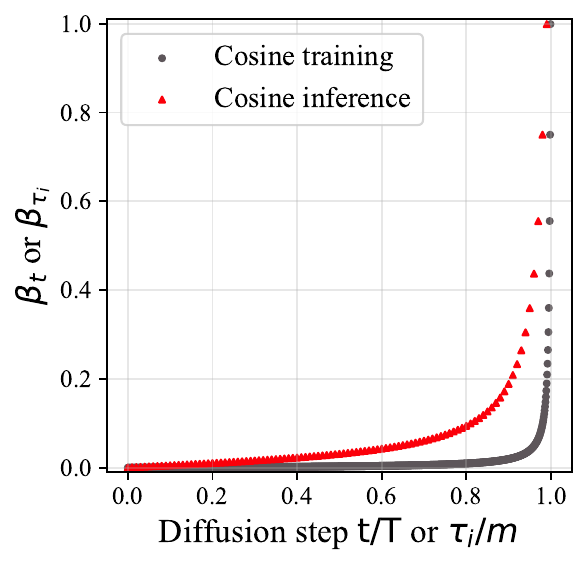}\label{fig:beta_subfig2}}
  \vspace{-1mm}
  \caption{Noise schedules in the training and inference process.}
  \vspace{-2mm}
  \label{fig:Cosine Noise Schedule}
\end{figure}

\begin{figure*}[!htbp]
  \centering
  \subfloat[Linear noise schedule.]
  {\includegraphics[width=1.0\textwidth]{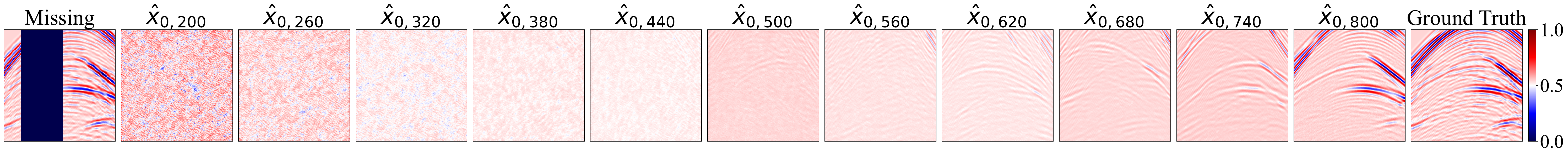}\label{fig:linear_subfig1}}
      \hspace{8mm}
  \subfloat[Cosine noise schedule.]
  {\includegraphics[width=1.0\textwidth]{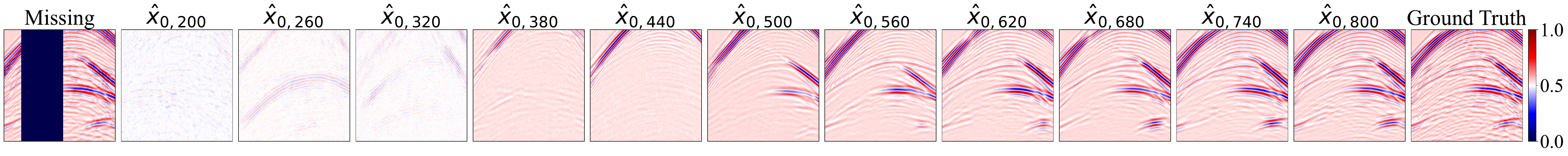}\label{fig:cosine_subfig2}}
  \vspace{-2mm}
  \caption{Seismic interpolation visualization in the reverse diffusion process with different noise schedules.}
  \label{fig:Seismic interpolation Cosine Noise Schedule}
\end{figure*}

DDPM \cite{ho2020denoising} applies the linear noise schedule for $\beta$, where noise increases at a constant rate as the diffusion process proceeds. Since the primary concern in seismic data interpolation is the fidelity of the generated signal, as opposed to diversity, expediting the transition through the stage of high noise can facilitate the reconstruction of unknown areas. We adopt the following cosine schedule \cite{nichol2021improved} 
\begin{equation}
\bar{\alpha}_t=\frac{f(t)}{f(0)}, \quad f(t)=\cos \left(\frac{t / T+s}{1+s} \cdot \frac{\pi}{2}\right)^2,
\end{equation}
where the offset $s=0.008$ is used to prevent $\beta_t$ from being too small near $t=0$. The gray and blue dots in Fig. \ref{fig:alpha_subfig1} display the changing trend of $\bar{\alpha}_t$ in the training process. Compared with the linear noise schedule, the cosine noise schedule can decelerate the global rate of information decay. Meanwhile,  the gray dots in Fig. \ref{fig:beta_subfig2} show the changing trend of $\beta_t$ concerning diffusion steps during the training process. The reduction of the strong noise states is observable, and it can facilitate the learning of seismic signals. To intuitively observe the differences between the generation processes of different noise schedules, Fig. \ref{fig:Seismic interpolation Cosine Noise Schedule} illustrates the seismic data interpolation results $\hat{\boldsymbol{x}}_{0,t}$ at some middle timestamps during the reverse diffusion process. The interpolated content at intermediate timestamps under the linear noise schedule may deviate significantly from the ground truth distribution in Fig. \ref{fig:linear_subfig1}. In contrast, the differences in distribution between each timestamp are much smaller under the cosine noise schedule, as shown in Fig. \ref{fig:cosine_subfig2}. This phenomenon occurs since the cosine noise schedule quickly passes through the high noise phase. Increased availability of known valid information facilitates the generation of missing regions, ensuring consistent alignment between the interpolated content and the ground truth.

\subsubsection{Loss Function}
The log-likelihood can be improved in the log domain by parameterizing the variance $\mathbf{\Sigma}_\theta\left(\boldsymbol{x}_t, t\right) = \sigma_t^2 \mathbf{I}$ with the following interpolation between $\beta_t$ and $\tilde{\beta}_t$ \cite{nichol2021improved}
\begin{equation*}
\mathbf{\Sigma}_\theta\left(\boldsymbol{x}_t, t\right)=\exp \left(v \log \beta_t+(1-v) \log \tilde{\beta}_t\right),
\end{equation*}
where $v$ can be concatenated on another channel of $\boldsymbol{\epsilon}_\theta\left(\boldsymbol{x}_t, t\right)$, serving as the output of the model. Finally, the loss function of our model is set to
\begin{equation}\label{hybridloss}
L_{\text {hybrid}}=L_{\text {simple}}+\lambda_\text {1} L_{\text {vlb}},
\end{equation}
where $L_{\text {simple}}$ and $L_{\text {vlb}}$ are defined in Eq. (\ref{equ:l_simple}) and Eq. (\ref{equ:vlb}), respectively. We follow the setting in \cite{nichol2021improved} and adopt $\lambda_\text {1} = 0.001$ to avoid $L_{\text {vlb}}$ overwhelming $L_{\text {simple}}$.

\begin{figure*}[!t]
  \centering
  {\includegraphics[width=0.85\textwidth]{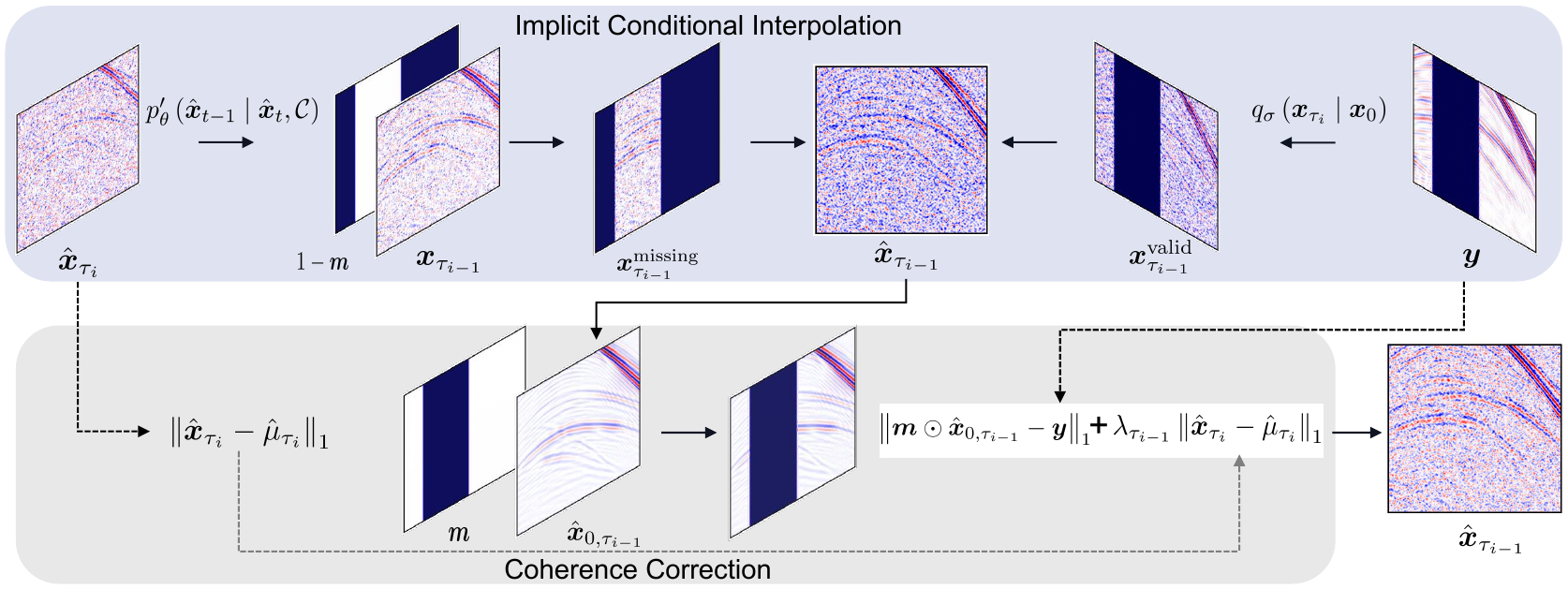}}
  \vspace{-3mm}
  \caption{The pipeline of the implicit conditional interpolation and coherence correction. The prediction result of the $i$-th step $\hat{\boldsymbol{x}}_{ \tau_{i}}$,  missing mask $\boldsymbol{m}$, and revealed seismic traces $\boldsymbol{y}$ are the inputs to the implicit conditional interpolation module and it outputs the preliminary prediction of the $i$-1-th step $\hat{\boldsymbol{x}}_{ \tau_{i-1}}$, integrating the known information. Then, the coherence correction further refines it to obtain the final prediction result of the $i$-1-th step.}
  \label{fig: inference}
  \vspace{-2mm}
\end{figure*}

\subsection{Seismic DDIM Inference}\label{Seismic_DDIM_Inference}
Let $\boldsymbol{x} \in \mathcal{R}^{n_r \times n_t}$ as the original complete seismic data, with $n_r$ and $n_t$ as the number of traces and time samples. The degradation process of observed seismic data can be formally expressed as
\begin{equation*}
\boldsymbol{y}=\boldsymbol{m} \odot \boldsymbol{x}, \quad \text { such that } \quad \boldsymbol{m} \left[ i,: \right]= \begin{cases} \boldsymbol{J}, & i \text { is valid} \\ \boldsymbol{0}, & \text { else }\end{cases}
\end{equation*}
where $\odot$ represents the element-wise multiplication, $\boldsymbol{J}$ is the all-ones matrix, and $\boldsymbol{0}$ denotes the zero matrix. The notation $\boldsymbol{m} \left[ i,: \right]$ indicates the missing mask of $i$-th trace data.
Our SeisDDIMCR model incorporates multiple parameterization processes to achieve stepwise approximation, following the paradigm 
\begin{equation}\label{equ:coherence object}
p_\theta^{\prime}\left(\hat{\boldsymbol{x}}_{0: T} \mid \mathcal{C} \right):=p_\theta^{\prime}\left(\hat{\boldsymbol{x}}_T \mid \mathcal{C}\right) \prod_{t=1}^T p_\theta^{\prime}\left(\hat{\boldsymbol{x}}_{t-1} \mid  \hat{\boldsymbol{x}}_t, \mathcal{C}\right),
\end{equation}
where the latent data $\boldsymbol{x}_{t}$ should obey the interpolation coherence constraint
\begin{equation}\label{equ:coherence constraint}
\mathcal{C}: \boldsymbol{m} \odot \hat{\boldsymbol{x}}_{0} = \boldsymbol{y}. 
\end{equation}
We denote $\hat{\boldsymbol{x}}_{0}$ as the approximation of $\boldsymbol{x}$. 
We decompose this stepwise approximation into three strategies to achieve different interpolation goals one by one, including implicit conditional interpolation, coherence correction, and resampling. 


\subsubsection{Implicit Conditional Interpolation}\label{Inference}
Starting from Gaussian noises, generating target seismic data requires a multi-step DDPM sampling process, incurring a significant computational burden. To address this issue, we adopt the DDIM \cite{song2020denoising} sampling strategy, which avoids the Markov assumption, thereby enhancing computational feasibility and improving interpolation quality.
Intuitively, it seems that the loss function of DDPM ultimately only depends on $q\left(\boldsymbol{x}_{t} \mid \boldsymbol{x}_0\right)$ and the sampling process is only related to  $p\left(\boldsymbol{x}_{t-1} \mid \boldsymbol{x}_t\right)$, from which Song \textit{et al.} \cite{song2020denoising} get inspiration for proposing denoising diffusion implict models (DDIM). They introduce the following non-Markovian inference 
\begin{equation}
q_\sigma\left(\boldsymbol{x}_{1: T} \mid \boldsymbol{x}_0\right):=q_\sigma\left(\boldsymbol{x}_T \mid \boldsymbol{x}_0\right) \prod_{t=2}^T q_\sigma\left(\boldsymbol{x}_{t-1} \mid \boldsymbol{x}_t, \boldsymbol{x}_0\right),
\end{equation}
with a real vector $\sigma = \left(\sigma_1, \ldots, \sigma_T \right) \in \mathbb{R}_{\geq 0}$. They choose 
\begin{equation}
\label{equ:ddim_q}
\begin{aligned}
&q_\sigma\left(\boldsymbol{x}_{t-1} \mid \boldsymbol{x}_t, \boldsymbol{x}_0\right) \\
=&\mathcal{N}\left(\sqrt{\bar{\alpha}_{t-1}} \boldsymbol{x}_0+\sqrt{1-\bar{\alpha}_{t-1}-\sigma_t^2} \cdot \frac{\boldsymbol{x}_t-\sqrt{\bar{\alpha}_t} \boldsymbol{x}_0}{\sqrt{1-\bar{\alpha}_t}}, \sigma_t^2 \boldsymbol{I}\right)
\end{aligned}
\end{equation}
to ensure $q_\sigma\left(\boldsymbol{x}_t \mid \boldsymbol{x}_0\right)$ remains consistent with the form in Eq. (\ref{equ:q_t}). 
Sampling from this non-Markovian generative process is focused on constructing $\sigma$ to improve sample generation and reduce sample steps. Intuitively, the sampling operation can be formulated as 
\begin{equation}
\label{equ:ddim_sampling}
\begin{aligned}
&\boldsymbol{x}_{t-1}=\sqrt{\bar{\alpha}_{t-1}}\left(\frac{\boldsymbol{x}_t-\sqrt{1-\bar{\alpha}_t} \epsilon_\theta\left(\boldsymbol{x}_t, t\right)}{\sqrt{\bar{\alpha}_t}}\right) \\
+ &\sqrt{1-\bar{\alpha}_{t-1}-\sigma_t^2} \cdot \epsilon_\theta\left(\boldsymbol{x}_t, t\right)+\sigma_t\boldsymbol{z},
\end{aligned}
\end{equation}
where the generative process becomes Markovian and equals DDPM if $\sigma_t=\sqrt{\left(1-\bar{\alpha}_{t-1}\right) /\left(1-\bar{\alpha}_t\right)} \sqrt{1-\bar{\alpha}_t / \bar{\alpha}_{t-1}}$ for all $t$. We can consider a sampling process of length less than $T$ when $q_\sigma\left(\boldsymbol{x}_t \mid \boldsymbol{x}_0\right)$ is fixed since the optimization result of DDPM essentially contains its optimization results for arbitrary subsequence parameters. Denoting the increasing time subsequence of the original time sequence $[1, \ldots, T]$ as $\boldsymbol{\tau}=\left[\tau_1, \tau_2, \ldots, \tau_{m}\right]$ with of length $m$ (the corresponding changes in $\bar{\alpha}_{\tau_{i}}$ and $\beta_{\tau_{i}}$ are shown in the red points of Figs. \ref{fig:alpha_subfig1} and \ref{fig:beta_subfig2}, respectively), the $\sigma_{\tau}$ used in accelerated sampling process follows
\begin{equation}
\sigma_{\tau_i}(\eta)=\eta \sqrt{\left(1-\bar{\alpha}_{\tau_{i-1}}\right) /\left(1-\bar{\alpha}_{\tau_i}\right)} \sqrt{1-\bar{\alpha}_{\tau_i} / \bar{\alpha}_{\tau_{i-1}}},
\end{equation}
where $\eta \geq 0$. In particular, the generative process is defined as DDIM if $\eta = 0$ for all $t$ since the variance $\sigma$ keeps zero, so that the deterministic forward process becomes an implicit probabilistic model.

Seismic data interpolation is a conditional generation task where unrevealed areas are inferred from the known signals. The DDIM sampling strategy, i.e., Eq. (\ref{equ:ddim_sampling}), has shortened the sampling length yet remains an unconditional process. Thus, we employ the approach in \cite{lugmayr2022repaint} to incorporate revealed traces into $\boldsymbol{x}_t$ and guide DDIM sampling using known information. Fig. \ref {fig: inference} illustrates the operational process of our implicit conditional interpolation.
Each step of the iterative reverse diffusion stage in the inference process uses the following implicit conditional interpolation formula
\begin{equation}
\label{equ:resampling}
\hat{\boldsymbol{x}}_{\tau_{i-1}}=\boldsymbol{m} \odot \boldsymbol{x}_{\tau_{i-1}}^{\text {valid}}+(1-\boldsymbol{m}) \odot \boldsymbol{x}_{\tau_{i-1}}^{\text {missing}},
\end{equation}
where $\boldsymbol{x}_{\tau_{i-1}}^{\text {valid}}$ is directly sampled from the forward diffusion process, i.e., Eq. (\ref{equ:q_t}), which adds known information to the reverse process, and $\boldsymbol{x}_{\tau_{i-1}}^{\text {missing}}$ is obtained by using the DDIM sampling formula Eq. (\ref{equ:ddim_sampling}). As a result, $\hat{\boldsymbol{x}}_{\tau_{i-1}}$ incorporates information from both known signals and model-predicted signals before forwarding it to the next reverse diffusion step. The added noisy known traces will guide the unrevealed parts closer to the target distribution. The recovery of missing seismic data is designed as an implicit conditional interpolation process based on valid seismic data. 

\subsubsection{Coherence Correction}
Eq. (\ref{equ:coherence object}) defines the optimization objectives for the inference process, and it can be formulated by taking the logarithm 
\begin{equation}\label{equ:coherence object logarithm}
\begin{aligned}
& \log p_\theta^{\prime}\left(\hat{\boldsymbol{x}}_{\tau_{1}:\tau_{m}} \mid \mathcal{C} \right) \\
=&\log p_\theta^{\prime}\left(\hat{\boldsymbol{x}}_{\tau_{m}} \mid \mathcal{C}\right) + \sum_{i=1}^m \log p_\theta^{\prime}\left(\hat{\boldsymbol{x}}_{\tau_{i-1}} \mid  \hat{\boldsymbol{x}}_{\tau_{i}}, \mathcal{C}\right),
\end{aligned}
\end{equation}
where 
\begin{equation*}\label{equ:logarithm xT}
\begin{aligned}
&\log p_\theta^{\prime}\left(\hat{\boldsymbol{x}}_{\tau_{m}} \mid \mathcal{C}\right) \\
= &  \log p_\theta\left(\hat{\boldsymbol{x}}_{\tau_{m}}\right) + \log \left(p_\theta^{\prime}\left(\boldsymbol{m} \odot\hat{\boldsymbol{x}}_0=\boldsymbol{y} \mid \hat{\boldsymbol{x}}_{\tau_{m}}\right)\right) +C,
\end{aligned}
\end{equation*}
and
\begin{equation*}\label{equ:logarithm xt}
\begin{aligned}
&\log p_\theta^{\prime}\left(\hat{\boldsymbol{x}}_{\tau_{i-1}} \mid  \hat{\boldsymbol{x}}_{\tau_{i}}, \mathcal{C}\right) \\
= &  \log p_\theta\left(\hat{\boldsymbol{x}}_{\tau_{i-1}} \mid \hat{\boldsymbol{x}}_{\tau_{i}}\right) + \log \left(p_\theta^{\prime}\left(\boldsymbol{m} \odot\hat{\boldsymbol{x}}_0=\boldsymbol{y} \mid \hat{\boldsymbol{x}}_{\tau_{i-1}}\right)\right) +C.
\end{aligned}
\end{equation*}
Eq. (\ref{equ:pre_x0}) demonstrates that one-step estimation of $\hat{\boldsymbol{x}}_0$ from $\hat{\boldsymbol{x}}_t$ is tractable. Although this estimate may not be sufficiently accurate, the task can be accomplished through stepwise sampling to ensure successive coherence corrections as   
\begin{equation}\label{equ:seismic coherence constraint}
\mathcal{C}: \boldsymbol{m} \odot \hat{\boldsymbol{x}}_{0,\tau_{i}} \to \boldsymbol{y}, \quad \forall i \in [1, \ldots, m],
\end{equation}
where $\hat{\boldsymbol{x}}_{0,\tau_{i}}$ is calculated by Eq. (\ref{equ:pre_x0}).
We follow the final optimization formulas and the greedy optimization procedure provided in \cite{zhang2023towards} to maximize Eq. (\ref{equ:coherence object logarithm}). They use the theoretical L2 norm, yet in practice we found that the L1 norm can achieve better recovery quality in seismic data interpolation.
First, after sampling $\boldsymbol{x}_{\tau_{m}}$, we correct $\boldsymbol{x}_{\tau_{m}}$ by minimizing 
\begin{equation}\label{Coherence Correction T}
\left\|y-\boldsymbol{m} \odot\hat{\boldsymbol{x}}_{0, \tau_{m}}\right\|_1 + \lambda_{\tau_{m}}\left\|\hat{\boldsymbol{x}}_{\tau_{m}}\right\|_1, 
\end{equation}
where $\lambda_{\tau_{m}}$ is the weight parameter. Then, we gradually perform implicit conditional interpolation from $\boldsymbol{x}_{\tau_{i}}$  to $\boldsymbol{x}_{\tau_{i-1}}$ and then conduct optimization as 
\begin{equation}\label{Coherence Correction t}
 \left\|y-\boldsymbol{m} \odot\hat{\boldsymbol{x}}_{0, \tau_{i-1}}\right\|_1 + \lambda_{\tau_{i-1}}\left\|\hat{\boldsymbol{x}}_{\tau_{i-1}}-\hat{\mu}_{\tau_{i}}\right\|_1, 
\end{equation}
where 
\begin{equation*}
\hat{\mu}_{\tau_{i}} = \sqrt{\alpha_{\tau_{i-1}}} \hat{\boldsymbol{x}}_{0, \tau_{i}}+\sqrt{1-\alpha_{\tau_{i-1}}-\sigma_{\tau_{i}}^2} \cdot \frac{\hat{\boldsymbol{x}}_{\tau_{i}}-\sqrt{\alpha_{\tau_{i}}} \hat{\boldsymbol{x}}_{0, \tau_{i}}}{\sqrt{1-\alpha_{\tau_{i}}}}.
\end{equation*}
The values from $\lambda_{\tau_{m}}$ to $\lambda_{\tau_{1}}$ are gradually increased to incrementally raise the weight of the DDIM prior regularization. It should be noted that both of the aforementioned gradient descent can be performed multiple times, denoted by $G$. As $t$ decreases, the estimation error of $\hat{\boldsymbol{x}}_{0, \tau_{i}}$ will decrease, allowing the optimization process to gradually correct the interpolation error with increasing accuracy. To demonstrate this correction process more clearly, Fig. \ref{fig: inference} visualizes the calculation flow of the correction loss. Upon completion of the condition generation, the coherence constraint optimizes the distribution of the interpolated traces.

\subsubsection{Resampling}
Conditional interpolation and coherence correction have advanced the interpolation process, bringing it closer to the target distribution. However, relying solely on the known signal as the condition is insufficient. Self-consistency and continuity still require further enhancement. Thus, we introduce the resampling strategy \cite{lugmayr2022repaint}. 
After sampling $\boldsymbol{x}_{\tau_{i-1}}$ in the inverse diffusion process, the forward diffusion sampling is performed again to generate $\boldsymbol{x}_{\tau_{i}}$, with the difference being that $\boldsymbol{x}_{\tau_{i}}$ now contains the information from $\boldsymbol{x}_{\tau_{i-1}}^{\text {missing}}$, thereby promoting consistency with known signals. Naturally, this kind of resampling operation cannot be performed only once. We define the travel length, denoted as $L$, to set how many times to backtrack for each resampling process, and we define the travel height, denoted as $H$, which determines the interval between time steps before and after two different resampling processes. Although resampling extends inference time, our experiments will show that combining it with coherence correction achieves a better trade-off between computational efficiency and interpolation quality.  

The seismic DDPM training and DDIM interpolation are two key processes in our SeisDDIMCR model.
Algorithm \ref{alg:SeisDDIMR_training} and Algorithm \ref{alg:SeisDDIMR_interpolation} list the overview of our training and inference procedure, respectively.
\begin{algorithm}[htbp]
\caption{Training Seismic DDPM}
\label{alg:SeisDDIMR_training}
\hspace*{0.02in} {\bf Input:} Complete training data $\{\boldsymbol{x}_0^i\}_{i=1}^n$ with total number $n$;\\
Specifying the parameters of DDPM, i.e., diffusion steps $T$;
Batch size $K$; The number of iterations $N$.
\begin{algorithmic}[1]
\State Randomly initialize the noise matching network;
\For{$j=1, \ldots, N$}
    \State Sample batch data $\{x_0^i\}_{i=1}^K$ from training data;
    \State Sample $\{t_i\}_{i=1}^K$ from $\operatorname{Uniform}(\{1, \ldots, T\})$;
    \State Sample $\{\epsilon_{t_i}\}_{i=1}^K$ from $\mathcal{N}(\mathbf{0}, \mathbf{I})$;
    \State Get $\left\{\epsilon_\theta\left(\mathbf{x}_{t_i}^i, t_i\right)\right\}_{i=1}^K$ from the noise matching \hspace*{0.243in}network;
    \State Update the noise matching network with $ L_\text{hybrid}$;\\
\textbf{end for}
\EndFor
\end{algorithmic}
\end{algorithm}

\begin{algorithm}[htbp]
\caption{Implicit Conditional Interpolation with Coherence-corrected Resampling}
\label{alg:SeisDDIMR_interpolation}
\hspace*{0.02in} {\bf Input:}
Missing seismic data $\boldsymbol{x}_0$;
Corresponding missing mask $\boldsymbol{m}$;
Trained seismic DDPM model; Diffusion sampling steps $\boldsymbol{\tau}=\left[\tau_1, \tau_2, \ldots, \tau_{m}\right]$ with length of $m$;
Travel length $L$; Travel height $H$; Gradient descent number of coherence correction $G$.
\begin{algorithmic}[1]
\State $\boldsymbol{x}_{\tau_{m}}^{\text {missing}} \sim \mathcal{N}(\mathbf{0}, \mathbf{I})$;
\State Sample $\boldsymbol{x}_{\tau_{m}}^{\text {valid}}$ from Eq. (\ref{equ:q_t});
\State Get $\boldsymbol{x}_{\tau_{m}}$ from Eq. (\ref{equ:resampling});
\State $\tau_{i}=\tau_{m}$;
\While {$\tau_{i} >\tau_{1}$}
    \For{$h=1, \ldots, H$}
        \State $\epsilon \sim \mathcal{N}(\mathbf{0}, \mathbf{I})$ if $\tau_{i}>\tau_{1}$, else $\epsilon=\mathbf{0}$; 
        \State Sample $\boldsymbol{x}_{\tau_{i-1}}^{\text {valid}}$ from Eq. (\ref{equ:q_t});
        \State$ \boldsymbol{z} \sim \mathcal{N}(\mathbf{0}, \mathbf{I})$ if $\tau_{i}>\tau_{1}$, else $\boldsymbol{z}=\mathbf{0}$;
        \State Get $\boldsymbol{x}_{\tau_{i-1}}^{\text {missing}}$ from Eq. (\ref{equ:ddim_sampling});
        \State Get $\hat{\boldsymbol{x}}_{\tau_{i-1}}$ from Eq. (\ref{equ:resampling}), $\tau_{i} =\tau_{i-1}$;
        \State Correct $\hat{\boldsymbol{x}}_{\tau_{i-1}}$ to minimize Eq. (\ref{Coherence Correction T}) and Eq. (\ref{Coherence Correction t}) \Statex\hspace{\algorithmicindent}\hspace{\algorithmicindent}by G-step gradient descent;
        \EndFor
        \State \textbf{end for} 
    \If{$\tau_{i} >\tau_{1}$}
        \For {$l=1, \ldots, L-1$}
            \State Repeat 6-12;
            \For{$h=1, \ldots, H$}
                \State Get $\hat{\boldsymbol{x}}_0$ from Eq. (\ref{equ:pre_x0}), $\tau_{i} =\tau_{i+1}$;
                \State Sample $\hat{\boldsymbol{x}}_{\tau_{i}}$ from Eq. (\ref{equ:q_t}), where $\boldsymbol{x}_0 = \hat{\boldsymbol{x}}_0$;
            \EndFor
            \State \hspace*{0.01in}\textbf{end for} 
        \EndFor 
        \State \hspace*{0.01in}\textbf{end for}
    \EndIf 
    \State \hspace*{0.01in}\textbf{end if}
\EndWhile
\State \textbf{end while}
\end{algorithmic}
\hspace*{0.02in} {\textbf{Output:} Interpolated data $\boldsymbol{x}_{\tau_{1}}$.}
\end{algorithm}

\section{Experiments}\label{secExperiments}

\subsection{Evaluation Metrics}
We choose three metrics, i.e., MSE, signal-to-noise ratio (SNR), and peak signal-to-noise ratio (PSNR), to compare the fidelity of the interpolated seismic data. MSE between the interpolated seismic data  $\{\hat{\boldsymbol{x}}^j\}_{j=1}^n$ and the ground truth $\{\boldsymbol{x}^j_\text{gt}\}_{j=1}^n$ is calculated using 
\begin{equation}
\mathrm{MSE}=\frac{1}{n}\sum_{j=1}^n\|{\hat{\boldsymbol{x}}^j-\boldsymbol{x}^j_\text{gt}}\|^{2}_{F},
\end{equation}
where $\|\cdot\|_F$ represents the Frobenius norm. Its value closer to 0 implies a higher fidelity of the interpolation result. The SNR for a single interpolated sample is defined as
\begin{equation}
\mathrm{SNR}=10\log_{10}\frac{\|\boldsymbol{x}_\text{gt}\|^{2}_{F}}{\|\boldsymbol{x}_\text{gt}-\hat{\boldsymbol{x}}\|^{2}_{F}}.
\end{equation}
PSNR is calculated by the following formula as
\begin{equation}
\mathrm{PSNR}=10\log_{10}\frac{\mathrm{MAX}_{\boldsymbol{x}_\text{gt}}^2}{\mathrm{MSE}},
\end{equation}
where $\mathrm{MAX}_{\boldsymbol{x}_\text{gt}}$ refers to the highest value of $\boldsymbol{x}_\text{gt}$. Obviously, larger SNR and PSNR both symbolize higher interpolation fidelity.
The quality of the texture of the interpolation is evaluated using structural similarity (SSIM) \cite{zhou1284395}, which is widely used in the field of image generation following the formula
\begin{equation}
\begin{aligned}
\operatorname{SSIM}(\boldsymbol{x}_\text{gt}, \hat{\boldsymbol{x}}) & =L(\boldsymbol{x}_\text{gt}, \hat{\boldsymbol{x}}) \cdot C(\boldsymbol{x}_\text{gt}, \hat{\boldsymbol{x}}) \cdot S(\boldsymbol{x}_\text{gt}, \hat{\boldsymbol{x}}).
\end{aligned}
\end{equation}
Separately, $L(\cdot)$, $C(\cdot)$, and $S(\cdot)$ indicate similarities in luminance, contrast, and structure, and they are each defined as
\begin{equation*}
L({\boldsymbol{x}_\text{gt}}, \hat{\boldsymbol{x}})=\frac{2 \mu_{\boldsymbol{x}_\text{gt}} \mu_{\hat{\boldsymbol{x}}}+c_1}{\mu_{\boldsymbol{x}_\text{gt}}^2+\mu_{\hat{\boldsymbol{x}}}^2+c_1}, 
\end{equation*}
\begin{equation*}
C({\boldsymbol{x}_\text{gt}}, \hat{\boldsymbol{x}})=\frac{2 \sigma_{\boldsymbol{x}_\text{gt}} \sigma_{\hat{\boldsymbol{x}}}+c_2}{\sigma_{\boldsymbol{x}_\text{gt}}^2+\sigma_{\hat{\boldsymbol{x}}}^2+c_2}, 
\end{equation*}
\begin{equation*}
S({\boldsymbol{x}_\text{gt}}, \hat{\boldsymbol{x}})=\frac{\sigma_{{\boldsymbol{x}_\text{gt}} {\hat{\boldsymbol{x}}}}+c_3}{\sigma_{\boldsymbol{x}_\text{gt}} \sigma_{\hat{\boldsymbol{x}}}+c_3},
\end{equation*}
where $\mu_{\boldsymbol{x}_\text{gt}}(\mu_{\hat{\boldsymbol{x}}})$, $\sigma_{\boldsymbol{x}_\text{gt}}(\sigma_{\hat{\boldsymbol{x}}})$, and $\sigma_{{\boldsymbol{x}_\text{gt}} \hat{\boldsymbol{x}}}$ denote the mean value and standard deviation, and covariance, respectively. Constants $c_1$, $c_2$, and $c_3$ are typically set close to zero to prevent numerical instability. Thus, a higher SSIM implies a more similar texture.

\subsection{Data Set}
We validate our method over one open synthetic dataset provided by the Society of Exploration Geophysicists (SEG) C3 and one field dataset Mobil Avo Viking Graben Line 12 (MAVO), extensively used in seismic interpolation \cite{Yu9390348} and denoising \cite{yu2019deep}, \cite{Meng9775677}. The SEG C3 dataset consists of 45 shots, each with a 201$\times$201 receiver grid, 625 time samples per trace, and a sampling interval of 8 ms. 
For each shot, we randomly select 20 slices along the inline for validation and testing, while the remaining 161 slices are utilized to create the training set. For each slice, we extract the first 2.4 s of records and then randomly sample patches with dimensions of 128$\times$128 in both time and trace. In total, we generate 30,000 patches for training, 6,000 for validation, and 6,000 for testing.
MAVO dataset comprises a 1001$\times$120 receiver grid with 1500 time samples per trace. It is collected at a time rate of 4 ms and a spatial rate of 25 m. We intercept records within a 4.0 s window, select 100 slices each for validation and testing, and allocate the remaining 801 slices for training. Subsequently, we randomly extract patches with dimensions of 256$\times$112 for both time and trace. The final counts of patches for the training, validation, and test sets are 20,000, 4,000, and 4,000, respectively. All seismic patches are first normalized within the interval $[0,1]$ by applying min-max normalization. 
\subsection{Implementation Details}
The diffusion step for the Seismic DDPM model is set to 1000. 
We train the seismic DDPM model on the training sets of SEG C3 and MAVO separately, as described in Algorithm \ref{alg:SeisDDIMR_training}, with N iterations of 300,000. The noise matching network is optimized by AdamW with a learning rate of 1e-4. The batch size is set to 50 for the SEG C3 dataset and the MAVO dataset. Our SeisDDIMCR test is conducted by using Algorithm \ref{alg:SeisDDIMR_interpolation}, where we adopt diffusion sampling step $m$=100, travel length $L$=2, and travel height $H$=1. The gradient descent number $G$ is set to 1 for the SEG C3 dataset and to 2 for the MAVO dataset, respectively. $\lambda_{\tau_{m}}$ to $\lambda_{\tau_{1}}$ starts at 1e-4 and increases by a factor of 1.01. We compare our experimental results with 5 currently popular methods, including DD-CGAN \cite{Chang2020}, cWGAN-GP \cite{wei2022big}, PConv-UNet \cite{pan2020partial}, ANet \cite{Yu9390348}, and Coarse-to-Fine \cite{wei2022hybrid}. All of the experiments are implemented using Pytorch 1.12.1 and NVIDIA A100 Tensor Core GPU.

\subsection{Experimental Results}\label{section:Experimental Results}
We conduct Algorithm \ref{alg:SeisDDIMR_interpolation} to accomplish our model testing. Interpolation reconstructions are performed on three missing categories of seismic data, and the experimental results are displayed below, followed by a comparison to other methods. It is worth noting that our SeisDDIMCR model is trained only once on each dataset, whereas other comparison methods are trained multiple times according to various trace missing forms. The details of the model parameters remain consistent with their respective original papers. To ensure fairness in comparison, we strive to achieve the best possible training results for each model by using various training techniques.

\subsubsection{Random Missing Traces}
\begin{figure*}[!htbp]
\vspace{-4mm}	
  \centering
  \subfloat[DD-CGAN.]
  {\includegraphics[height=0.13\textheight]{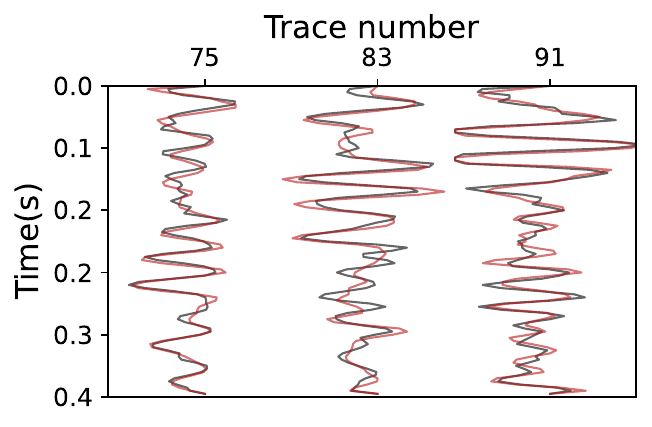}\label{fig:mavorandom_subfig1}\hspace{6.0mm}}
  \subfloat[cWGAN-GP.]
  {\includegraphics[height=0.13\textheight]{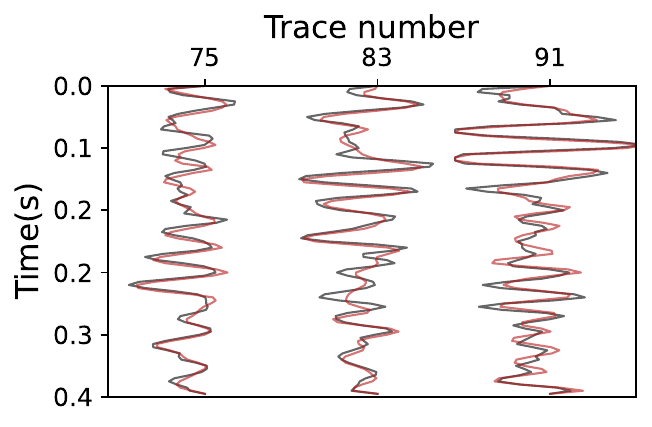}\label{fig:mavorandom_subfig2}\hspace{6.0mm}}
  \subfloat[PConv-UNet.]
  {\includegraphics[height=0.13\textheight]{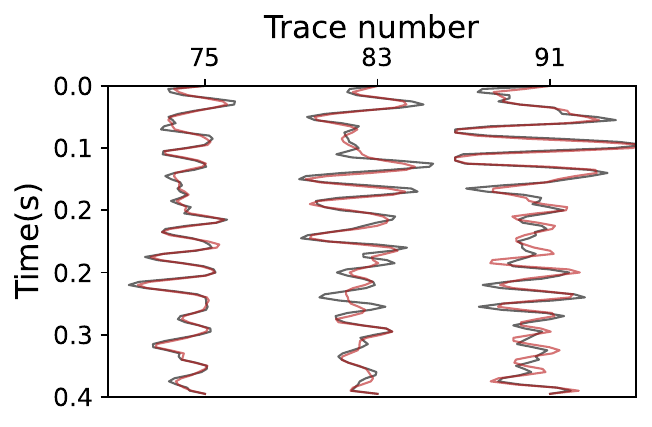}\label{fig:mavorandom_subfig3}}
         \clearpage
  \subfloat[ANet.]
  {\includegraphics[height=0.13\textheight]{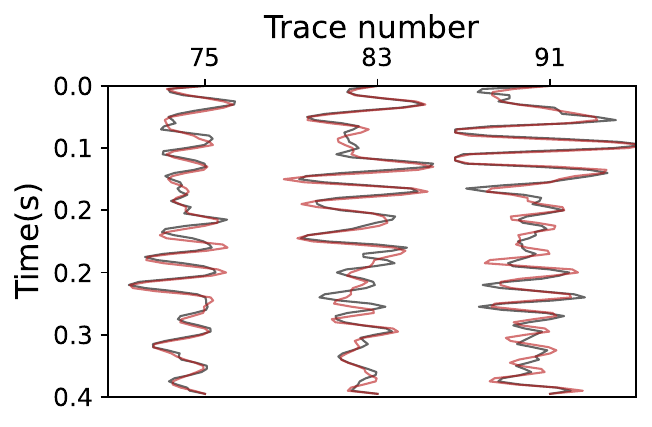}\label{fig:mavorandom_subfig4}\hspace{6.0mm}}
  \subfloat[Coarse-to-Fine.] 
    {\includegraphics[height=0.13\textheight]{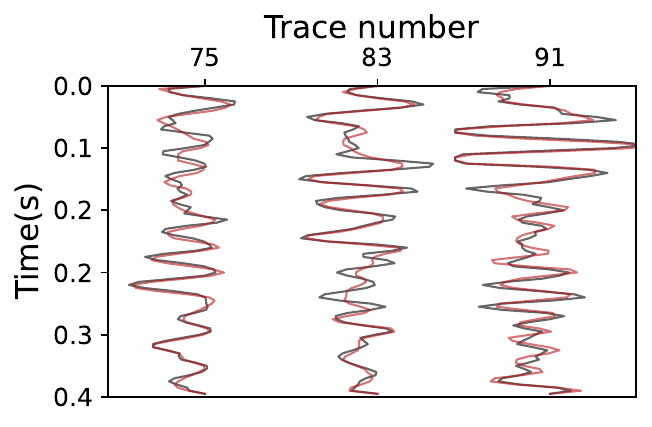}\label{fig:mavorandom_subfig5}\hspace{6.0mm}}
  \subfloat[Ours.]
    {\includegraphics[height=0.13\textheight]{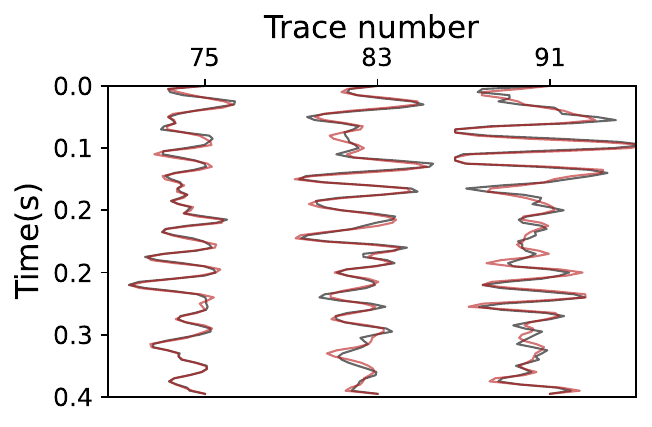}\label{fig:mavorandom_subfig6}}
  
  \caption{Interpolation results of MAVO test data with random missing traces on different methods. Several randomly missing traces are chosen in wiggle plots to demonstrate the performance of interpolation, where the red and black wiggly lines represent the interpolation result and the ground truth, respectively.}
  \label{fig:mavo random}
\end{figure*}

\begin{figure*}[!htbp]
  \centering
    \subfloat[Ground Truth.]
  {\includegraphics[height=0.151\textwidth]{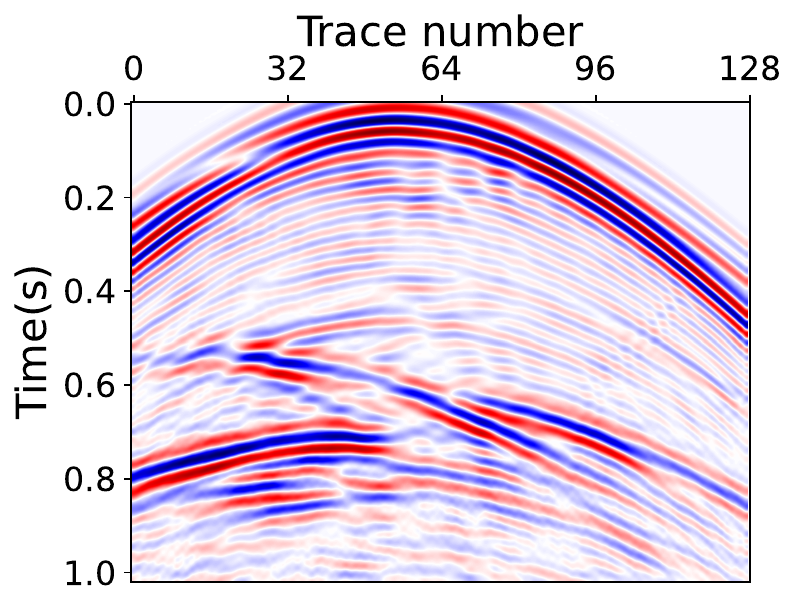}\label{fig:SEGC3regular_subfig1}\hspace{1.0mm}}
  \subfloat[Regular missing data.]
  {\includegraphics[height=0.151\textwidth]{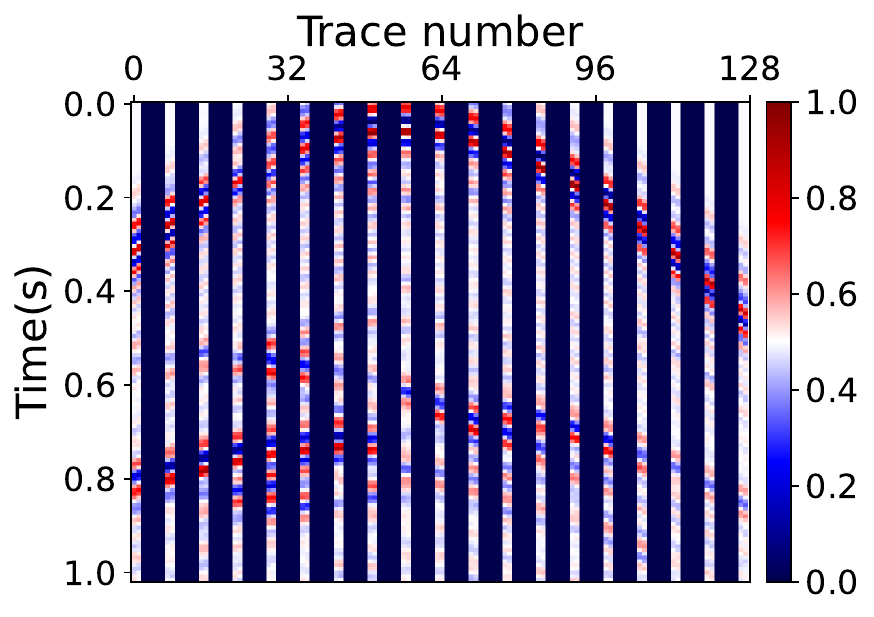}\label{fig:SEGC3regular_subfig2}}
  \subfloat[$f\text{-}k$ spectra of (a).]
  {\includegraphics[height=0.151\textwidth]{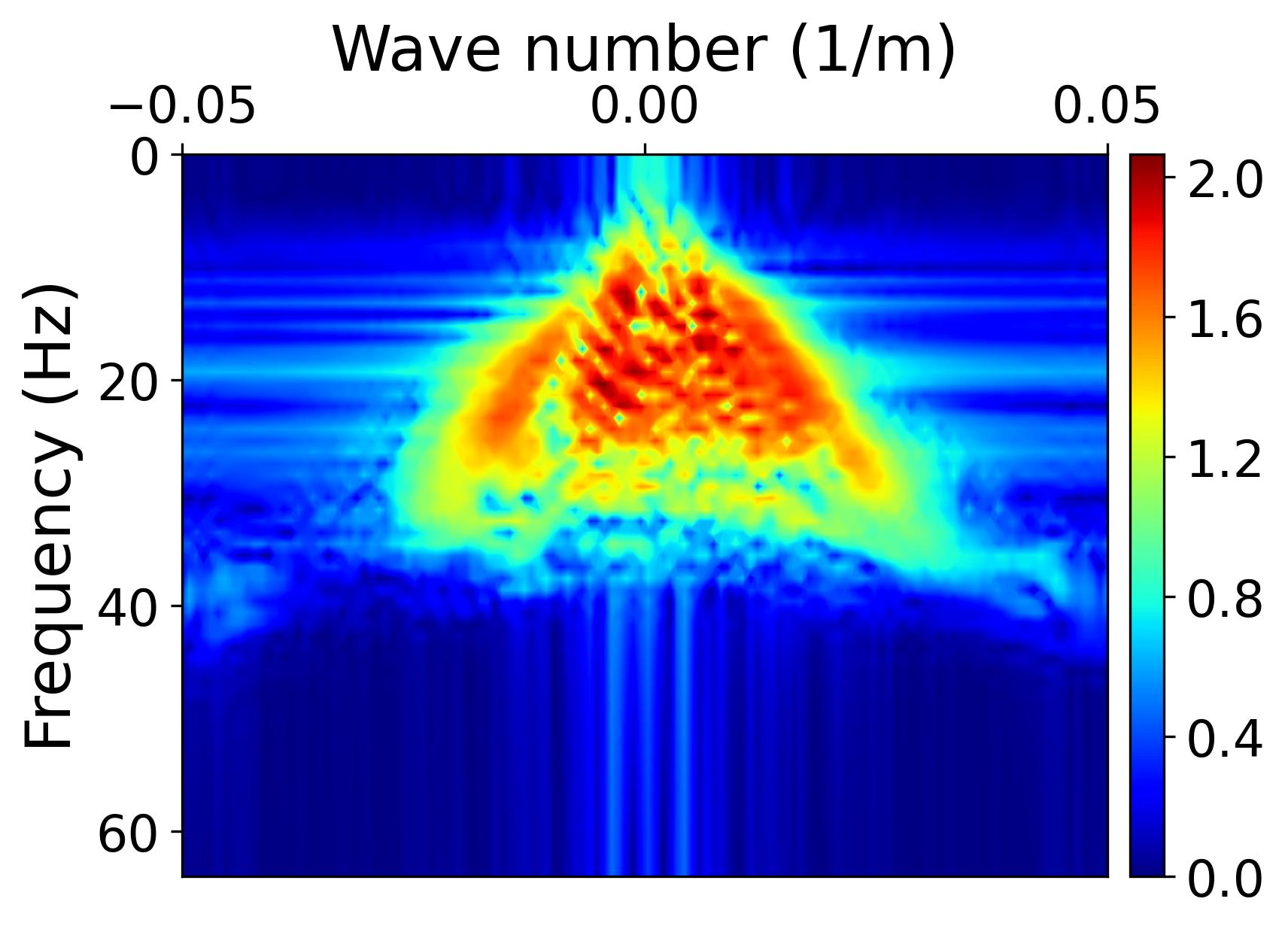}\label{fig:SEGC3regular_subfig3}}
  \subfloat[$f\text{-}k$ spectra of (b).]
  {\includegraphics[height=0.151\textwidth]{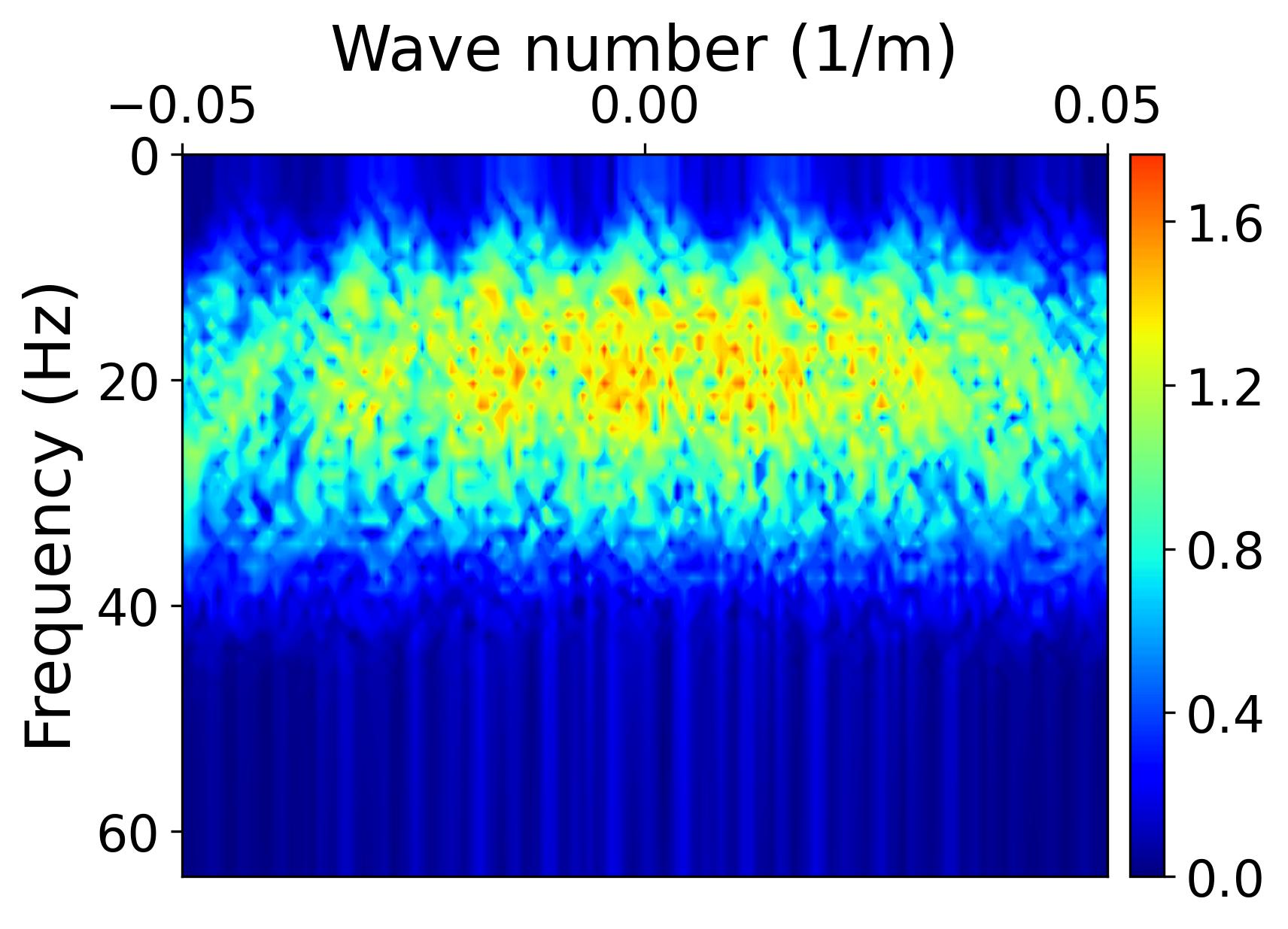}\label{fig:SEGC3regular_subfig4}}
  \subfloat[DD-CGAN.]
  {\includegraphics[height=0.151\textwidth]{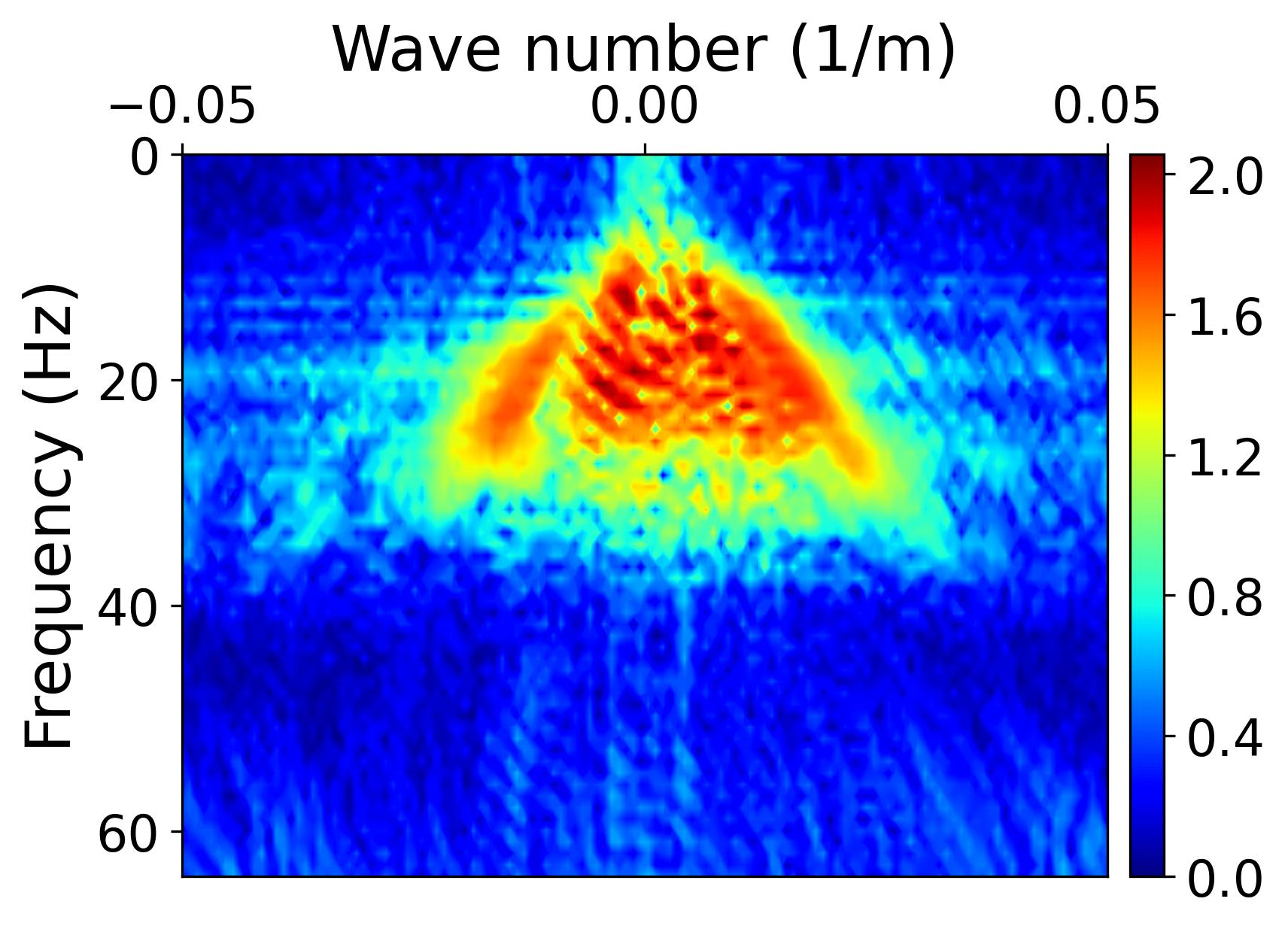}\label{fig:SEGC3regular_subfig5}}
    \clearpage
  \subfloat[cWGAN-GP.]
  {\includegraphics[height=0.151\textwidth]{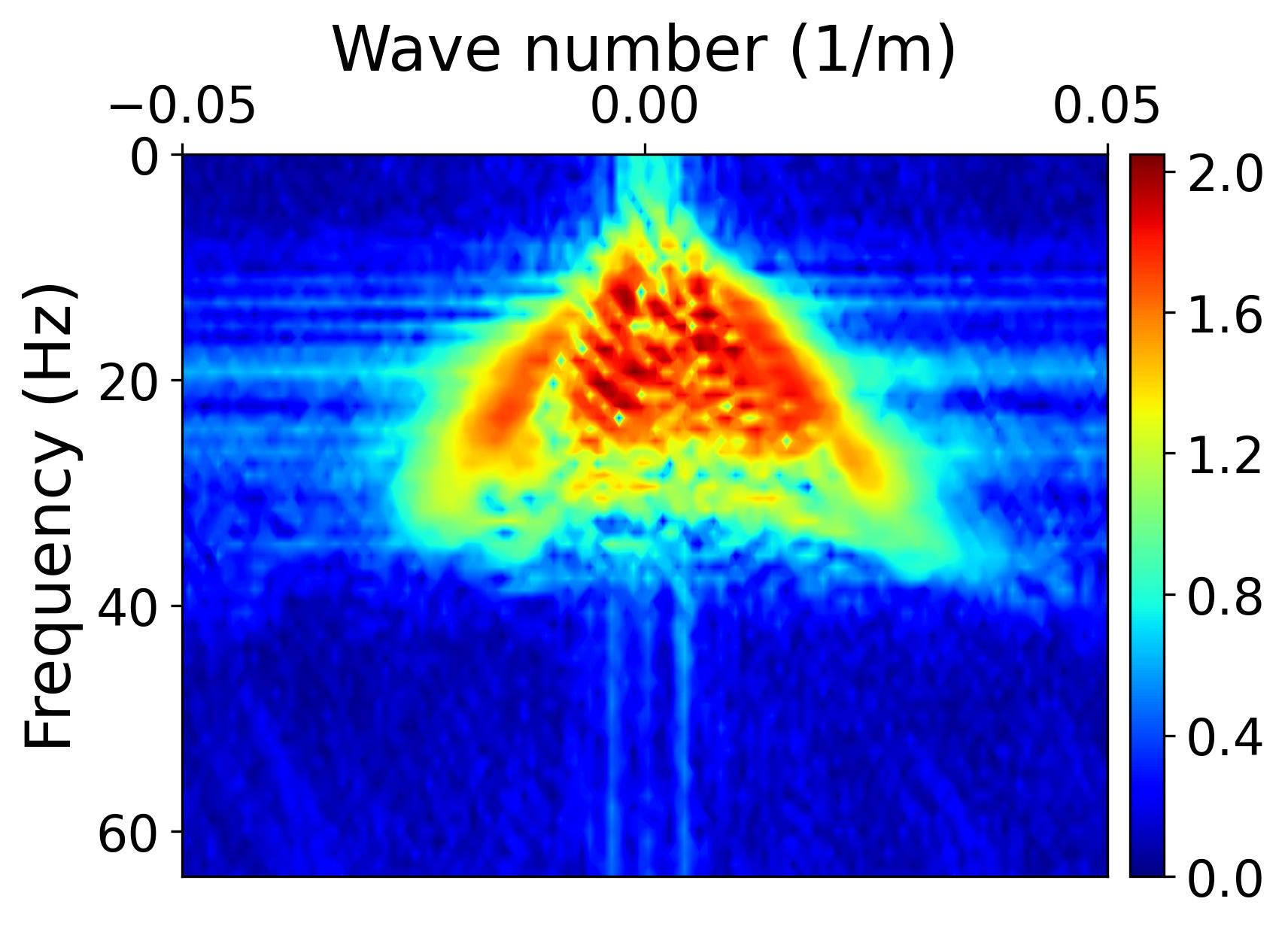}\label{fig:SEGC3regular_subfig6}}
  \subfloat[PConv-UNet.]
  {\includegraphics[height=0.151\textwidth]{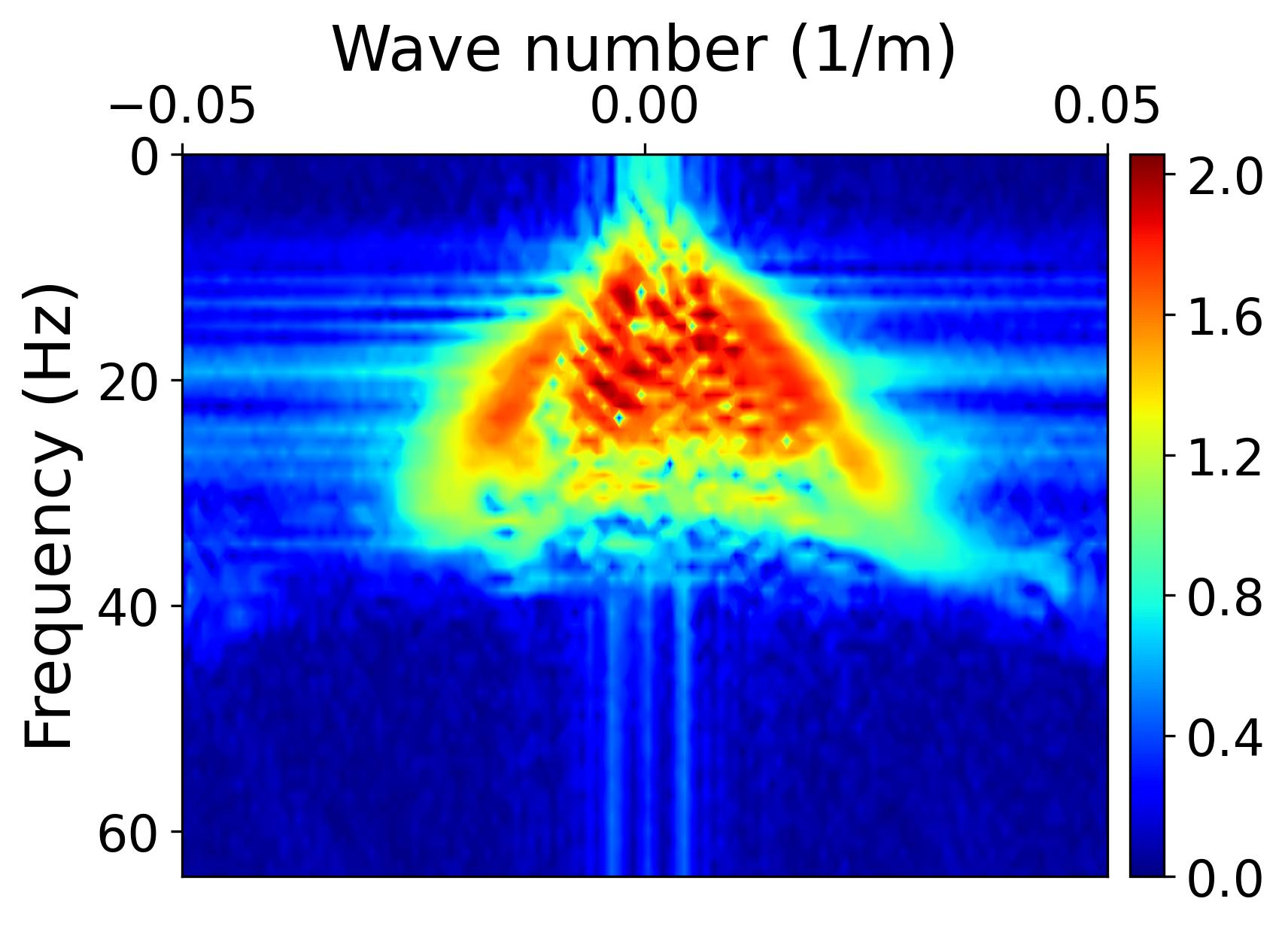}\label{fig:SEGC3regular_subfig7}}
  \subfloat[ANet.]
  {\includegraphics[height=0.151\textwidth]{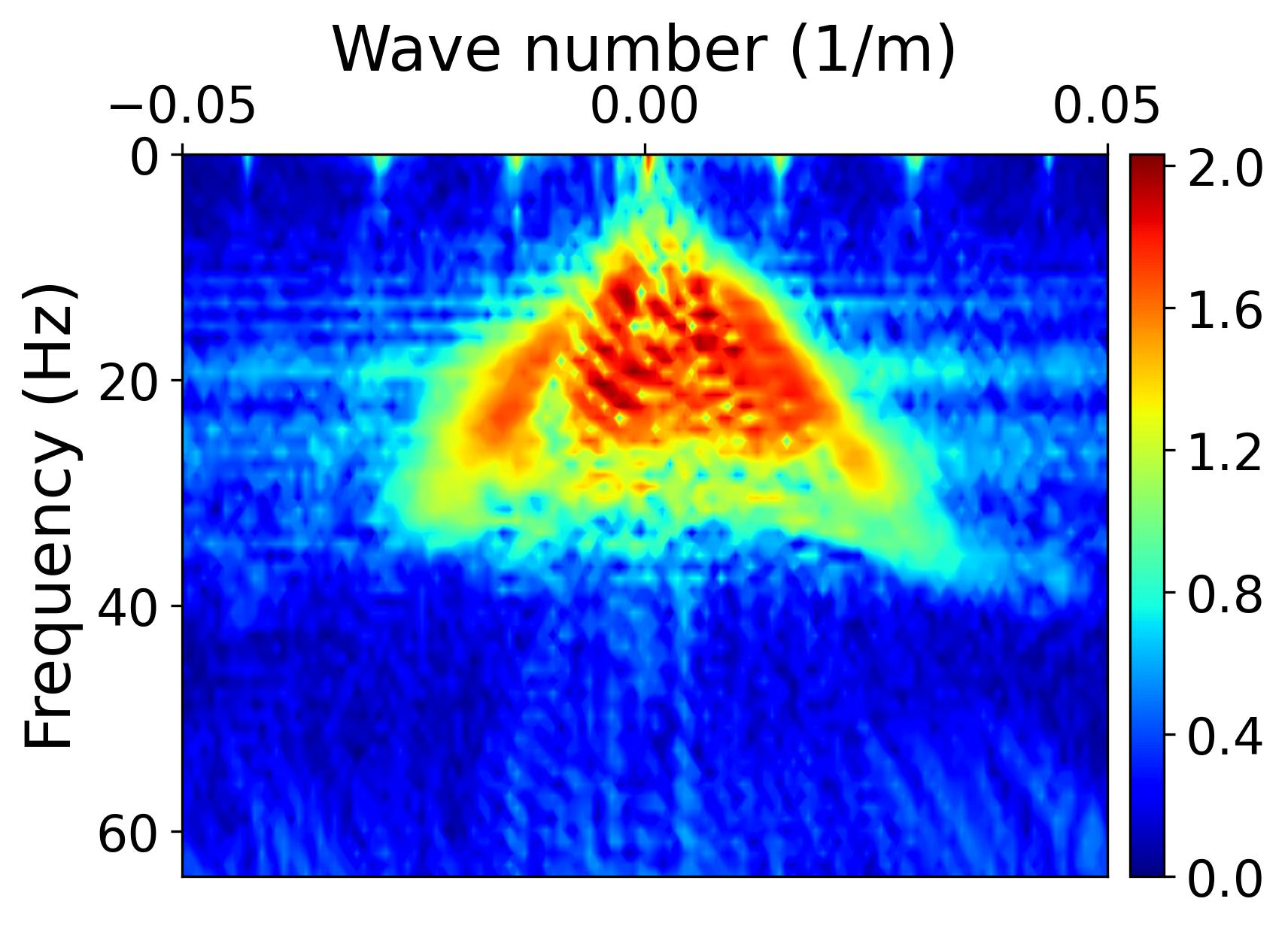}\label{fig:SEGC3regular_subfig8}}
  \subfloat[Coarse-to-Fine.] 
    {\includegraphics[height=0.151\textwidth]{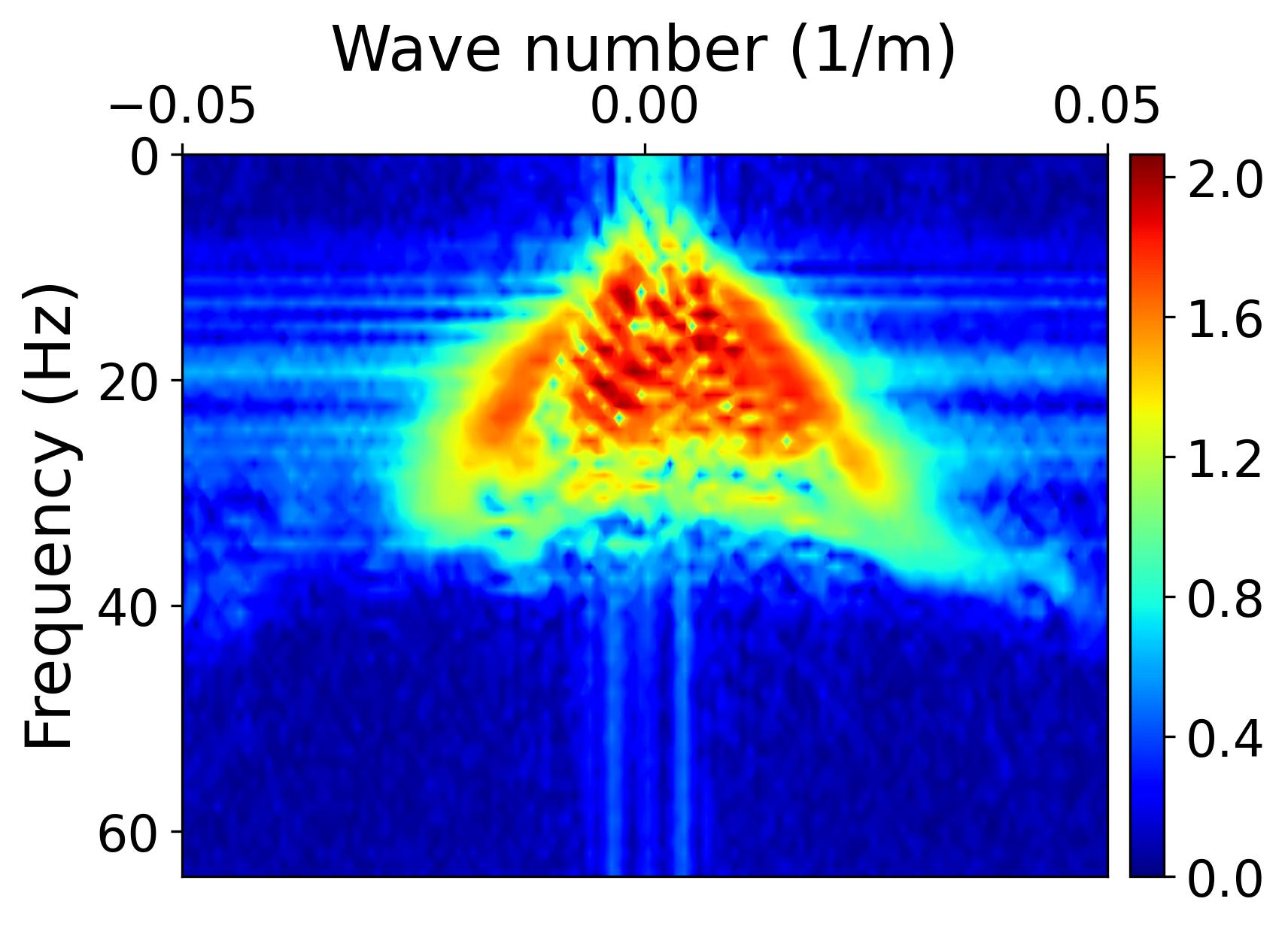}\label{fig:SEGC3regular_subfig9}}
  \subfloat[Ours.]
    {\includegraphics[height=0.151\textwidth]{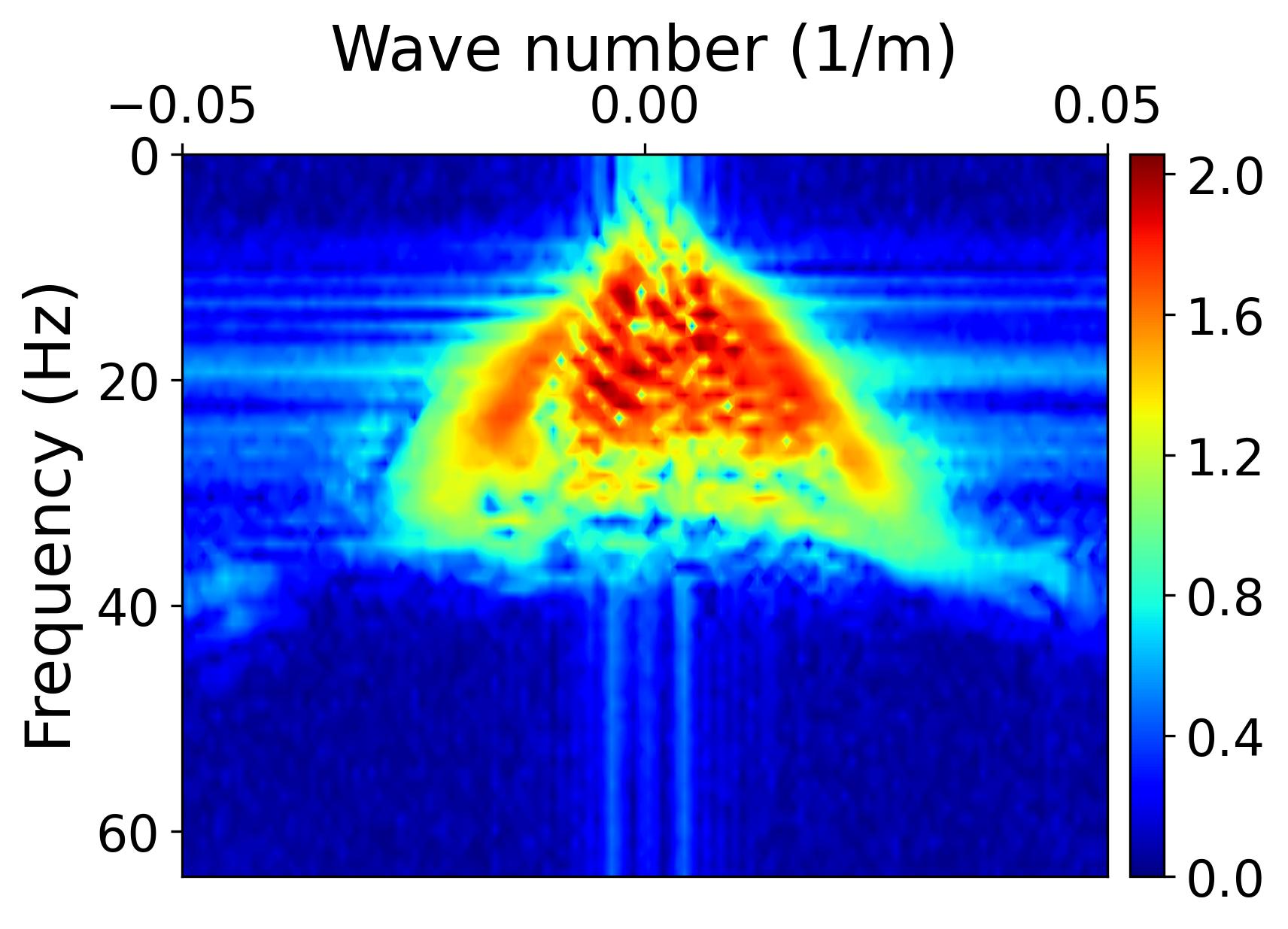}\label{fig:SEGC3regular_subfig1}}
  
  \caption{The $f\text{-}k$ spectra of SEG C3 test data interpolation results with 70\% regular missing traces on different methods.}
  \label{fig:SEGC3 regular}
  \vspace{-2mm}
\end{figure*}

For each patch in the test sets of SEG C3 and MAVO, we design random missing phenomena with missing rates ranging from 0.2 to 0.8. The initial values of the missing traces are set to 0. The experimental results of random missing interpolation are listed on the left side of Tab. \ref{tab:segc3missing} and Tab. \ref{tab:mavomissing}. Except for a slightly lower SSIM on MAVO test data compared to PConv-UNet, other results indicate that our model has better fidelity. 
Fig. \ref{fig:mavo random} shows the interpolated traces of the random missing MAVO test data. It can be seen that our method achieves the best performance both on amplitudes and phases. As a special case of random missing seismic data, the regular missing scenario will cause a serious aliasing problem. It usually appears as excessive artifacts in the high-frequency band of $f\text{-}k$ spectra caused by erroneous estimation or interpolation of the missing data frequency. Fig. \ref{fig:SEGC3 regular} compares the $f\text{-}k$ spectra of SEG C3 test data with 70\% regular missing traces. Severe aliasing can be noticed in Fig. \ref{fig:SEGC3regular_subfig4}. It is obvious that the $f\text{-}k$ spectra of the DD-CGAN, cWGAN-G, and ANet are all accompanied by significant high-frequency artifacts. Comparisons between the performance of all methods indicate that our model gains the most consistent $f\text{-}k$ spectra with the ground truth.

\begin{table*}[htbp]
\scriptsize
\caption{Comparison of different methods on the test set of the SEG C3 dataset with various missing types. The best performance is highlighted in bold.}
\centering
\vspace{-2mm}
\begin{tabular}{lrrrrrrrrrrrrrrrr}
\toprule
\textbf{Missing type} & \multicolumn{4}{c}{ Random } & \multicolumn{4}{c}{Consecutive} & \multicolumn{4}{c}{ Multiple } \\
Model & MSE$\downarrow$ & SNR$\uparrow$ & PSNR$\uparrow$ & SSIM $\uparrow$ & MSE$\downarrow$ & SNR$\uparrow$ & PSNR$\uparrow$ & SSIM $\uparrow$ & MSE$\downarrow$ & SNR$\uparrow$ & PSNR$\uparrow$& SSIM $\uparrow$  \\
\midrule  \textbf{DD-CGAN}\cite{Chang2020} & 3.207e-04 & 28.986 & 34.940 & 0.931 & 8.689e-04 & 24.656  & 30.610 & 0.866 & 7.142e-04 & 25.508 & 31.462 & 0.880\\ 
\textbf{cWGAN-GP}\cite{wei2022big} & 9.913e-05 & 34.084 & 40.038 & 0.981 & 3.193e-04 & 29.004 & 34.958 & 0.935 & 2.782e-04 & 29.603 & 35.556 & 0.949\\
\textbf{PConv-UNet}\cite{pan2020partial} & 7.004e-05 & 35.593 & 41.547 & 0.986 & 4.037e-04 & 27.985 & 33.939 & 0.932 & 2.789e-04 & 29.592 & 35.546 & 0.955\\
\textbf{ANet}\cite{Yu9390348}& 1.618e-04 & 31.957 & 37.911 & 0.970 & 5.204e-04 & 26.883 & 32.837 & 0.926 & 4.323e-04 & 27.689 & 33.642 & 0.937\\
\textbf{Coarse-to-Fine}\cite{wei2022hybrid} & 7.639e-05 & 35.216 & 41.170 & 0.984 & 2.279e-04 & 30.469 & 36.423 & 0.960 & 1.870e-04 & 31.327 & 37.281 & 0.968\\
\textbf{Ours} & \bf{5.425e-05} & \bf{36.934} & \bf{42.889} & \bf{0.988} & \bf{1.635e-04} & \bf{32.050} & \bf{38.004} & \bf{0.972} & \bf{1.601e-04} & \bf{32.120} & \bf{38.074} & \bf{0.977}\\ 
\bottomrule
\end{tabular}
  \label{tab:segc3missing}
  \vspace{-2mm}
\end{table*}

\begin{table*}[htbp]
\scriptsize
\caption{Comparison of different methods on the test set of the MAVO dataset with various missing types. The best performance is highlighted in bold.}
\centering
\vspace{-2mm}
\begin{tabular}{lrrrrrrrrrrrrrrrr}
\toprule 
\textbf{Missing type} & \multicolumn{4}{c}{ Random } & \multicolumn{4}{c}{Consecutive}  & \multicolumn{4}{c}{  Multiple } \\
Model & MSE$\downarrow$ & SNR$\uparrow$ & PSNR$\uparrow$ & SSIM $\uparrow$ & MSE$\downarrow$ & SNR$\uparrow$ & PSNR$\uparrow$ & SSIM $\uparrow$ & MSE$\downarrow$ & SNR$\uparrow$ & PSNR$\uparrow$ &SSIM $\uparrow$  \\
\midrule  \textbf{DD-CGAN}\cite{Chang2020} & 3.390e-04 & 29.167 & 34.698 & 0.941 & 4.350e-04 & 28.083 & 33.615 & 0.923 & 5.510e-04 & 27.057 & 32.589 & 0.912\\ 
\textbf{cWGAN-GP}\cite{wei2022big} & 1.991e-04 & 31.478 & 37.010 & 0.967 & 2.167e-04 & 31.110 & 36.642 & 0.960 & 3.212e-04 & 29.400 & 34.932 & 0.949\\ 
\textbf{PConv-UNet}\cite{pan2020partial} & 1.345e-04 & 33.182 & 38.714 & \bf{0.975} & 1.589e-04 & 32.458 & 37.990 & 0.972 & 2.151e-04 & 31.141 & 36.673 & 0.965\\ 
\textbf{ANet}\cite{Yu9390348}& 2.142e-04 & 31.161 & 36.693 & 0.968 & 2.463e-04 & 30.553 & 36.085 & 0.961 & 3.477e-04 & 29.056 & 34.588 & 0.950\\ 
\textbf{Coarse-to-Fine}\cite{wei2022hybrid} & 1.676e-04 & 32.224 & 37.756 & 0.970 & 1.450e-04 & 32.854 & 38.386 & 0.972 & 2.175e-04 & 31.093 & 36.625 & 0.962\\ 
\textbf{Ours} & \bf{1.308e-04} & \bf{33.432} & \bf{38.965} & 0.972 & \bf{1.079e-04} & \bf{34.274} & \bf{39.802} & \bf{0.977} & \bf{1.671e-04} & \bf{32.416} & \bf{37.944} & \bf{0.969}\\
\bottomrule
\end{tabular}
  \label{tab:mavomissing}
\end{table*}

\subsubsection{Consecutive Missing Traces}\label{exp:Consecutive Missing Traces}
\begin{figure*}[!htbp]
  \centering
  \vspace{-4mm}
  \subfloat[Consecutive missing seismic data.]
    {\includegraphics[height=0.137\textheight]{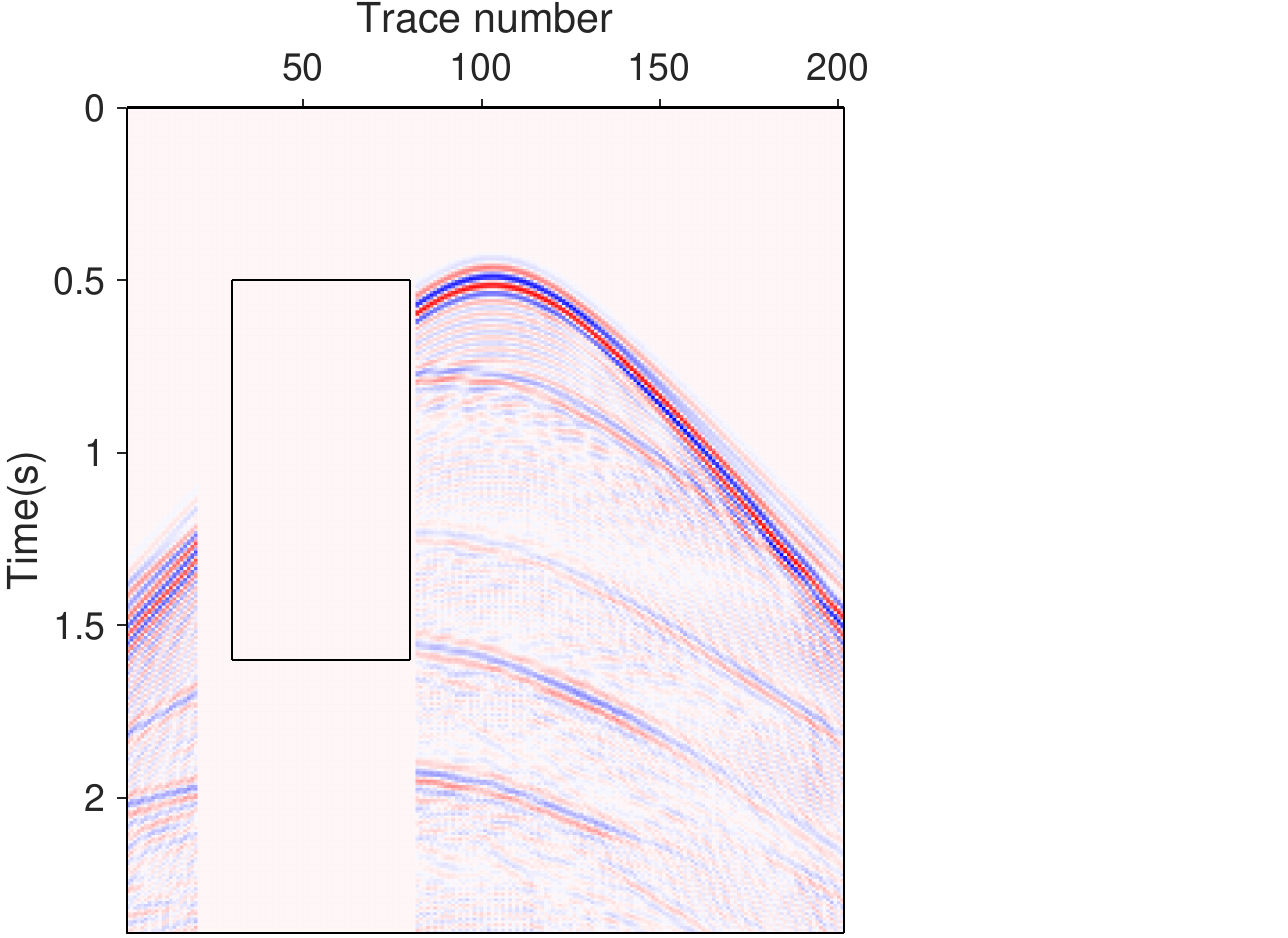}\label{fig:Segc3continus_subfig8}\hspace{-0.3mm}} 
  \subfloat[Ground truth.]
  {\includegraphics[height=0.137\textheight]{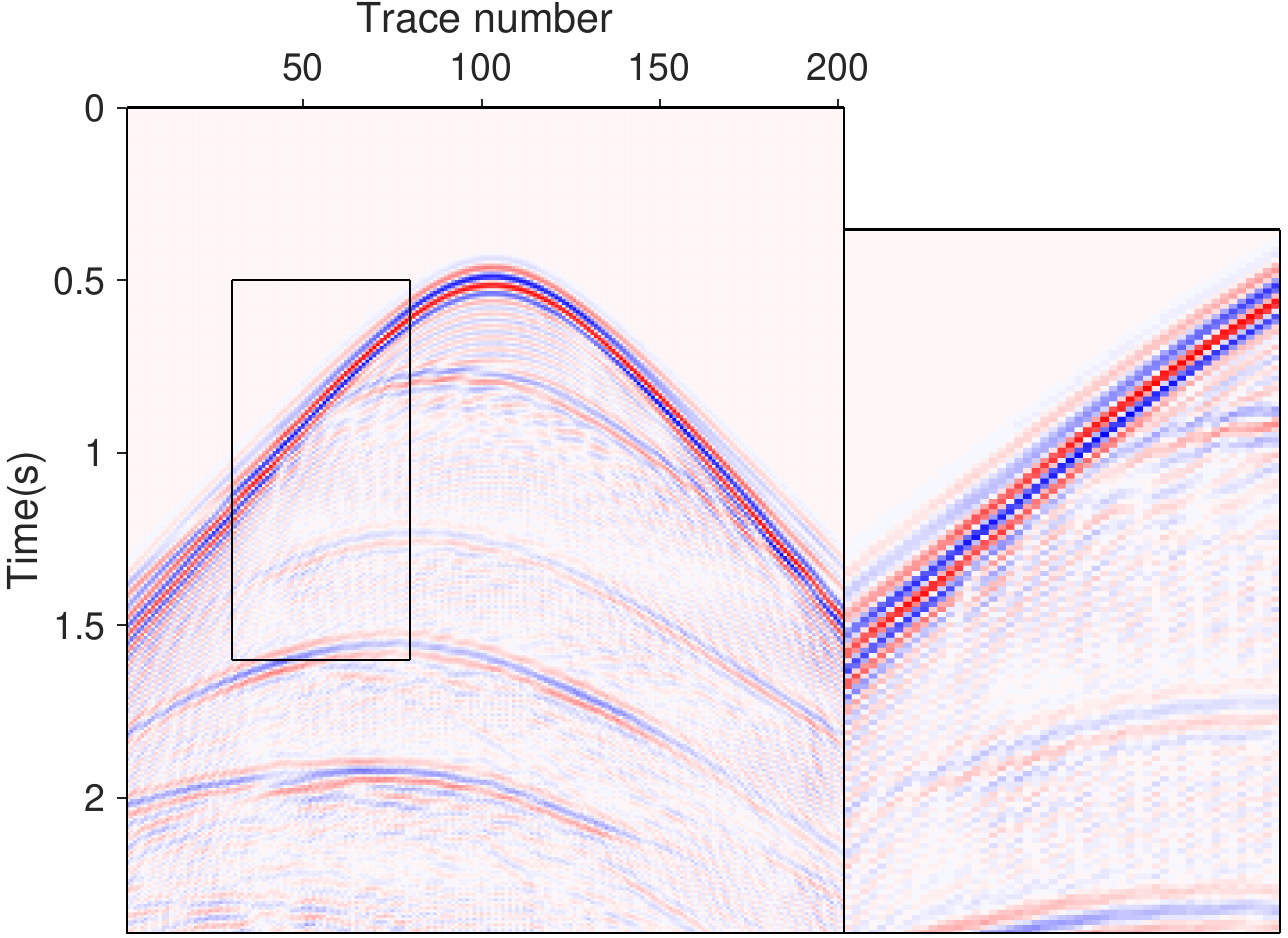}\label{fig:mavocontinus_ori}}
  \subfloat[DD-CGAN.]
  {\includegraphics[height=0.137\textheight]{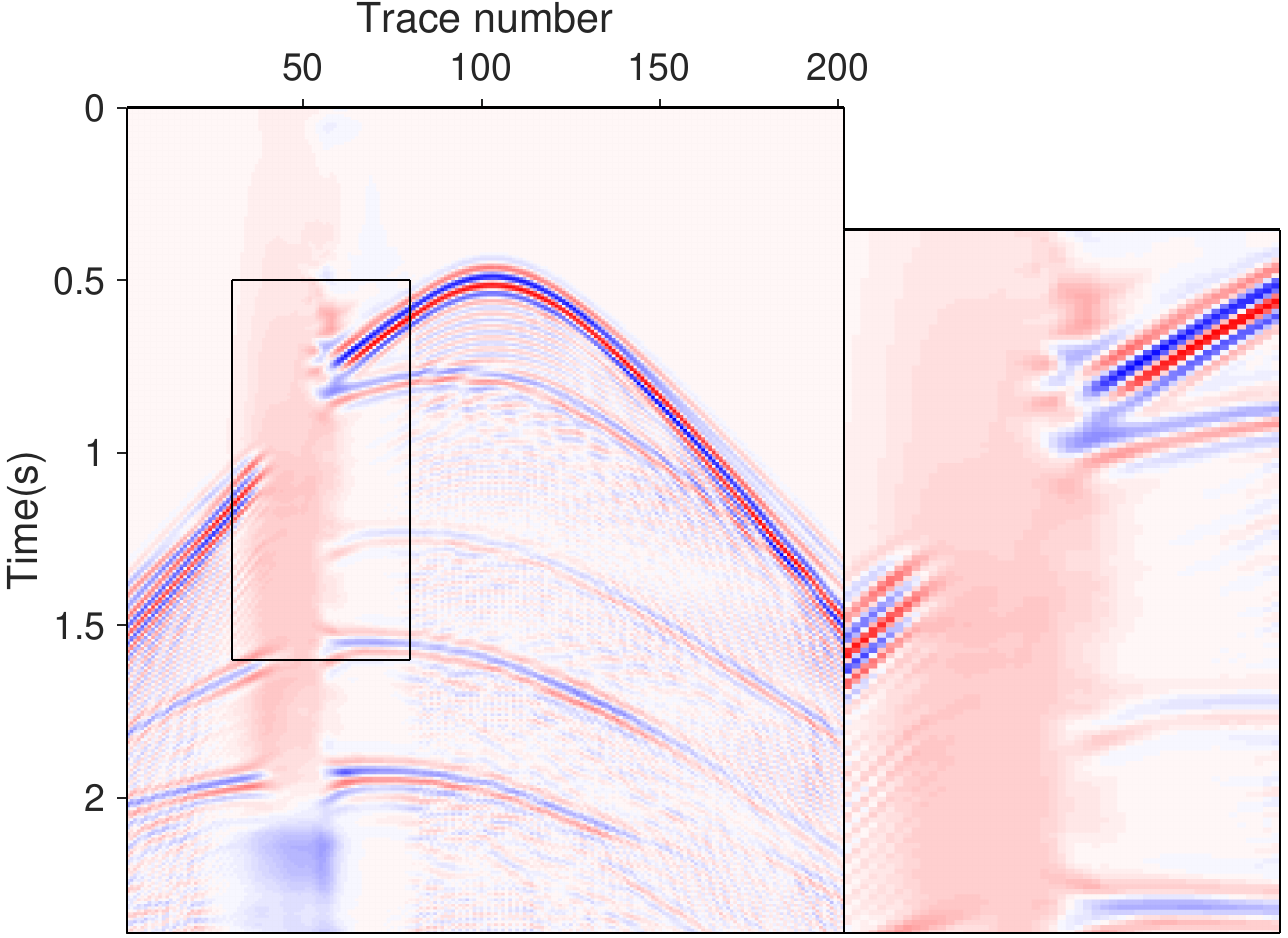}\label{fig:mavocontinus_ori}}
  \subfloat[cWGAN-GP.]
    {\includegraphics[height=0.137\textheight]{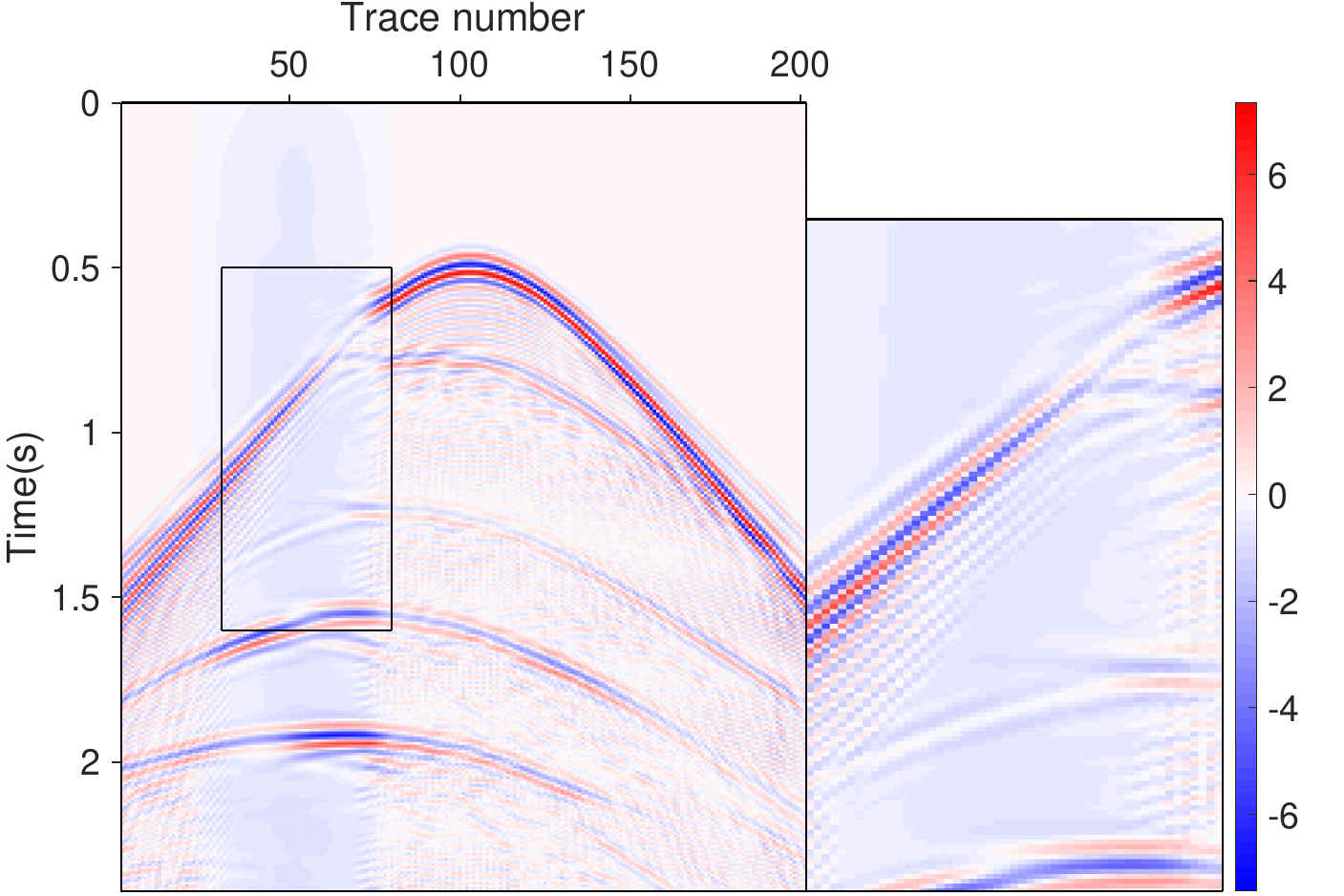}\label{fig:Segc3continus_subfig8}}
    \clearpage
  \subfloat[PConv-UNet.]
  {\includegraphics[height=0.137\textheight]{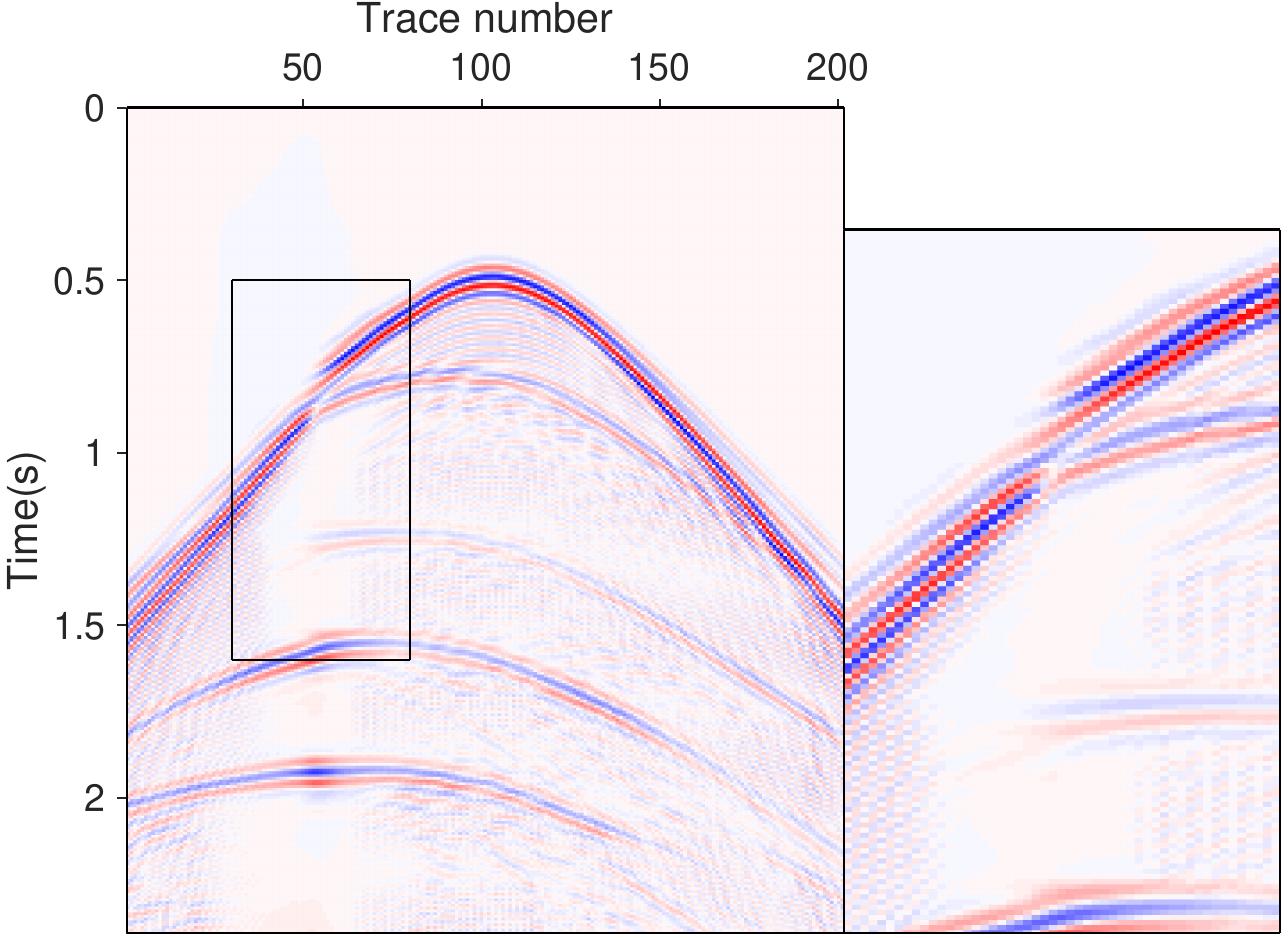}\label{fig:mavocontinus_ori}}
  \subfloat[ANet.]
  {\includegraphics[height=0.137\textheight]{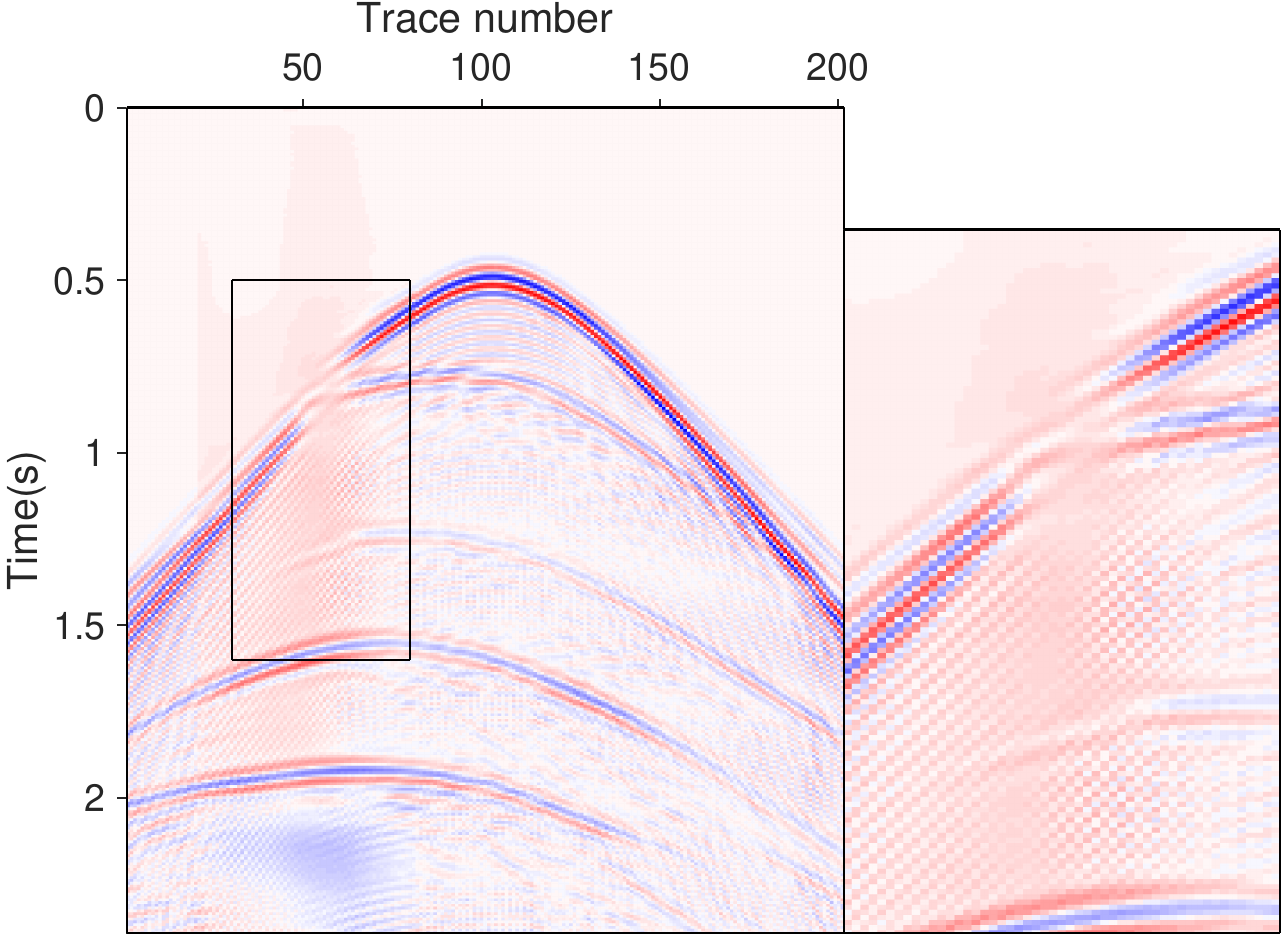}\label{fig:mavocontinus_ori}}
  \subfloat[Coarse-to-Fine.]
  {\includegraphics[height=0.137\textheight]{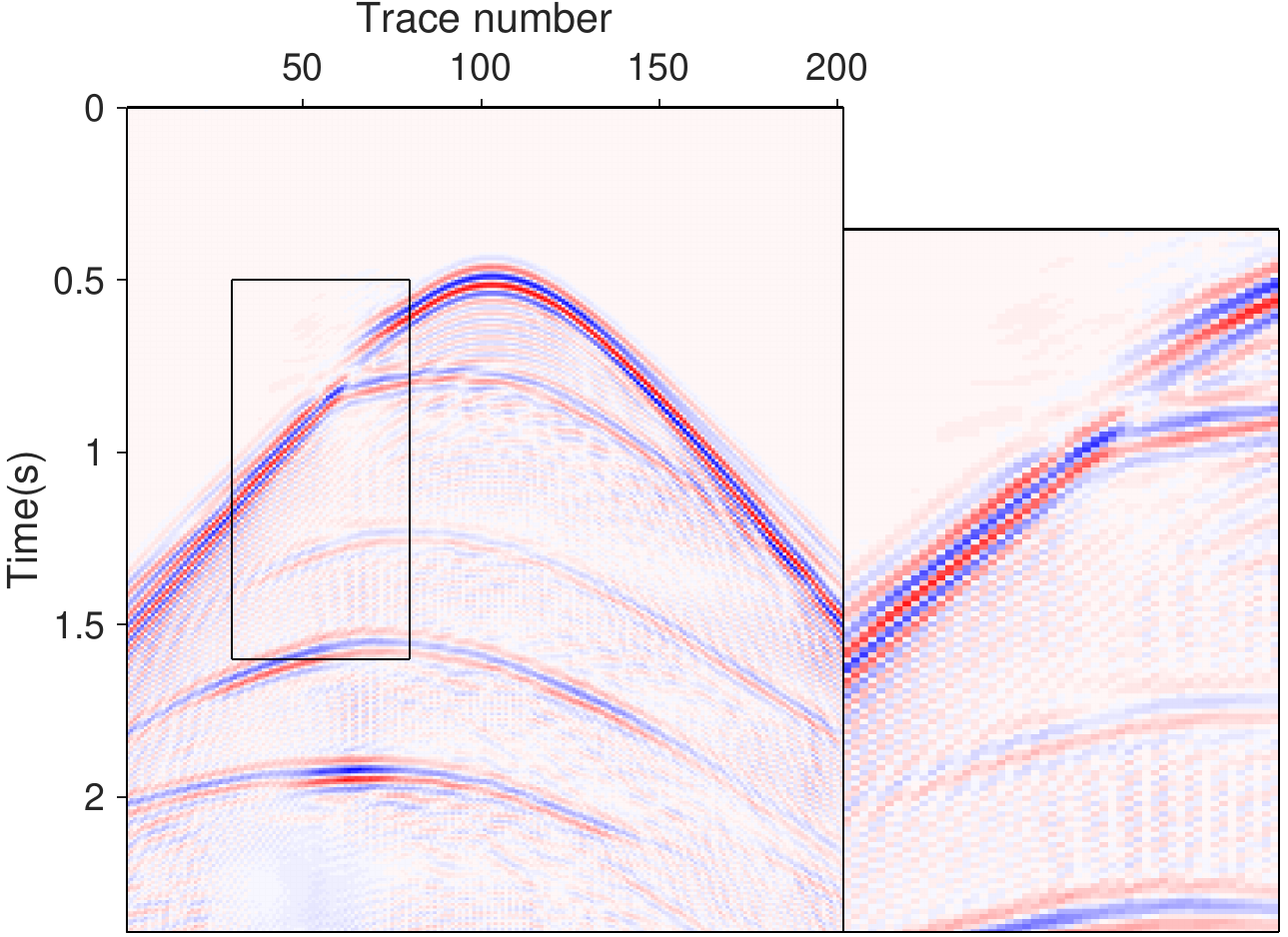}\label{fig:mavocontinus_ori}}
  \subfloat[Ours.]
    {\includegraphics[height=0.137\textheight]{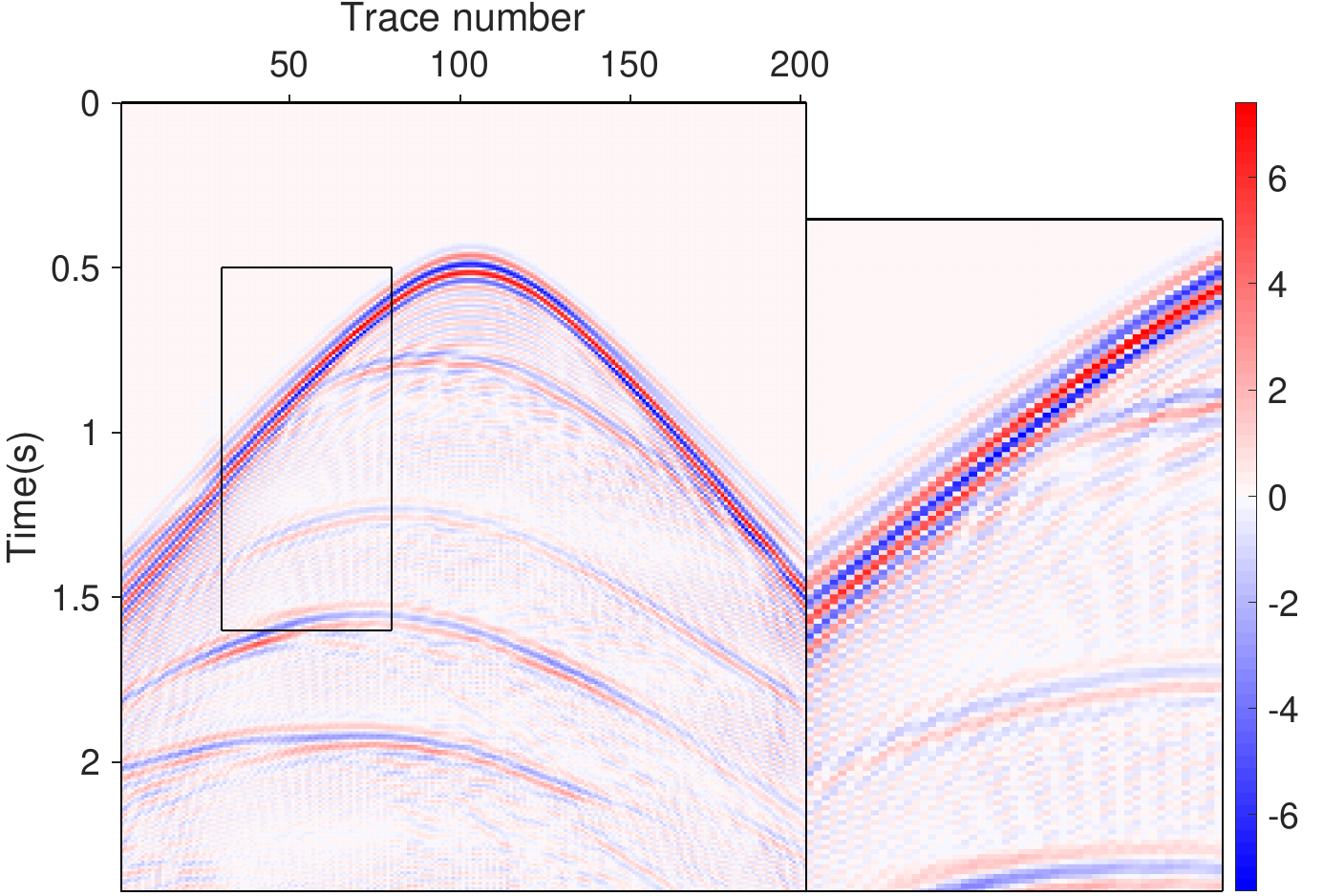}\label{fig:Segc3continus_subfig8}}
  \vspace{-1mm}
  \caption{Interpolation results of the SEG C3 complete test slice with 30\% consecutive missing traces on different methods. The seismic data is denormalized to their original amplitude range and we apply the gain method to display weak amplitude details clearly. The reconstruction region within the box is magnified in the bottom right corner to allow for a more detailed observation of the interpolation.}
  \label{fig:Segc3 continus}
  \vspace{-5mm}
\end{figure*}

We randomly create consecutive missing masks, with rates of missing data ranging from 0.1 to 0.6 (not including edge traces), and apply them to the patches in the SEG C3 and MAVO datasets. The value of missing traces is initialized to 0. The interpolation results of the middle four columns of Tab. \ref{tab:segc3missing} and Tab. \ref{tab:mavomissing} indicate that our model consistently surpasses other methods over these two datasets. We provide the comparisons of full slices via color plots from the SEG C3 complete test slices, as shown in Fig. \ref{fig:Segc3 continus}. We use a sliding window with overlap to crop patches for conducting interpolation and then splice them back into the original slice. The sliding step size is 8, and the overlapping partial prediction values are averaged. The ground truth data suffers from a consecutive missing of 30\%, resulting in degenerate missing data. DD-CGAN obviously cannot handle large-interval interpolation, and large-area artifacts manifest in the results of cWGAN-GP and ANet. PConv-UNet, based on valid feature similarity to conducting interpolation, produces inconsistent predictions for interpolating content from two directions. Among these, Coarse-to-Fine model demonstrates a high continuity in strong amplitude regions while its predictions are still inadequate at intersections with multiple strong amplitudes. Our model can consistently improve the performances over both strong and weak amplitudes, and keep anisotropy and spatial continuity of signals. 

\subsubsection{Multiple Missing Traces}
\begin{figure*}[!htbp]
  \centering
  \subfloat[Multiple missing seismic data.]
    {\includegraphics[height=0.175\textheight]{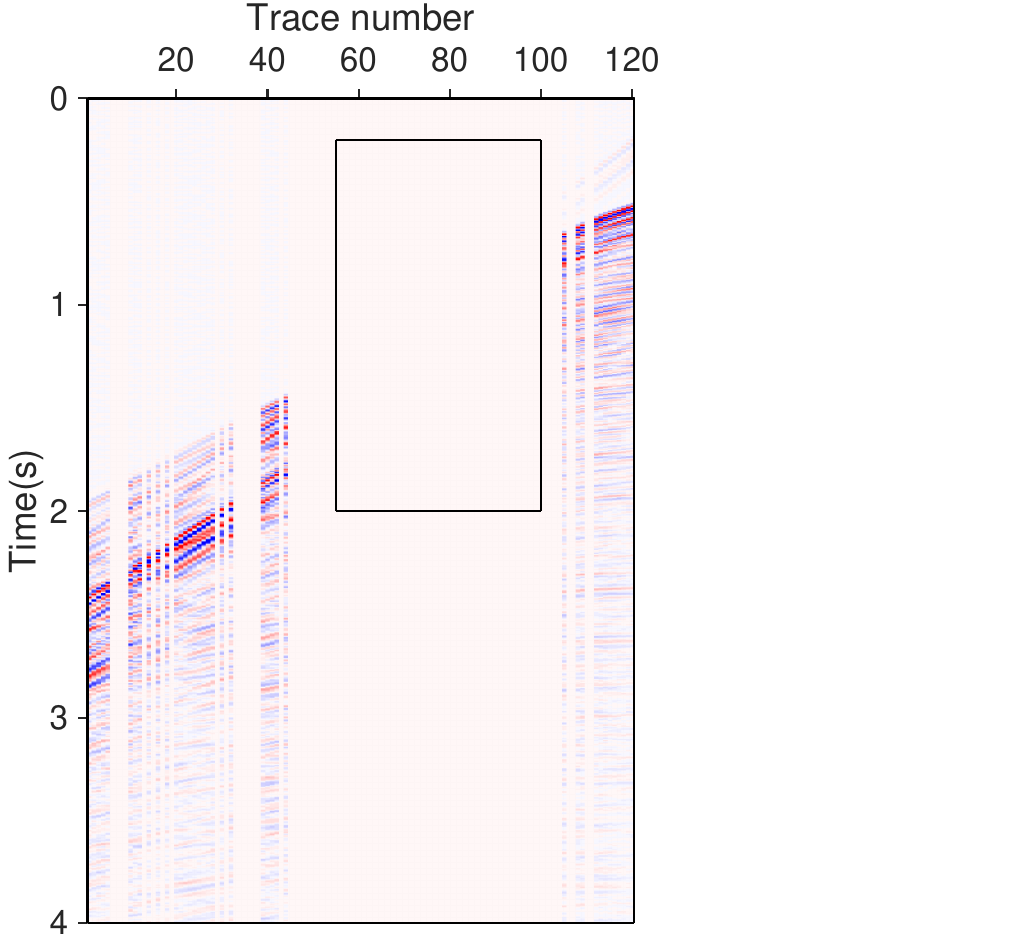}\label{fig:Segc3continus_subfig8}} 
  \subfloat[Ground truth.]
  {\includegraphics[height=0.175\textheight]{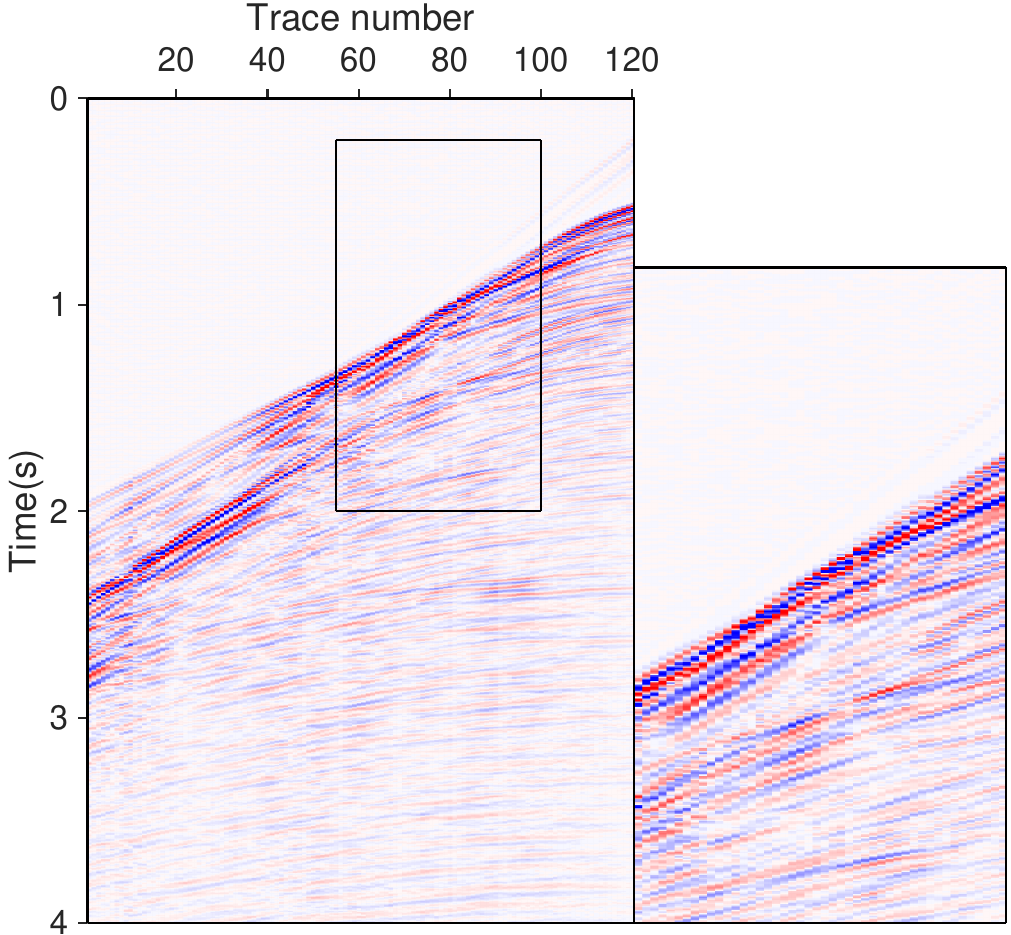}\label{fig:mavocontinus_ori}}  
  \subfloat[DD-CGAN.]
    {\includegraphics[height=0.175\textheight]{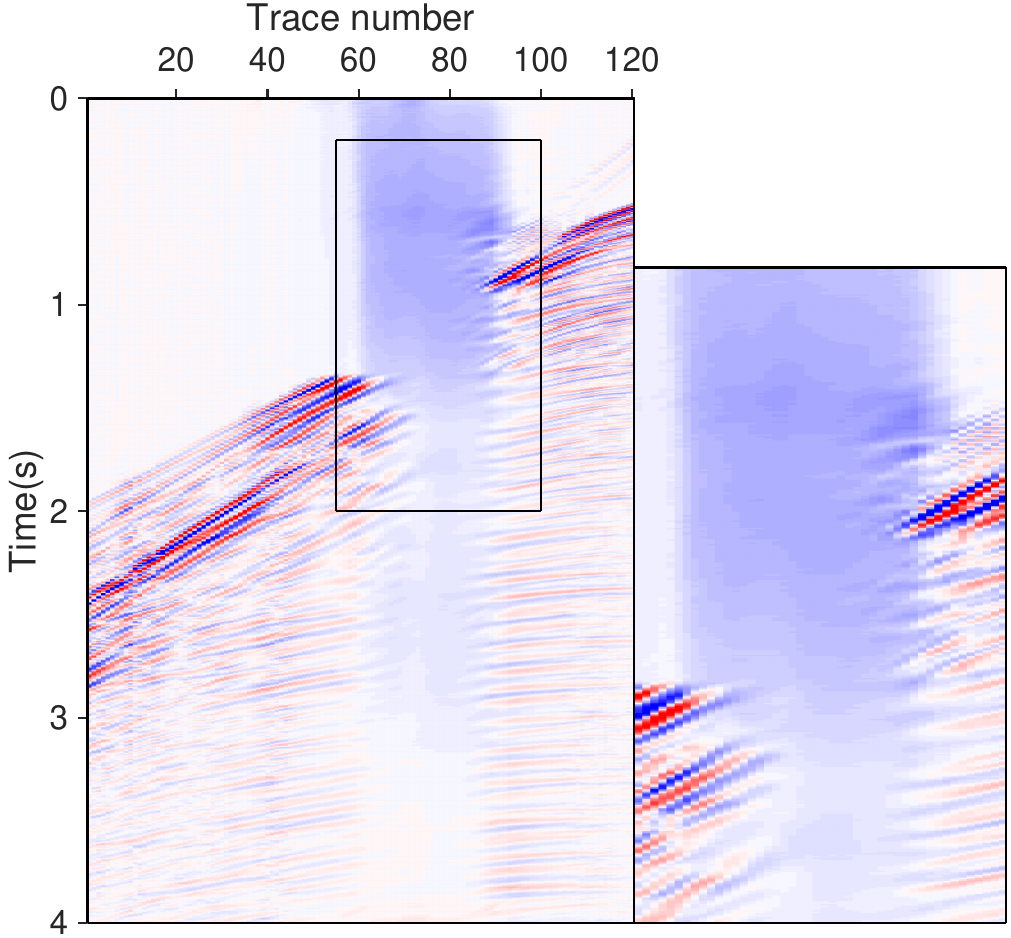}\label{fig:Segc3continus_subfig8}}
  \subfloat[cWGAN-GP.]
    {\includegraphics[height=0.175\textheight]{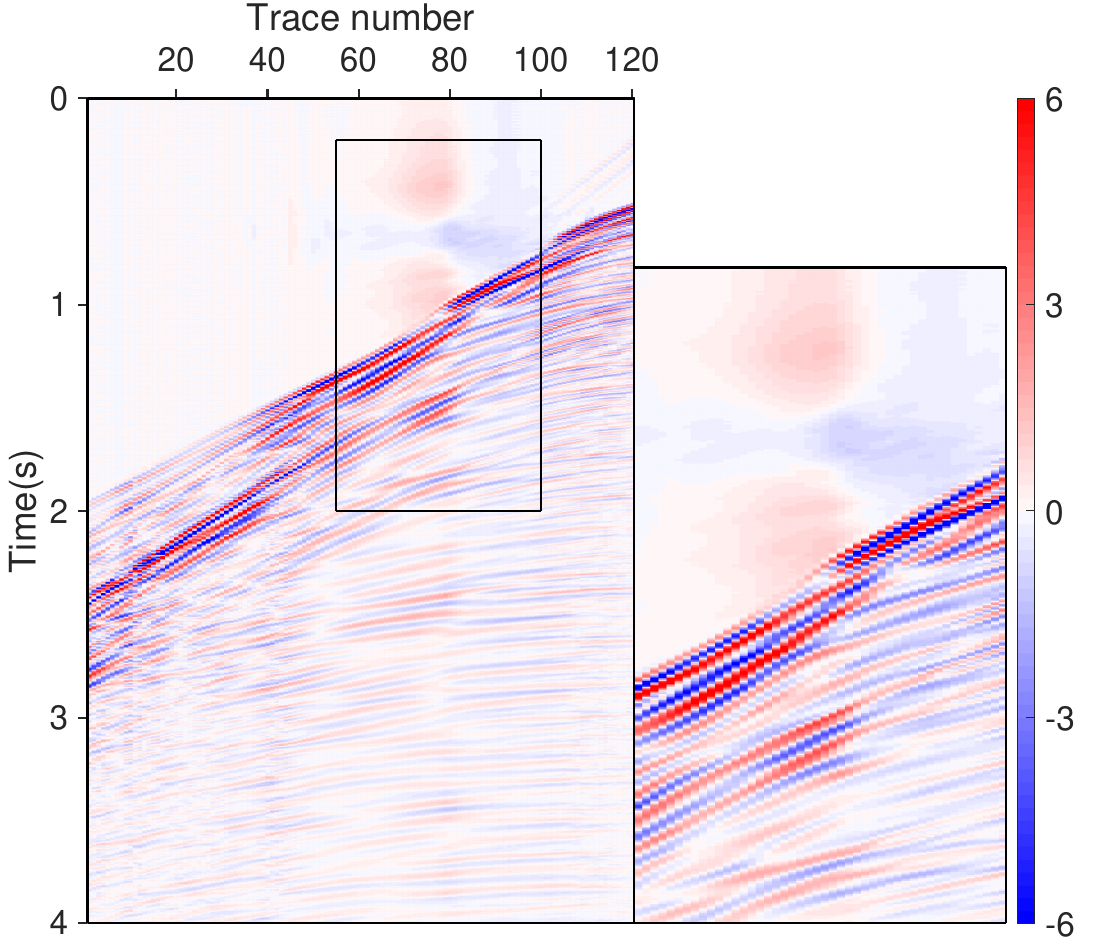}\label{fig:Segc3continus_subfig8}}
    \clearpage
  \subfloat[PConv-UNet.]
    {\includegraphics[height=0.175\textheight]{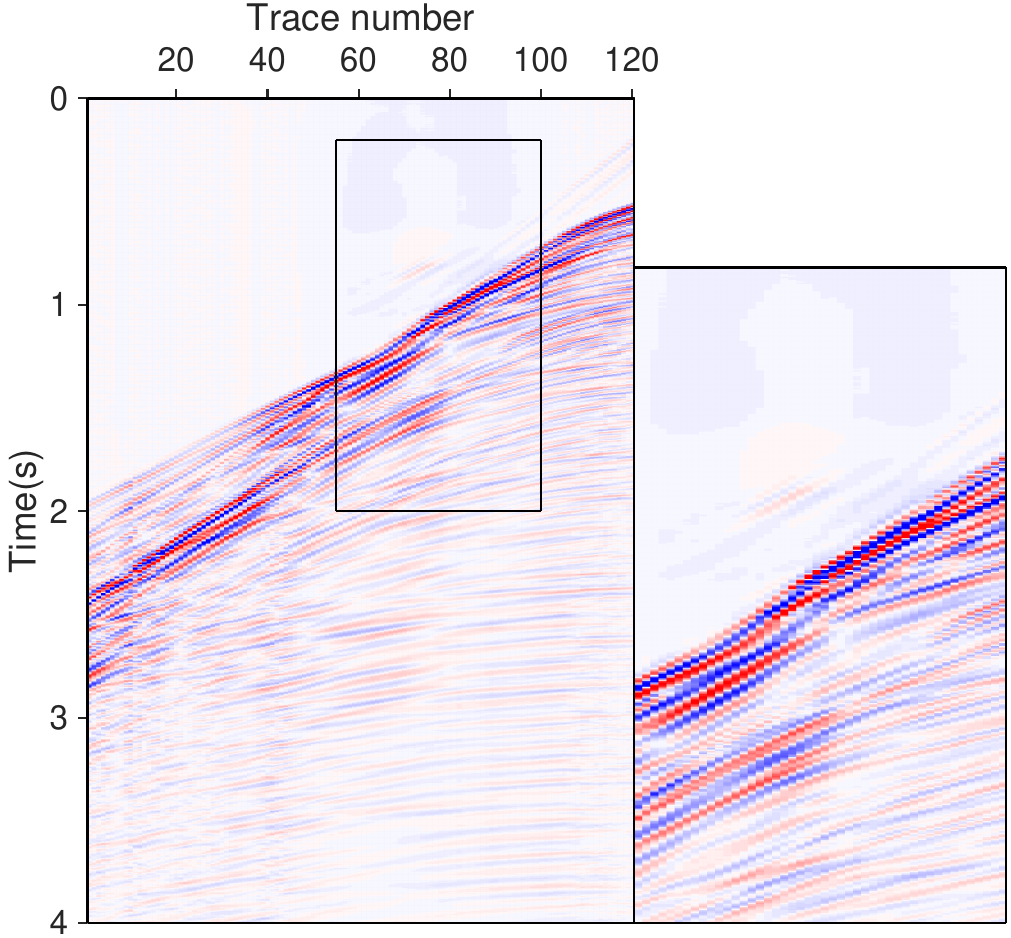}\label{fig:Segc3continus_subfig8}}
  \subfloat[ANet.]
    {\includegraphics[height=0.175\textheight]{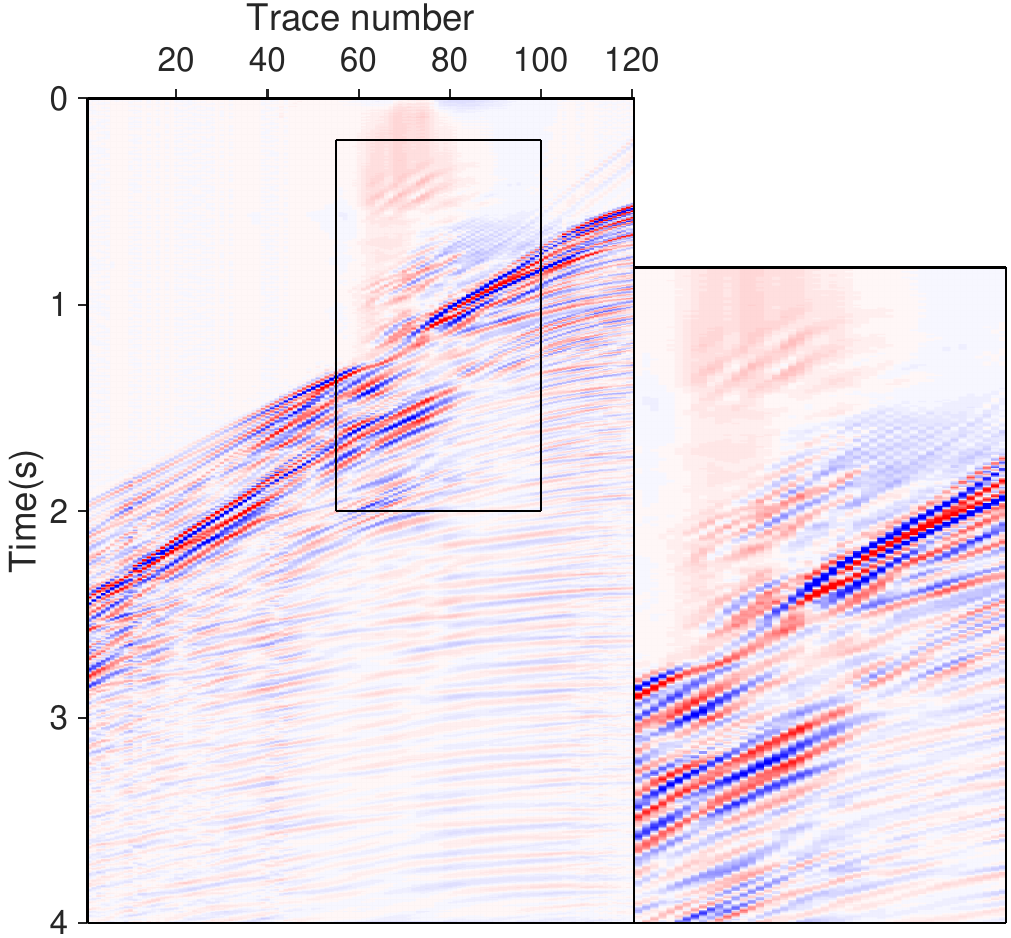}\label{fig:Segc3continus_subfig8}}
  \subfloat[Coarse-to-Fine.]
    {\includegraphics[height=0.175\textheight]{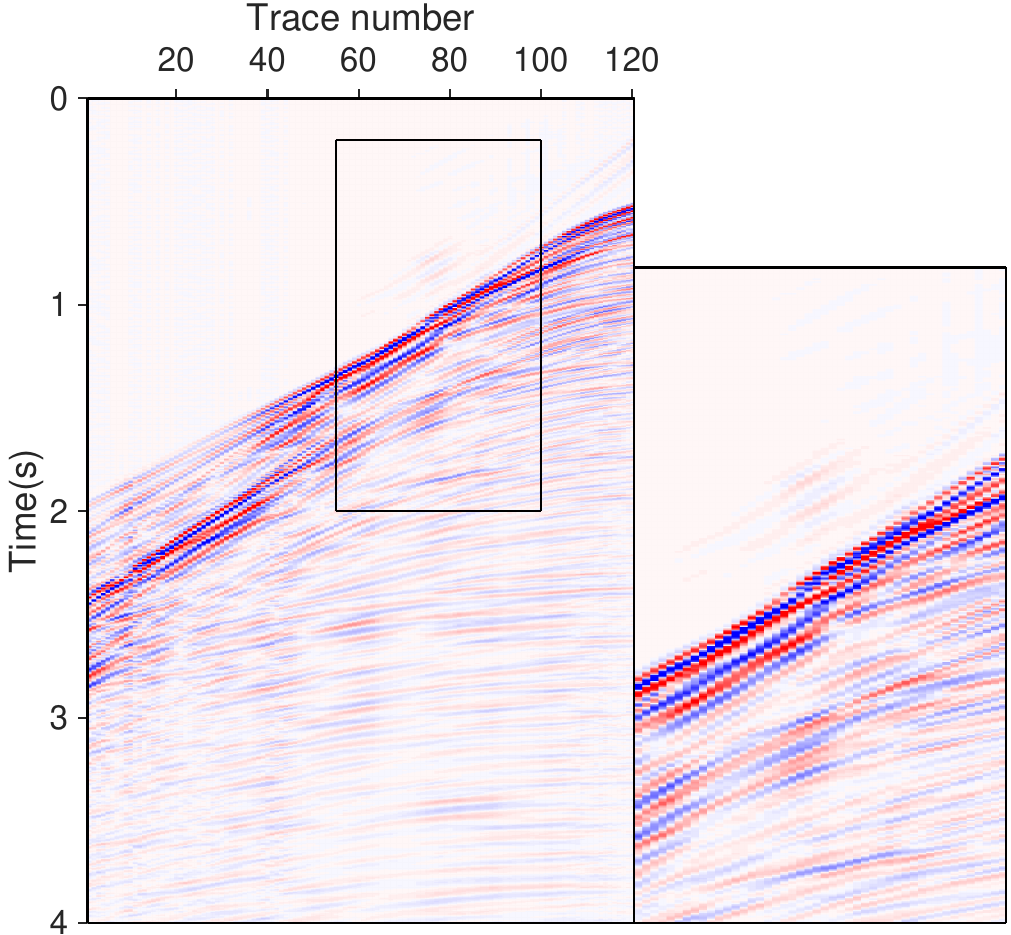}\label{fig:Segc3continus_subfig8}}
  \subfloat[Ours.]
    {\includegraphics[height=0.175\textheight]{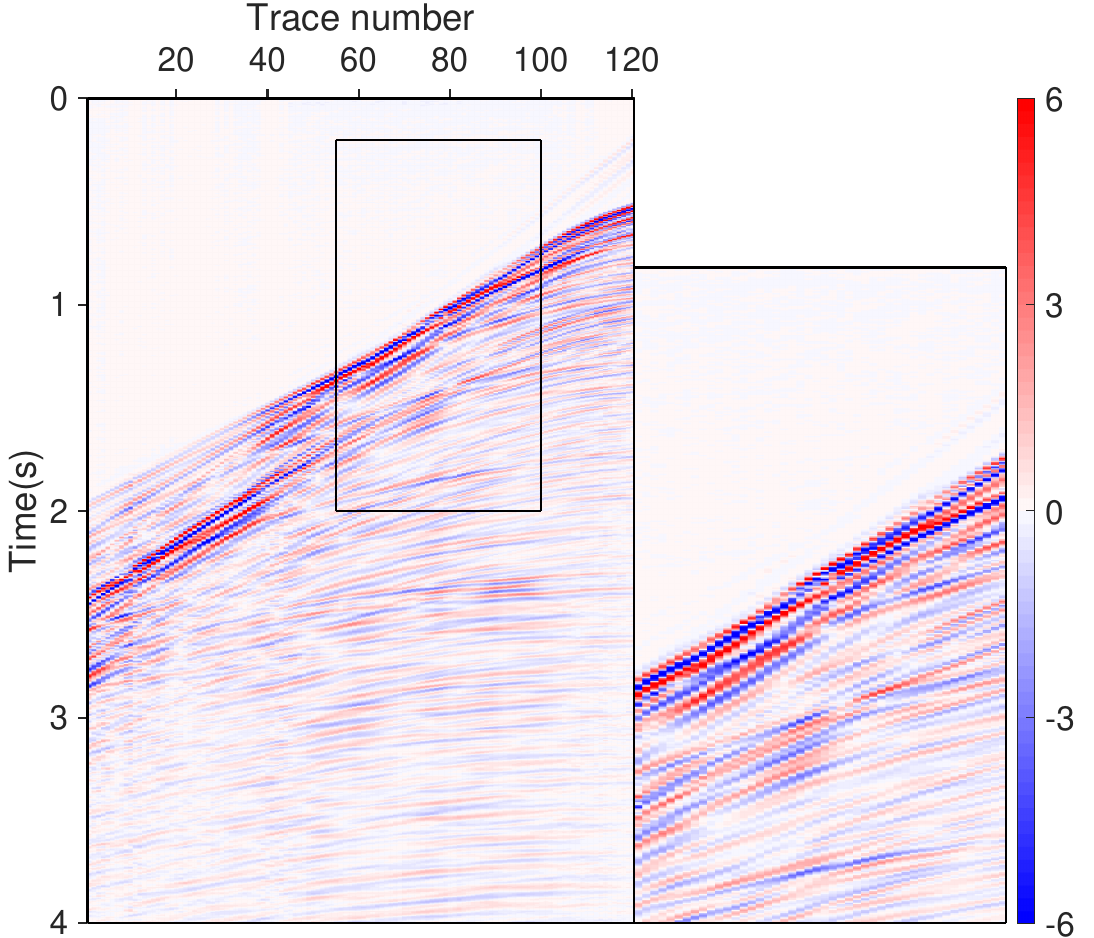}\label{fig:Segc3continus_subfig8}}
  
  \caption{Interpolation results of the MAVO complete test slice with 67.5\% multiple missing types on different methods. We restore the seismic data to their original amplitude range and employ the gain method to enhance the visibility of weak amplitude details. To facilitate a more detailed observation of the local interpolation, the reconstruction region within the box is magnified on the right side.}
  \label{fig:mavo multiple}
\end{figure*}

For the SEG C3 and MAVO datasets, we construct multiple missing data scenarios with both consecutive and random missing cases and the range of the total missing rate is $[0.1, 0.9]$. The missing traces are also initialized with a value of 0. The corresponding quantitative comparison results are listed in the right four columns of Tab. \ref{tab:segc3missing} and Tab. \ref{tab:mavomissing}, where our model consistently outperforms other methods on four metrics. Fig. \ref{fig:mavo multiple} exhibits the interpolation results on a multiple missing example with a total missing rare 67.5\% from the MAVO complete test slices. The interpolation method for complete track sets is the same as described in Section \ref{exp:Consecutive Missing Traces}, using sliding window predictions followed by stitching back into the original slice. Our model produces artifact-free results, while other methods generally result in the ubiquity-wide areas of artifacts, especially for DD-CGAN, cWGAN-G, and ANet, failing to provide reliable recovery. The Coarse-to-Fine method generates spurious signals in the marked region. In comparison, the amplitudes predicted by our model are more accurate and consistent with the ground truth.
Our model is capable of handling most cases of seismic missing trace reconstruction.

\subsection{Model Generalization}
\subsubsection{Different Missing Types}
In order to study the impact of changes in the missing form on model capability, we evaluate the performance of different methods under the mismatched training and testing mask patterns on the SEG C3 dataset, as shown in Tab. \ref{tab:onetraining}. 
First, when testing on the unseen consecutive mask pattern, the performance of the models trained on the random mask type has decreased significantly compared to those consecutive missing reconstruction results in Tab. \ref{tab:segc3missing}. Second, although the model trained on the multiple mask form exhibits interpolation capability on consecutive and random missing types, their results are still worse than those trained on the same mask pattern, as demonstrated in Tab. \ref{tab:segc3missing}.
Third, we can see that the consecutive missing model fails to interpolate random missing data, which is likely due to the significant differences in learning patterns between consecutive missing form and random missing form. 
It can be concluded that the effectiveness of generative models, which may be based on GAN or feature similarity, is sensitive to the constructed mask formula in training data. It seems better if the training missing construction can be closer to the missing form of the test data, although there easily exist gaps in the field scenarios. 
In contrast, our model training does not require rigorous construction of missing scenes and can complete the interpolation of any missing form with just one training session, while maintaining performance advantages. It demonstrates better generalization ability across different types of missing patterns. 

\begin{table}[htbp]
\scriptsize
\vspace{-1mm}
 \caption{Generalization comparison of different models on the test set of the SEG C3 dataset under mismatched training and testing mask patterns. The result of top performance is masked in bold.}
  \centering
  \vspace{-3mm}
\begin{tabular}{llrrrrr}
\toprule
Type & Model  & MSE & SNR &PSNR &SSIM\\
\midrule & DD-CGAN\cite{Chang2020} & 1.537e-03 & 22.179 & 28.132 & 0.827  \\
Random to& cWGAN-GP\cite{wei2022big} & 1.123e-03 & 23.544 & 29.498 & 0.883\\
 consecutive & PConv-UNet\cite{pan2020partial}& 1.134e-03 & 23.501 & 29.455 & 0.884\\
& ANet\cite{Yu9390348} & 1.132e-03 & 23.506 & 29.460 & 0.885 \\
& Coarse-to-Fine\cite{wei2022hybrid} & 9.363e-04 & 24.332 & 30.286 & 0.900\\
\midrule 
& DD-CGAN\cite{Chang2020} & 9.901e-04 & 24.090 & 30.043 & 0.859 \\
Multiple to & cWGAN-GP\cite{wei2022big} & 4.368e-04 & 27.644 & 33.598 & 0.926 \\
consecutive & PConv-UNet\cite{pan2020partial}& 4.539e-04 & 27.477 & 33.431 & 0.930\\
& ANet\cite{Yu9390348} & 6.750e-04 & 25.753 & 31.707 & 0.913\\
& Coarse-to-Fine\cite{wei2022hybrid} & 2.783e-04 & 29.600 & 35.554 & 0.954\\ 
\midrule 
& Ours  &\bf{1.635e-04} & \bf{32.050} & \bf{38.004} & \bf{0.972}\\
\midrule & DD-CGAN\cite{Chang2020} & 3.777e-02 & 8.275 & 14.228 & 0.471 \\
Consecutive to & cWGAN-GP\cite{wei2022big} & 8.157e-03 & 14.931 & 20.885 & 0.502 \\
random & PConv-UNet\cite{pan2020partial}& 4.768e-02 & 7.263 & 13.217 & 0.398 \\
& ANet\cite{Yu9390348} & 6.804e-03 & 15.718 & 21.672 & 0.495\\
& Coarse-to-Fine\cite{wei2022hybrid} & 3.194e-02 & 9.002 & 14.956 & 0.417\\
\midrule & DD-CGAN\cite{Chang2020} & 3.836e-04 & 28.208 & 34.161 & 0.919 \\
Multiple to & cWGAN-GP\cite{wei2022big} & 1.049e-04 & 33.840 & 39.794 & 0.979\\
random & PConv-UNet\cite{pan2020partial}& 7.939e-05 & 35.049 & 41.002 & 0.984 \\
& ANet\cite{Yu9390348} & 2.189e-04 & 30.644 & 36.597 & 0.962\\
& Coarse-to-Fine\cite{wei2022hybrid} & 8.755e-05 & 34.624 & 40.577 & 0.982\\
\midrule 
& Ours &\bf{5.425e-05} & \bf{36.934} & \bf{42.889} & \bf{0.988}\\
\bottomrule
\end{tabular}
  \label{tab:onetraining}
  \vspace{-4mm}
\end{table}

\subsubsection{Different Noise Levels}
\begin{figure}[!htbp]
	\centering
	\vspace{-2mm}
\includegraphics[width=3.4in]{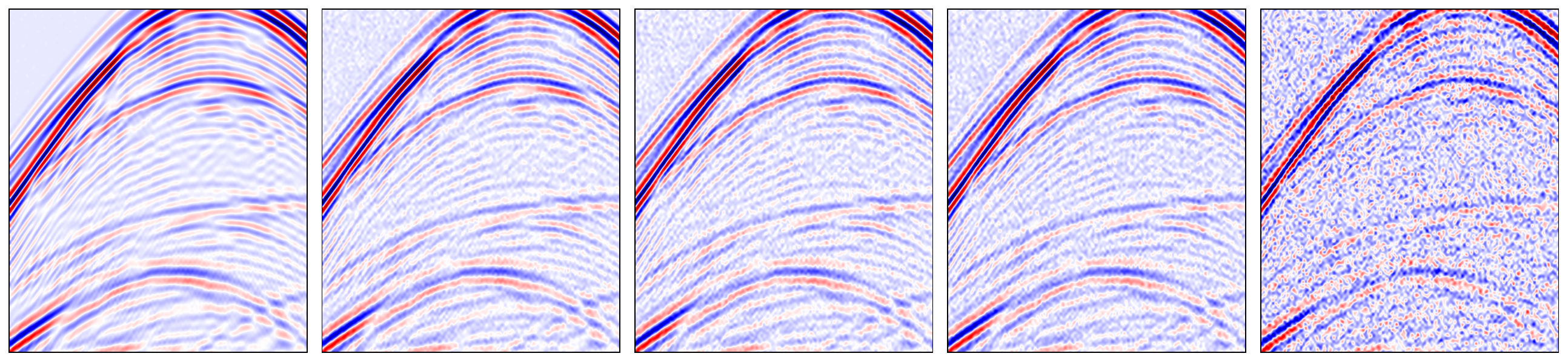}
\vspace{-3mm}
    \caption{Seismic data examples with increasing Gaussian noise from left to right.}
    \label{fig: noise typess}
    \vspace{-2mm}
\end{figure}

To examine our model's performance across various noise levels, we introduce Gaussian noise of differing intensities to the SEG C3 test data and evaluate the model's performance. Fig. \ref{fig: noise typess} visualizes seismic examples with noise addition, and Tab. \ref{tab:Generalization noise levels} summarizes the test results under multiple missing patterns. It should be noted that introducing noise reduces the SNR of the original signal. Our observations indicate that our model maintains stable and robust performance as long as the noise intensity does not overwhelm the seismic signal.

\begin{table}[htbp]
\scriptsize
\setlength{\tabcolsep}{10pt}
\vspace{-2mm}
\caption{Generalization ability evaluation under different noise levels on the SEG C3 test dataset with multiple missing traces.}
\centering
\vspace{-3mm}
\begin{tabular}{lrrrrr}
\toprule
\multicolumn{2}{c}{Gaussian Noise} & \multirow{2}{*}{MSE} & \multirow{2}{*}{SNR} & \multirow{2}{*}{PSNR} & \multirow{2}{*}{SSIM} \\
\cline{1-2}
$\mu$ & $\sigma$ & & & & \\
\midrule
0 & 1.0e-03 & 1.727e-04 & 31.899 & 37.854 & 0.976 \\
0 & 1.0e-02 & 2.584e-04 & 29.979 & 35.932 & 0.943 \\
1.0e-02 & 1.0e-02 & 2.536e-04 & 30.233 & 36.017 & 0.944 \\
1.0e-01 & 1.0e-02 & 2.596e-04 & 31.507 & 35.902 & 0.943 \\
1.0e-01 & 5.0e-02 & 1.586e-03 & 22.075 & 28.001 & 0.748 \\
\bottomrule
\end{tabular}
\label{tab:Generalization noise levels}
\vspace{-4mm}
\end{table}

\begin{figure*}[!t]
  \centering
  \subfloat[Ground truth.]
  {\includegraphics[height=0.11\textheight]{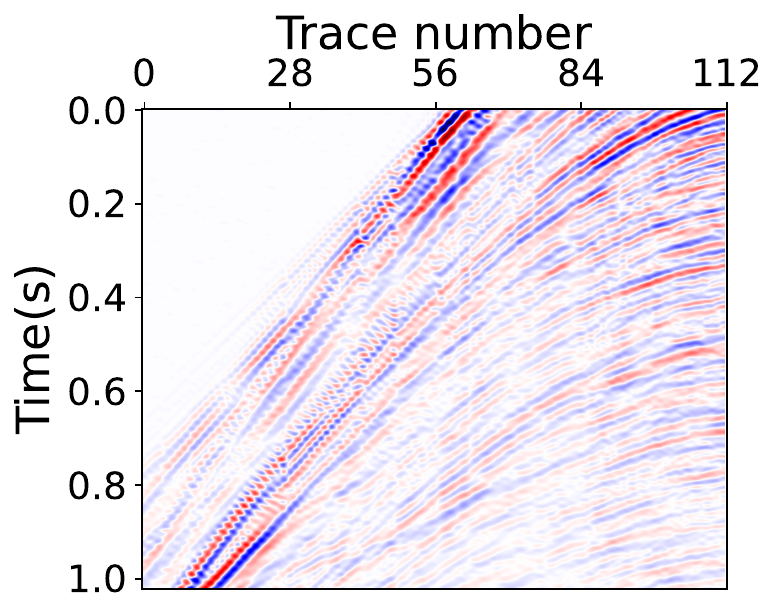}\label{fig:uncertainrandom_subfig1}}
  \subfloat[Random missing data.]
  {\includegraphics[height=0.11\textheight]{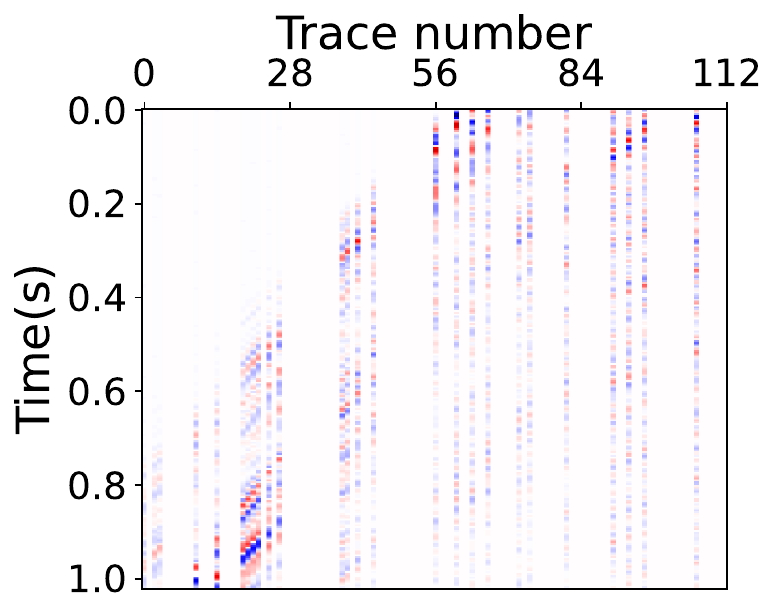}\label{fig:uncertainrandom_subfig2}}
  \subfloat[Interpolation result.]
  {\includegraphics[height=0.11\textheight]{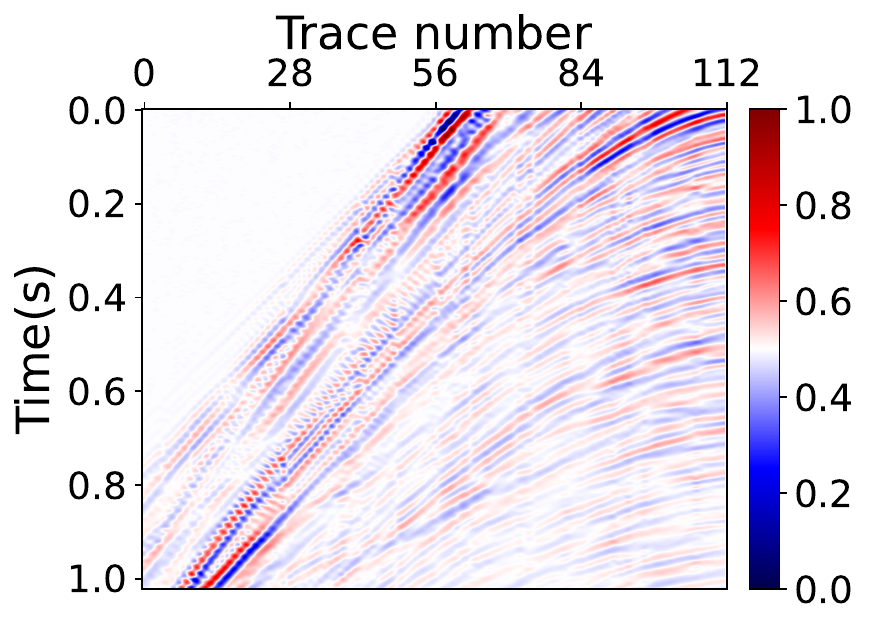}\label{fig:uncertainrandom_subfig3}}
  \subfloat[Uncertainty estimate.]
  {\includegraphics[height=0.11\textheight]{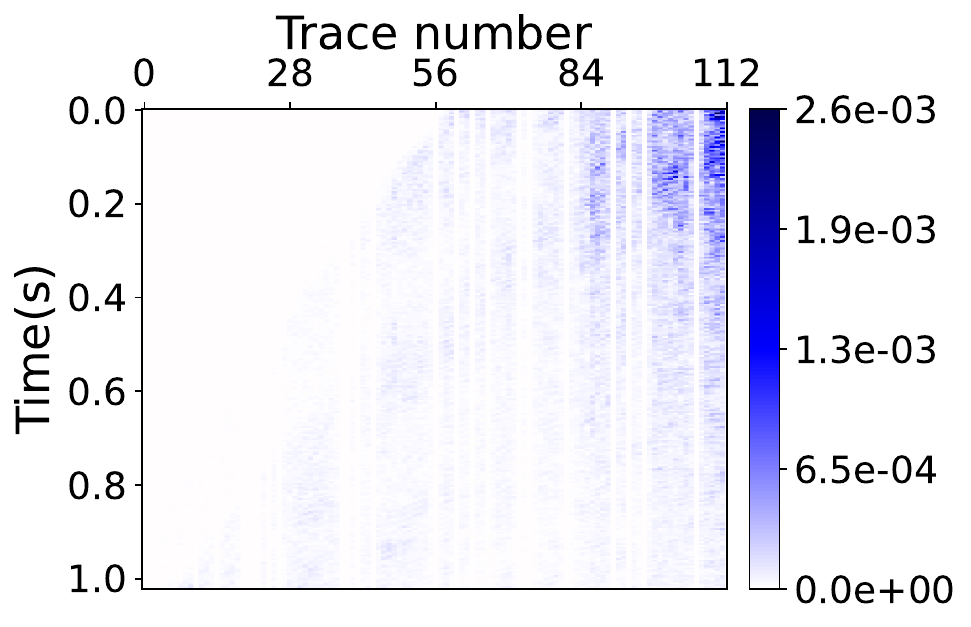}\label{fig:uncertainrandom_subfig4}}
  \subfloat[Absolute residual.]
  {\includegraphics[height=0.11\textheight]{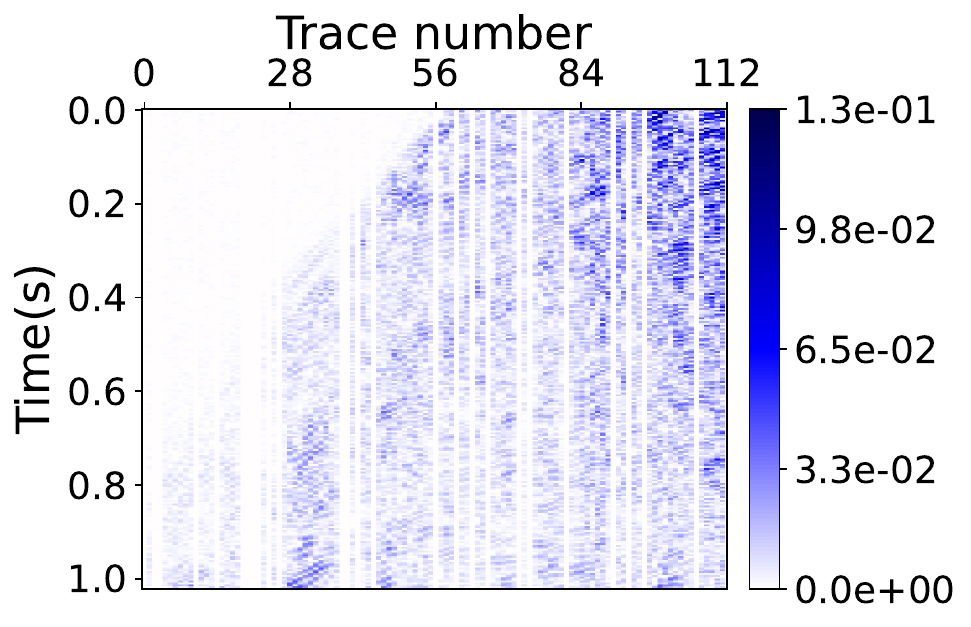}\label{fig:uncertainrandom_subfig5}} 
  \vspace{-1mm} 
  \caption{Uncertainty quantification on the interpolation result of the MAVO test data with 77\% random missing traces. The interpolation result in (c) is the uncertainty obtained from multiple test repetitions. The absolute value of the average residual is presented in (e) for comparison purposes. }
  \label{fig:uncertain random}
  \vspace{-6mm}
\end{figure*}

\begin{figure*}[!t]
  \centering
  \subfloat[Ground truth.]
  {\includegraphics[height=0.11\textheight]{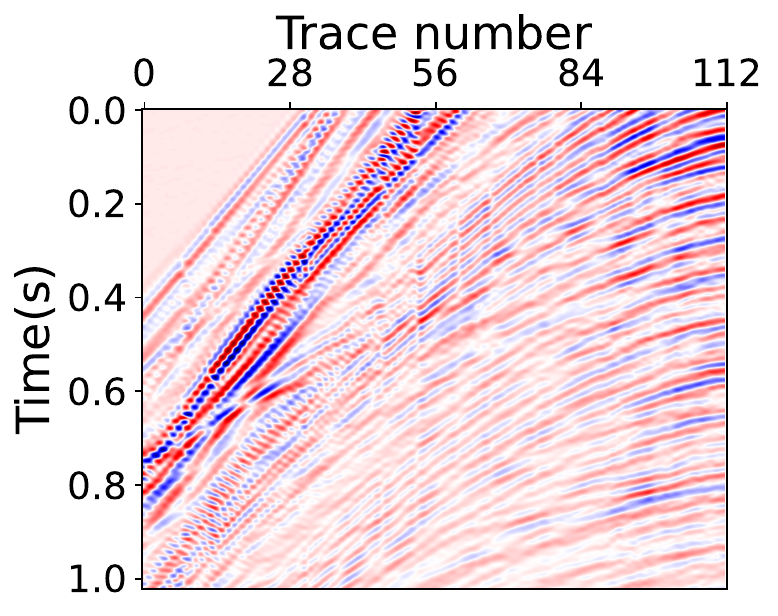}\label{fig:uncertaincontinus_subfig1}}
  \subfloat[Consecutive missing data.]
  {\includegraphics[height=0.11\textheight]{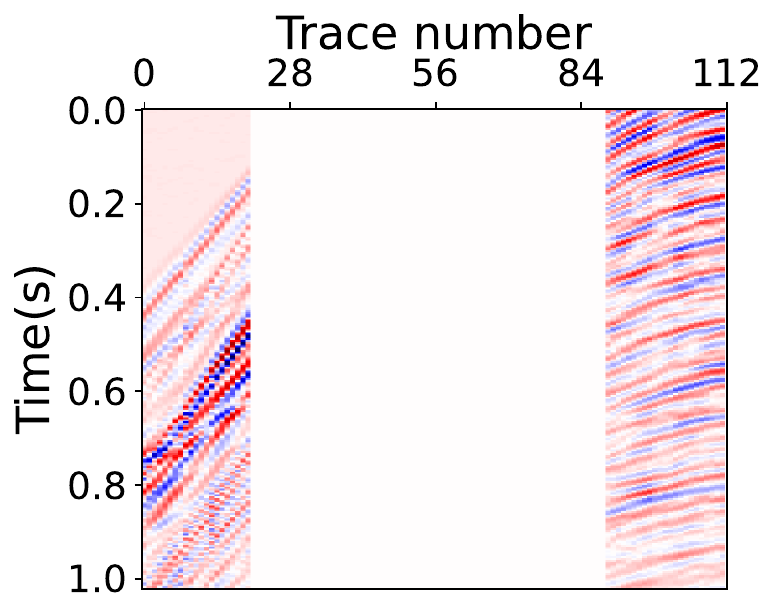}\label{fig:uncertaincontinus_subfig2}}
  \subfloat[Interpolation result.]
  {\includegraphics[height=0.11\textheight]{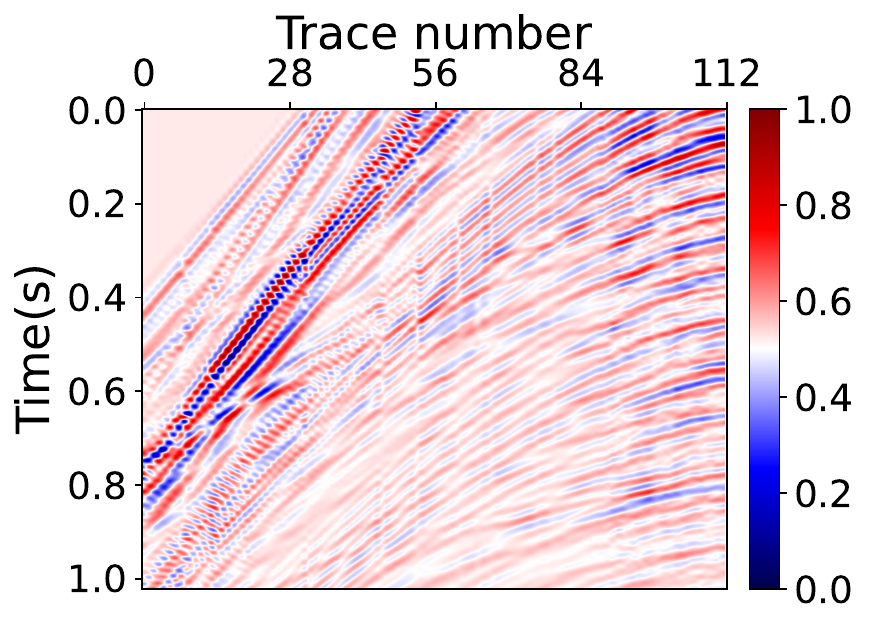}\label{fig:uncertaincontinus_subfig3}}
  \subfloat[Uncertainty estimate.]
  {\includegraphics[height=0.11\textheight]{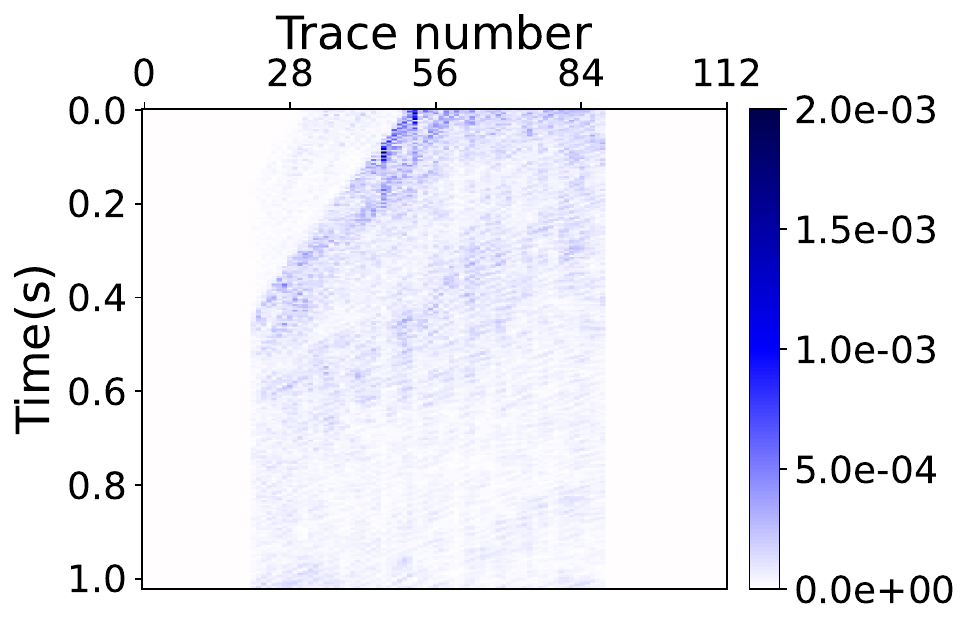}\label{fig:uncertaincontinus_subfig4}}
  \subfloat[Absolute residual.]
  {\includegraphics[height=0.11\textheight]{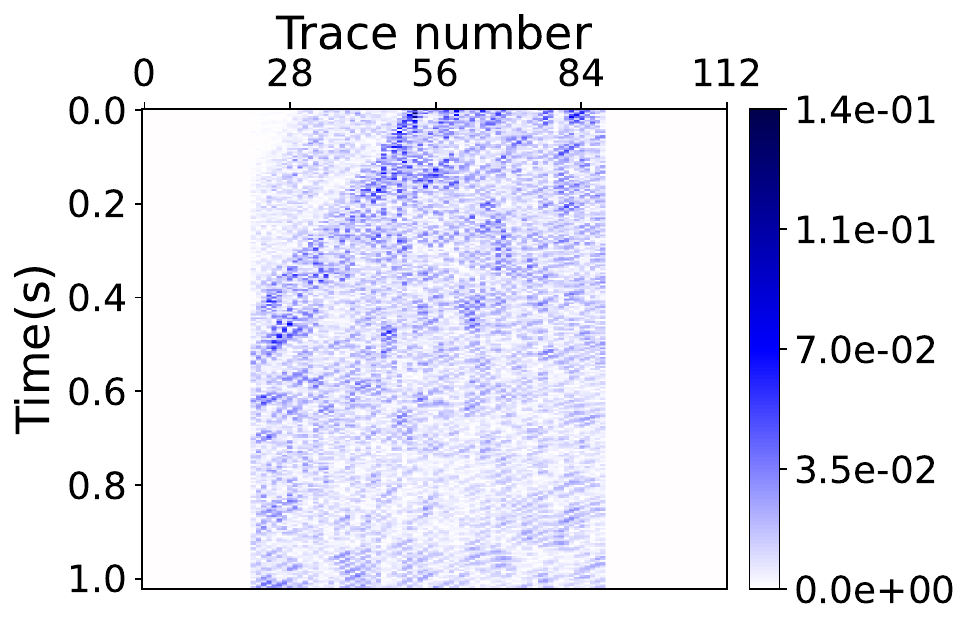}\label{fig:uncertaincontinus_subfig5}}  
  \vspace{-1mm}
  \caption{Uncertainty quantification on the interpolation result of the MAVO test data with 60\% consecutive missing traces. The interpolation result in (c) is the uncertainty obtained from multiple test repetitions. The absolute value of the average residual is presented in (e) for comparison purposes.}
  \label{fig:uncertain continus}
  \vspace{-6mm}
\end{figure*}

\begin{figure*}[!t]
  \centering
  \subfloat[Ground truth.]
  {\includegraphics[height=0.11\textheight]{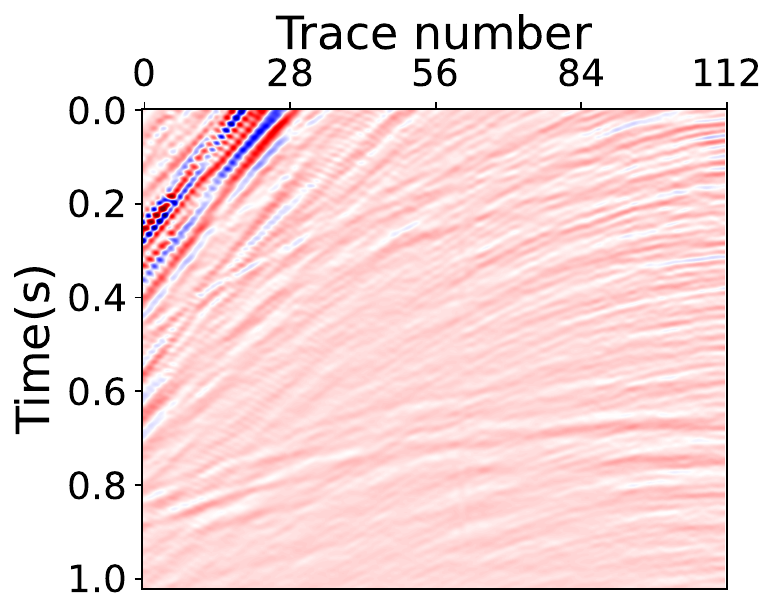}\label{fig:uncertainmultiple_subfig1}}
  \subfloat[Multiple missing data.]
  {\includegraphics[height=0.11\textheight]{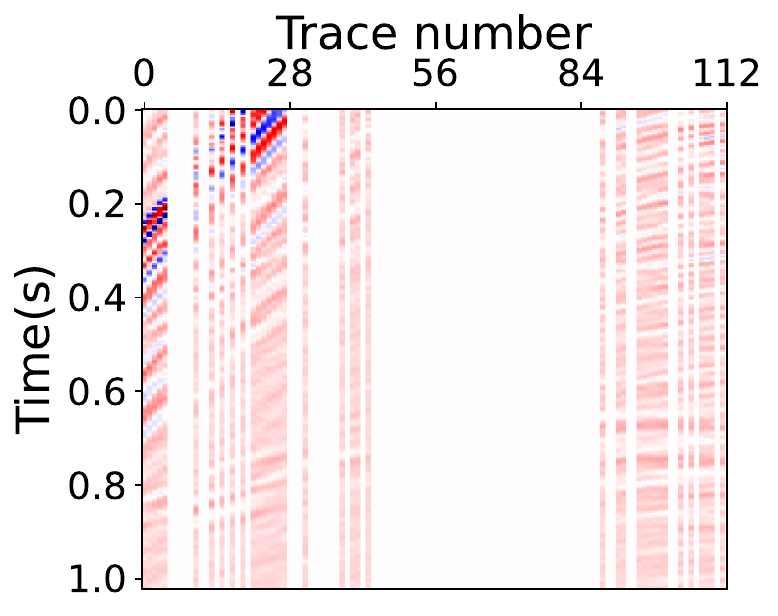}\label{fig:uncertainmultiple_subfig2}}
  \subfloat[Interpolation result.]
  {\includegraphics[height=0.11\textheight]{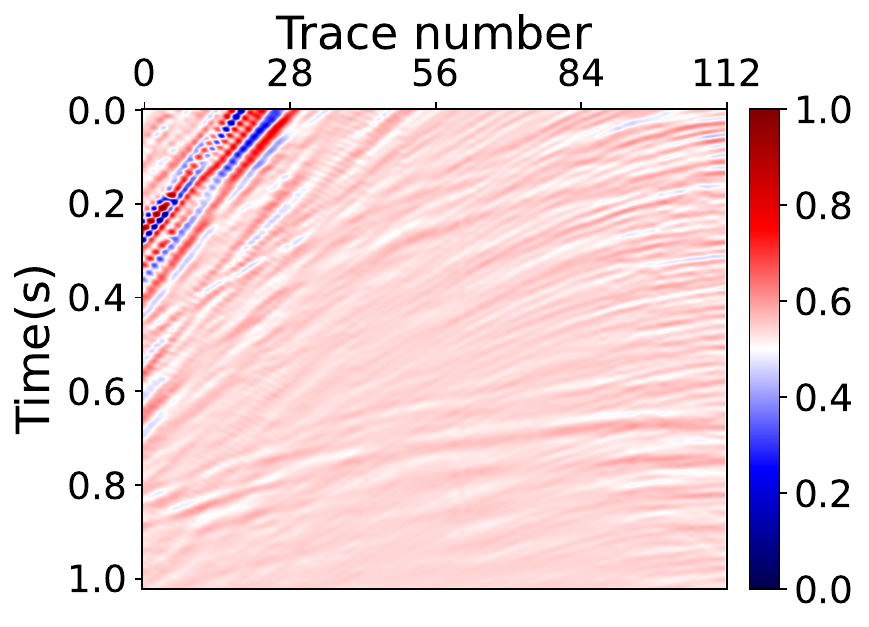}\label{fig:uncertainmultiple_subfig3}}
  \subfloat[Uncertainty estimate.]
  {\includegraphics[height=0.11\textheight]{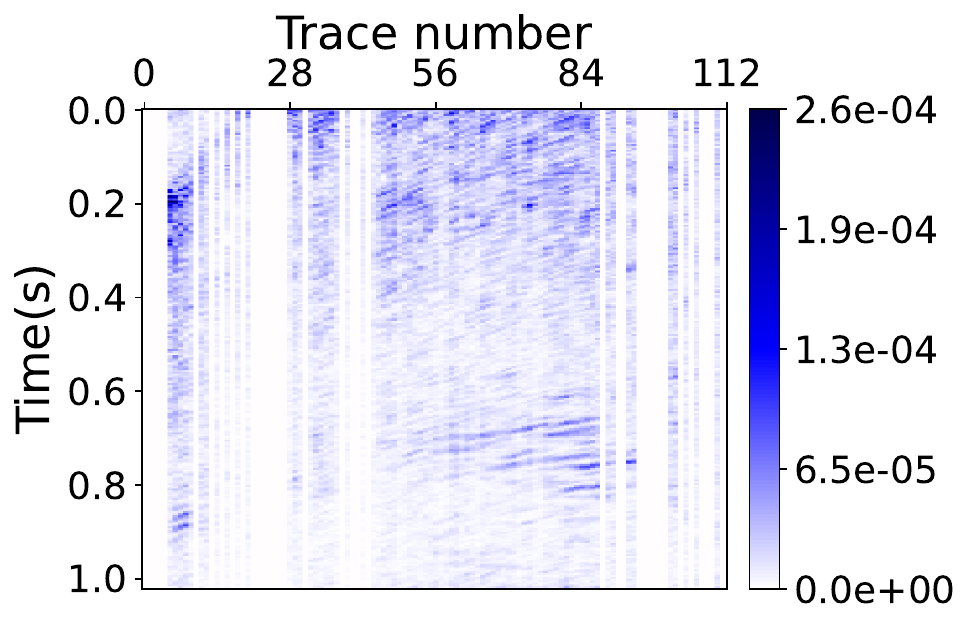}\label{fig:uncertainmultiple_subfig4}}
  \subfloat[Absolute residual.]
  {\includegraphics[height=0.11\textheight]{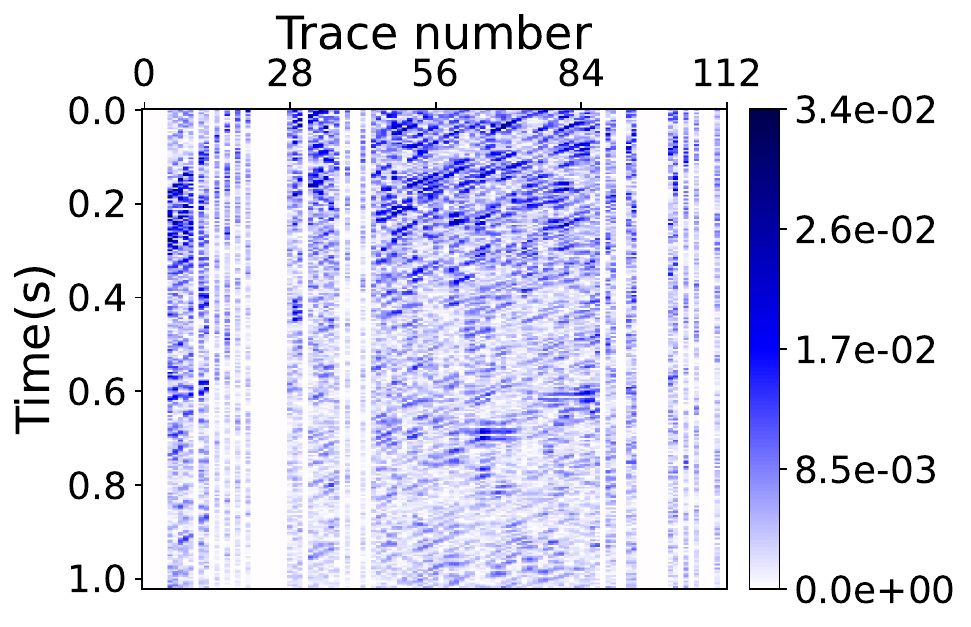}\label{fig:uncertainmultiple_subfig5}}  
  \vspace{-1mm}
  \caption{Uncertainty quantification on the interpolation result of the MAVO test data with 67\% multiple missing traces. The interpolation result in (c) is the uncertainty obtained from multiple test repetitions. The absolute value of the average residual is presented in (e) for comparison purposes.}
  \label{fig:uncertain multiple}
  \vspace{-6mm}
\end{figure*}
\subsection{Uncertainty Quantification}
Although various interpolation methods based on deep learning have accomplished promising results in the aforementioned publications, uncertainty quantification of the prediction is still absent subjecting to the fixed inference mode. However, providing measures of uncertainty for the predictions over or under confidence is important to improve the application security and avoid the cost of an error. The uncertainty in deep neural networks is divided into the reducible model uncertainty (also systemic or epistemic uncertainty) and irreducible data uncertainty (also statistical or aleatoric
uncertainty) \cite{gawlikowski2023survey}. The model uncertainty
is caused by inadequate models and unsuitable learning patterns, and data uncertainty is an inherent characteristic of data and cannot be reduced or eliminated by improving the subsequent model.

There are multiple random sampling operations in our SeisDDIMCR model as stated in Algorithm \ref{alg:SeisDDIMR_interpolation}, thus we adopt the approach deriving from uncertainty ensemble methods to capture the total uncertainty by calculating the standard deviation of the interpolation results obtained after multiple repetitions of Algorithm \ref{alg:SeisDDIMR_interpolation}. For a sample $\boldsymbol{x}$, the uncertainty is computed  as
\begin{equation*}
\frac{1}{n} \sum_{i=1}^n \|\hat{\boldsymbol{x}}_i-\hat{\boldsymbol{\mu}}_i\|^{2}_{F},
\end{equation*}
where $\hat{\boldsymbol{\mu}}_i = \frac{1}{n}\sum_{i=1}^n\hat{\boldsymbol{x}}_i$, $\hat{\boldsymbol{x}}_i$ is the interpolation result of a single test, and $n$ is the repetition test number. Fig. \ref{fig:uncertain random}-\ref{fig:uncertain multiple} visualize the uncertainty in the interpolation results of random, consecutive, and multiple missing traces, respectively. The average interpolation results $\hat{\boldsymbol{\mu}}_i$ and the absolute average residual $\left|\frac{1}{n} \sum_{i=1}^n\left(\hat{\boldsymbol{x}}_i-\boldsymbol{x}_{\text{gt}}\right)\right|$ ($\boldsymbol{x}_{\text{gt}}$ is the ground truth seismic data) are also exhibited to provide an intuitive reference. 
It seems that unreliable reconstruction results are more likely to occur in the missing areas with patch edges and strong lateral amplitude variations, due to limited information and highly curved events. Besides, areas with high interpolation uncertainty also acquire large absolute residuals. 

\subsection{Multi-shot Seismic Data Interpolation and Stacked Section}
\begin{figure*}[!t]
	\centering
	\vspace{-4mm}
	\subfloat[Ground truth.]
	{\includegraphics[height=0.165\textheight]{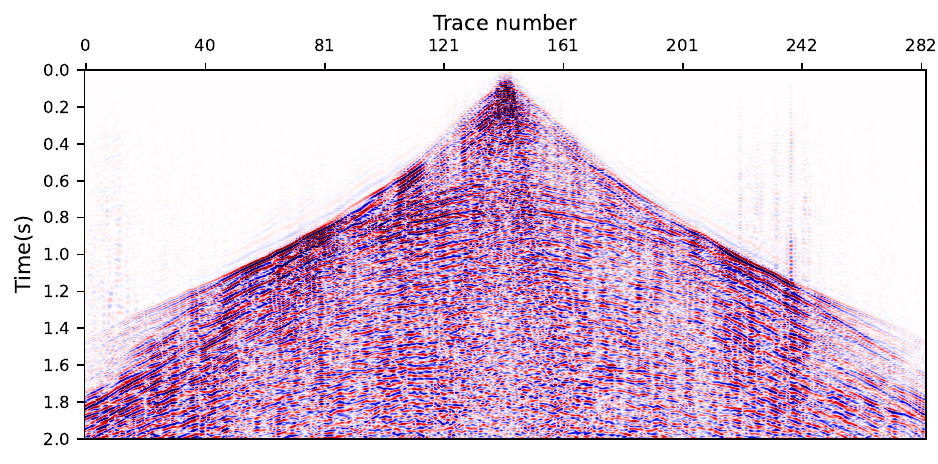}\label{fig:multishots_subfig1}}
	\vspace{-4mm}
	\subfloat[Random missing data.]
	{\includegraphics[height=0.165\textheight]{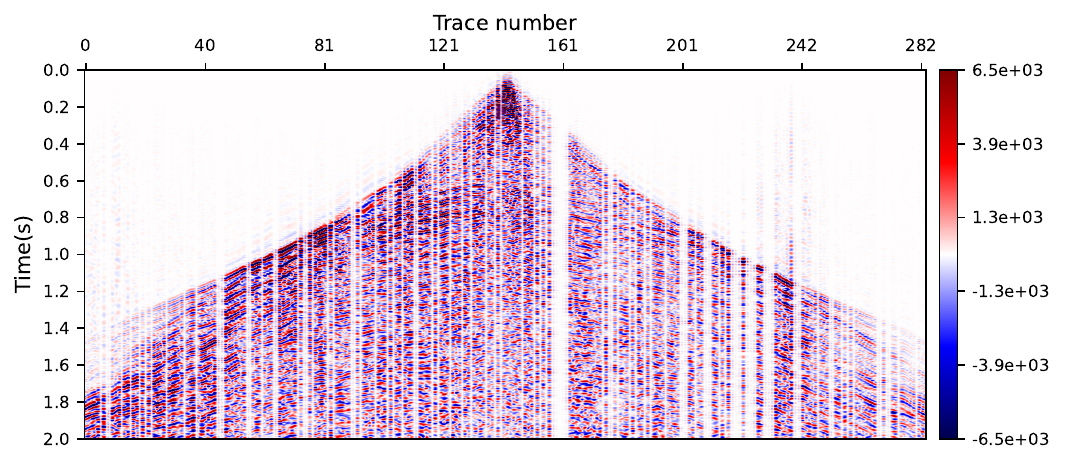}\label{fig:multishots_subfig2}} \\
	\subfloat[Interpolation result.]
	{\includegraphics[height=0.165\textheight]{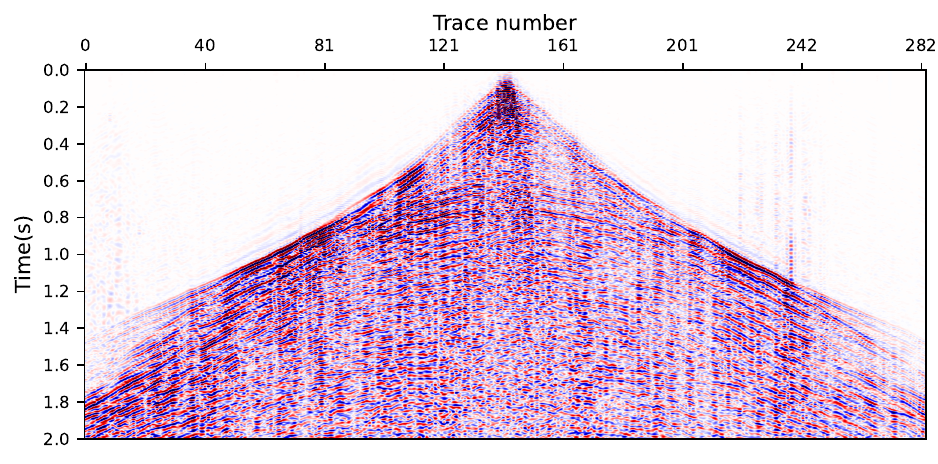}\label{fig:multishots_subfig3}}
	\vspace{-1mm}
	\subfloat[Residual.]
	{\includegraphics[height=0.165\textheight]{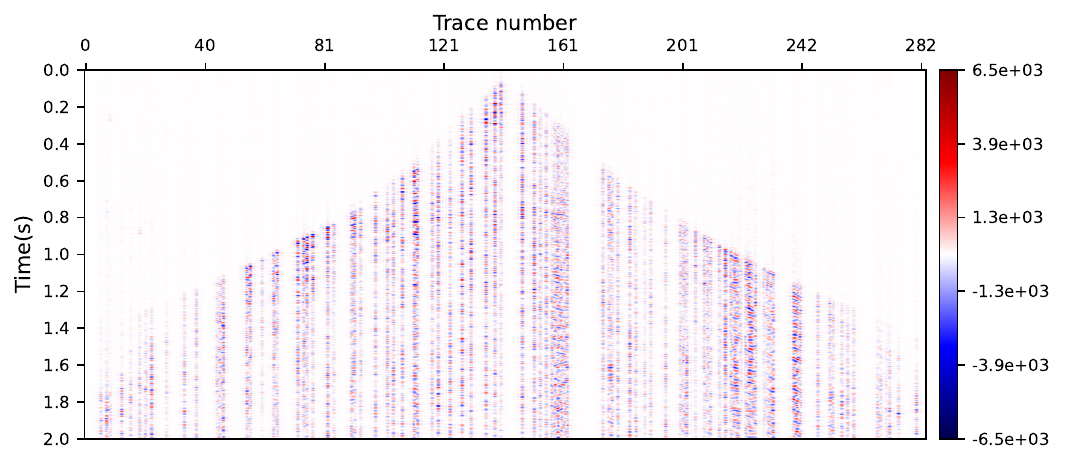}\label{fig:multishots_subfig4}}
	\caption{Interpolation results of a shot gather from the 2D Vibroseis Line 001 dataset with 34\% random missing traces. We recover the seismic data to their original amplitude range and utilize the gain method to enhance the visibility of weak amplitude details.}
	\label{fig:multishots}
	\vspace{-2mm}
\end{figure*}

\begin{figure*}[!t]
	\centering
	\vspace{-1mm}
	\subfloat[Ground truth.]
	{\includegraphics[height=0.13\textheight]{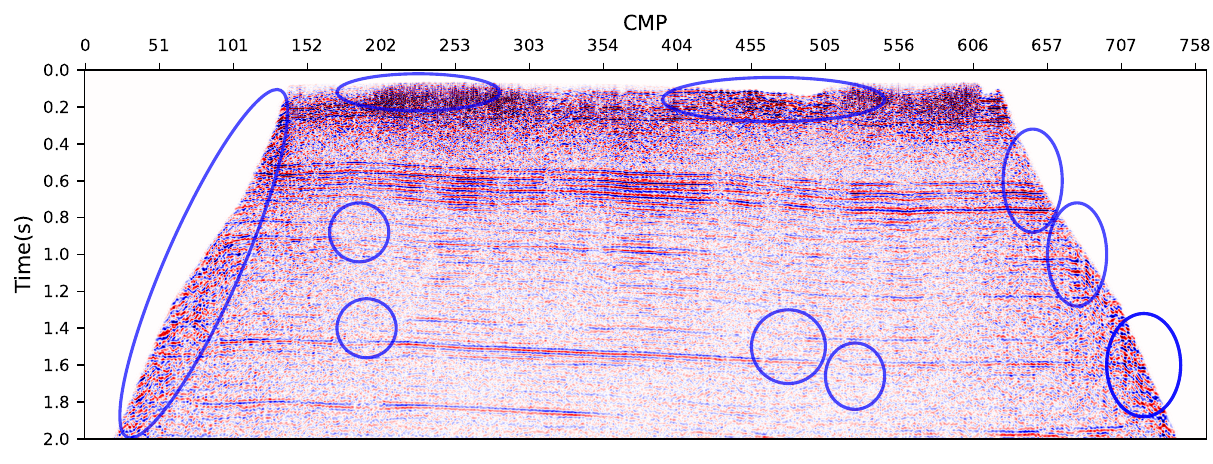}\label{fig:multishots_stacking_subfig1}}
	\hspace{0.1mm}
	\subfloat[Random missing data.]
	{\includegraphics[height=0.13\textheight]{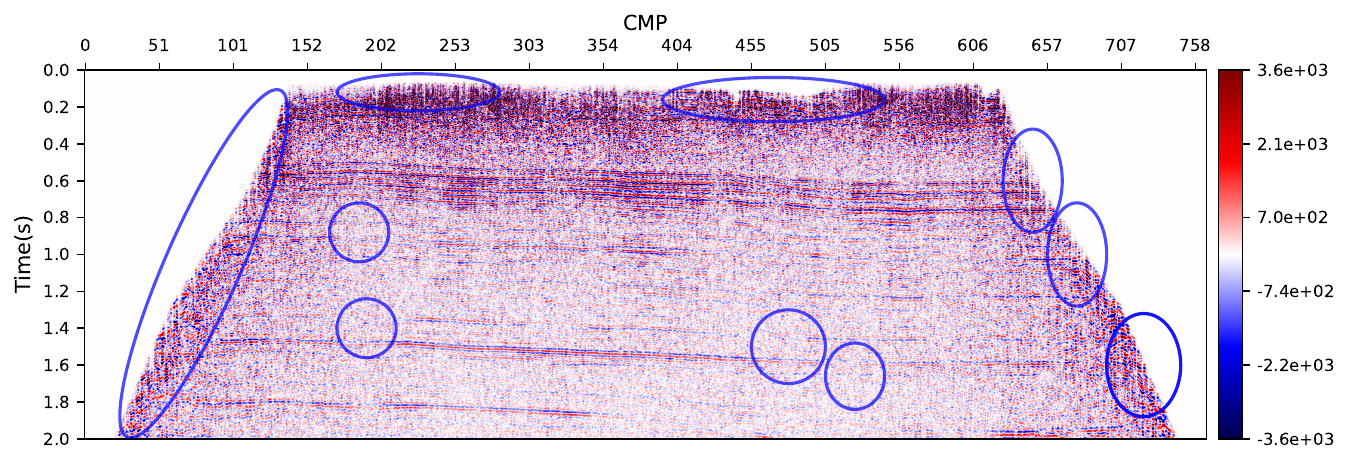}\label{fig:multishots_stacking_subfig2}} \\
	\vspace{-4mm}    
	\subfloat[Interpolation result.]
	{\includegraphics[height=0.13\textheight]{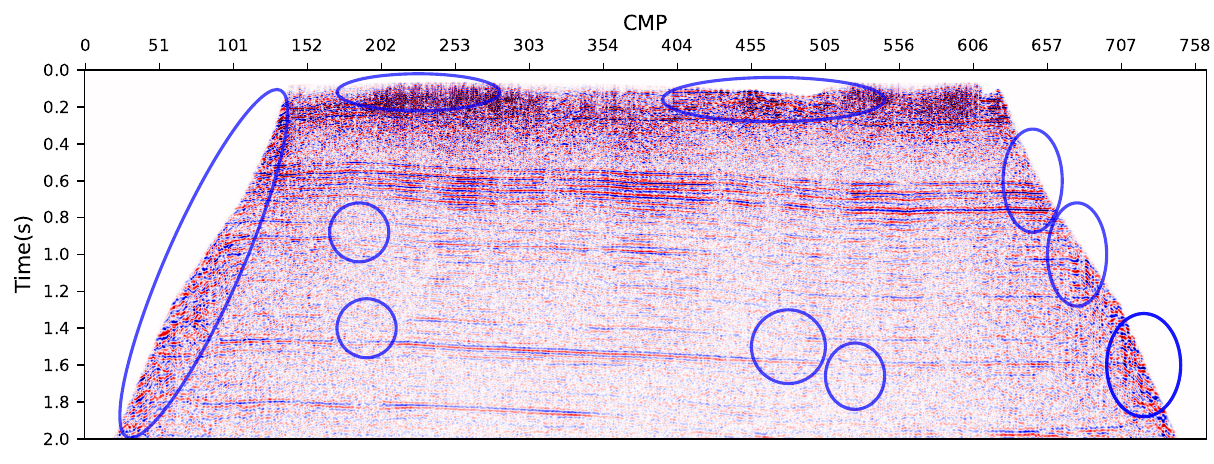}\label{fig:multishots_stacking_subfig3}}
	\vspace{-1mm}
	\subfloat[Residual.]
	{\includegraphics[height=0.13\textheight]{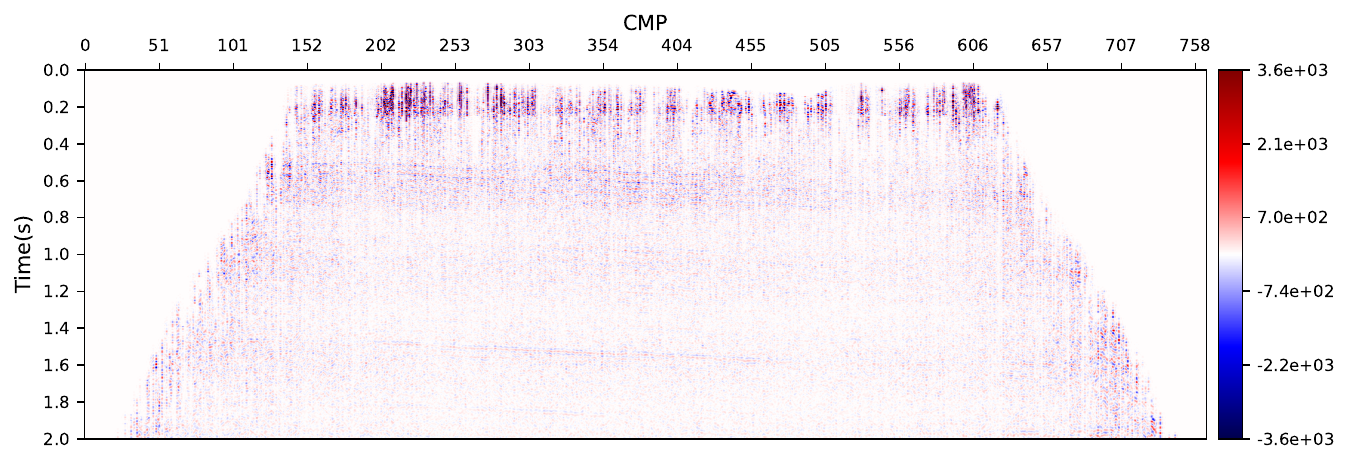}\label{fig:multishots_stacking_subfig4}}
	\caption{Stacked sections of the 120-shot 2D Vibroseis Line 001 dataset with random trace missing rates between 30\% and 40\%. The blue circles highlight areas with significant improvement after interpolation compared to the directly stacked sections of the missing seismic data. For ease of comparison, we also mark the same locations in the ground truth seismic data.}
	\label{fig:multishots_stacking}
	\vspace{-2mm}
\end{figure*}

To evaluate the effectiveness of interpolation and stacking on multi-shot data, we adopt the Poland 2D Vibroseis Line 001 field data obtained from SEG open datasets. It is widely used multi-shot prestack seismic data acquired in onshore environments using controlled-source technology \cite{liu2015signal}, \cite{li2024removing}. This dataset contains 251 shots, with 282 traces (2 auxiliary traces are omitted) per shot and 1,501 time samples per trace. The receiver interval and time interval are 25 m and 2 ms, respectively. Although the data size is small and convenient for testing, its signals are adequately complex and very challenging, coupled with several types of strong noise. To perform CMP-stacking (CMP is short for common midpoint), we form a rough but necessary processing flow including SPS loading, static correction, groundroll removing, strong near and far-offset linear energy removing, and strong low-velocity triangle zone noise attenuation, followed by surface consistent amplitude compensation and residual statics correction. These conventional processings simulate industrial interpolation scenarios. Limited by our computation power, we choose the first 120 consecutive shots and randomly select 40 shots of them for training, and total shots are used for prediction. We restrict the time sampling to 2,000 ms since the effective signal from deeper regions is limited. Training and prediction shots are split into 128$\times$128 patches along the trace and time direction by the sliding window with a stride of 32 and 64, respectively. There are 7,680 training patches in total. Prediction patches are merged into whole shots with overlapping regions averaged. Tab. \ref{tab:multiple-shots} summarizes the interpolation results for the 120 shots of the 2D Vibroseis Line 001 dataset, comparing them with the best comparison method, Coarse-to-Fine model. Our method consistently demonstrates superior performance across various missing data scenarios, including random missing rates from 0.2 to 0.8, consecutive missing rates from 0.1 to 0.3, and multiple missing rates from 0.2 to 0.55. To further provide a global view about the method's effectiveness by stacked sections, we conduct interpolation prediction with random missing trace rates between 0.3 and 0.4. Fig. \ref{fig:multishots} visualizes the interpolation results for single shot gather. Despite the high noise in the 2D Vibroseis Line 001 data, our method remains robust. Then, we stack our prediction results and compare it with the ground truth CMP-stacking section and the missing data CMP-stacking section in Fig. \ref{fig:multishots_stacking}. We circle some of the differences on the same part in three stacking sections. It can be clearly observed that our method aligns well with the ground truth. However, the section of missing data stacking shows numerous absences of seismic signals and obvious distortion of the seismic reflection interface. Furthermore, a residual CMP-stacking section between the ground truth and the prediction data also reveals high precision reconstruction and minor signal leakage.
\begin{table}[htbp]
	\scriptsize
	\vspace{-4mm}
	\caption{Interpolation results on the first 120 shots of the 2D Vibroseis Line 001 dataset. The result of top performance is masked in bold.}
	\centering
	\vspace{-2mm}
	\begin{tabular}{llrrrrr}
		\toprule
		Type & Model  & MSE & SNR &PSNR &SSIM\\
		\midrule 
		Random & Coarse-to-Fine\cite{wei2022hybrid} & 3.901e-03 & 18.304 & 24.088 & 0.766\\
		& Ours &\bf{3.716e-03} & \bf{19.130} & \bf{24.916} & \bf{0.775}\\
		\midrule 
		Consecutive & Coarse-to-Fine\cite{wei2022hybrid} & 2.488e-03 & 20.259 & 26.042 & 0.852\\
		& Ours &\bf{1.944e-03} & \bf{21.697} & \bf{27.482} & \bf{0.873}\\
		\midrule 
		Multiple & Coarse-to-Fine\cite{wei2022hybrid} & 3.233e-03 & 19.120 & 24.904 & 0.805\\
		& Ours &\bf{2.847e-03} & \bf{20.004} & \bf{25.790} & \bf{0.821}\\
		\bottomrule
	\end{tabular}
	\label{tab:multiple-shots}
	\vspace{-4mm}
\end{table}

\section{Ablation Study}\label{secabalation}
In this section, we will conduct a series of ablation studies on the key components, hyperparameters, and model analysis from four aspects, including inference strategy, seismic DDPM, convergence, and computational complexity.

\subsection{Seismic DDIM Inference}
To assess the efficacy of our proposed implicit conditional interpolation approach with coherence correction and resampling strategy, we execute Algorithm \ref{alg:SeisDDIMR_interpolation} under various configurations on the test set of the SEG C3 dataset with multiple missing traces. The interpolation results are presented in Tab. \ref{tab:ablationresampling}. Comparing the interpolation performance of Algorithm \ref{alg:SeisDDIMR_interpolation} based on DDPM and DDIM, it can be demonstrated that our proposed implicit interpolation significantly enhances the quality of signal recovery. When the conditional interpolation operation is removed, the performance of the model significantly decreases, demonstrating the critical role of this module. For the coherence correction mechanism, we examine whether to use it and the model's performance under different settings of the gradient descent number $G$ and the weight parameter $\lambda_{\tau_{m}}$. The results show that if we forego the use of coherence correction, there is a significant decline in model performance, confirming the module's effective corrective function. Furthermore, increasing the gradient descent number 
$G$ is undoubtedly beneficial, yet it also increases inference time, thus we continue to use the setting $G$=1. Lastly, changing the weight parameter $\lambda_{\tau_{m}}$ did not show a noticeable difference in performance.

\begin{figure}[!htbp]
	\centering
    \includegraphics[width=3.2in]{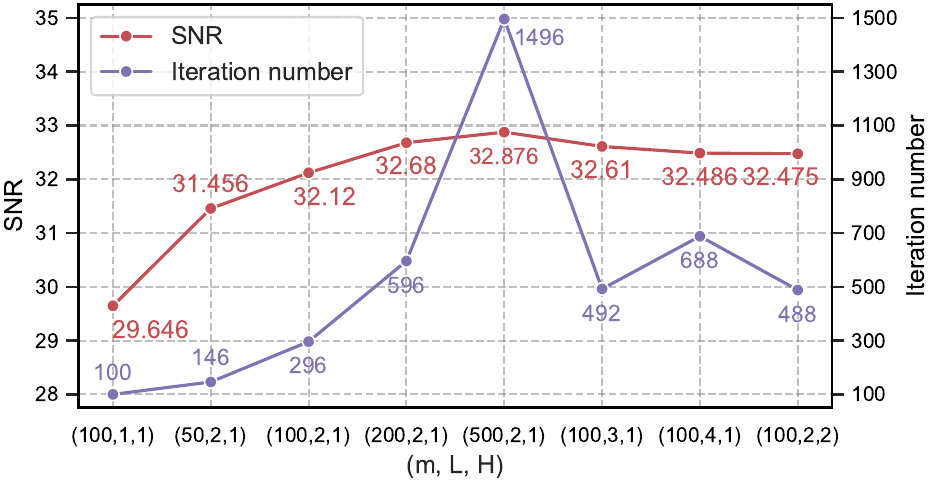}
    \vspace{-4mm}
	\caption{The model performance curves under different resampling settings. The red line illustrates the variation in SNR of the interpolation results and the purple line depicts the total number of iterations required in the inference process under different settings.}
	\label{fig: ablation resampling}
	\vspace{-4mm}
\end{figure}

It is infeasible to explore all potential scenarios for diffusion sampling steps $m$, travel length $L$, and travel height $H$. Therefore, we aim to identify the most feasible options by choosing some combinations. Fig. \ref{fig: ablation resampling} displays the model performance and the total number of iterations required in the inference process (Algorithm \ref{alg:SeisDDIMR_interpolation}) under different combinations of $m$, $L$, and $H$. Our goal is to achieve a trade-off between the model's interpolation capabilities and efficiency. First, the combination of (100,1,1) is equivalent to not using resampling, and we observe a significant decrease in performance, highlighting the importance of resampling. Second, increasing the sampling steps $m$, travel length $L$, and travel height $H$ can enhance the diffusion effect, yet they also result in a higher computational burden during the inference process. We chose the setting $m$=100, $L$=2, $H$=1 since it offers fewer iterations without sacrificing much interpolation accuracy. 

\begin{table}[htbp]
\scriptsize
\setlength{\tabcolsep}{4pt}
\vspace{-3mm}
\caption{Ablation of various settings of implicit conditional interpolation and coherence correction. The third and sixth rows are configured without coherence correction.}
\centering 
\vspace{-3mm}
\begin{tabular}{cccccccc}
\toprule  
Diffusion & \multirow{2}{*}{Condition} & \multicolumn{2}{c}{Coherence correction} & \multirow{2}{*}{MSE} & \multirow{2}{*}{SNR}  & \multirow{2}{*}{PSNR} & \multirow{2}{*}{SSIM}\\ %
\cline{3-4}
\\[-2.2mm] 
model & & $G$ & $\lambda_{\tau_{m}}$ & & & & \\
\midrule  
DDIM & \checkmark & 1 & 1.0e-04 & 1.601e-04 & 32.120 & 38.074 & 0.977\\
  \midrule 
  & \ding{55} & 1 & 1.0e-04 & 1.740e-03& 21.677& 27.631 & 0.845 \\
  & \checkmark & - & - & 4.028e-04 & 28.138 & 34.093 & 0.953 \\
DDIM  & \checkmark & 2 & 1.0e-04 & 1.323e-04 & 33.046& 39.000 & 0.979 \\
  & \checkmark & 1 & 1.0e-03 & 1.613e-04 & 32.058& 38.037 & 0.976 \\
\midrule 
DDPM & \checkmark & - & - & 5.433e-04 & 26.779 & 32.733  & 0.942\\
\bottomrule
\end{tabular}
  \label{tab:ablationresampling}
  \vspace{-4mm}
\end{table}

\subsection{Seismic DDPM}
The training of the Seismic DDPM is implemented by the process described in Algorithm \ref{alg:SeisDDIMR_training}. We selected three key components, i.e., diffusion steps $T$, loss function, 
and noise schedule, to validate the superiority of the adopted configuration. Seismic DDPM is trained on the SEG C3 dataset under different settings with the total iteration number $N$ = 300,000, respectively. Tab. \ref{tab:ablationSeismicDDPM} yields the interpolation results on the SEG C3 test dataset with multiple missing traces. First, the number of diffusion steps $T$ has a significant impact on the diffusion speed of our model. Increasing $T$ refines the model, but also causes additional computational burden. Achieving a balance between computational efficiency and model performance requires a compromise configuration of the diffusion steps. Second, Tab. \ref{tab:ablationSeismicDDPM} indicates that better interpolation results can be achieved by allowing the noise matching network to learn the noise variance $\sigma_t$ under the hybrid loss $L_{\text {hybrid}}$. Finally, 
training seismic DDPM with different noise schedules indicates that using a linear schedule suffers from significant performance degradation. This finding supports our decision to adopt the cosine schedule, which has demonstrated better performance.

\begin{table}[htbp]
\scriptsize
\setlength{\tabcolsep}{7pt}
\vspace{-2mm}
 \caption{Ablation of various settings in seismic DDPM. The result of the top performance is masked in bold. According to the first row, only one setting is changed per row.}
  \centering
  \vspace{-3mm}
\begin{tabular}{lrrrrrr}
\toprule
\multirow{2}{*}{$T$} &\multirow{2}{*}{$\lambda$} & Noise  & \multirow{2}{*}{MSE} & \multirow{2}{*}{SNR} & \multirow{2}{*}{PSNR} & \multirow{2}{*}{SSIM} \\ 
 & &schedule & &  &  &  \\
\midrule
 1000 & 0.001 & Cosine & \bf{1.601e-04} & \bf{32.120} & \bf{38.074} & \bf{0.977}\\ 
\hline \hline
\\[-2.2mm]
 500 &  &   & 1.792e-04 & 31.749 &37.561 & 0.972  \\
 100 &  &    & 2.237e-04 & 30.532 &36.529 & 0.962 \\
   &  0 & & 1.969e-04 & 31.081 &36.934 & 0.967 \\ 
  &  & Linear & 2.587e-04 & 30.043 &36.035 & 0.958 \\
\bottomrule
\end{tabular}
  \label{tab:ablationSeismicDDPM}
  \vspace{-5mm}
\end{table}

\begin{figure*}[!t]
	\centering
	\vspace{-1mm}
\includegraphics[width=7.0in]{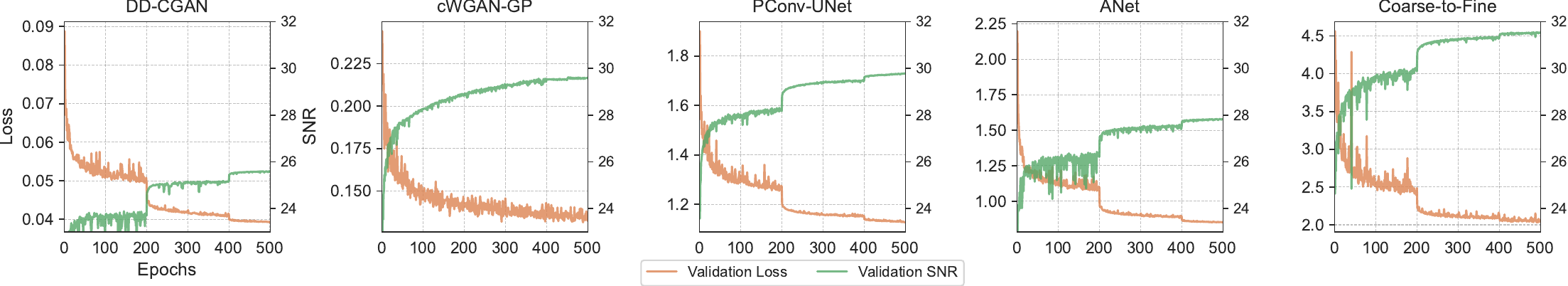}
\vspace{-3mm}
    \caption{The convergence of different methods. We employ a dual-axis display mechanism, where the left vertical axis depicts the validation loss, and the right vertical axis depicts the validation set SNR.}
    \vspace{-3mm}
    \label{fig: convex_5methods}
\end{figure*}

\subsection{Convergence}
\begin{figure}[!htbp]
	\centering
	\vspace{-2mm}
\includegraphics[width=3.35in]{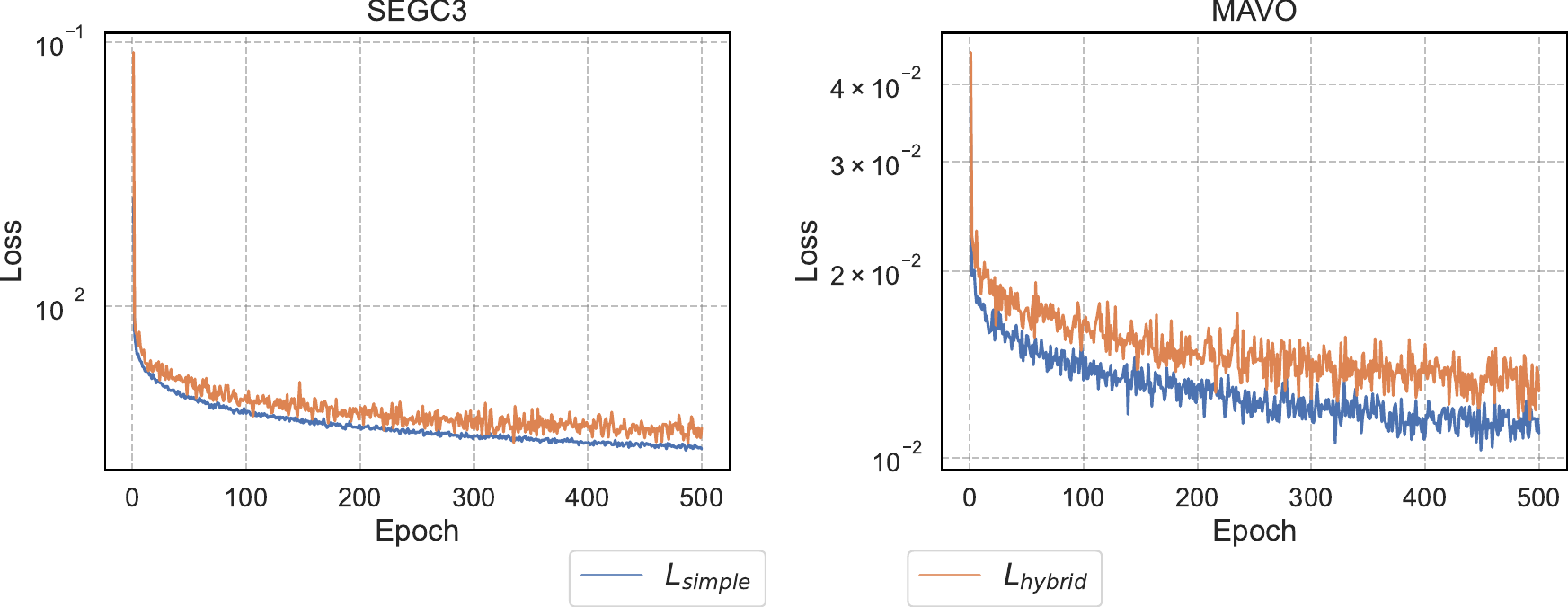}
	\vspace{-3mm}
    \caption{The convergence of our methods.}
    \label{fig: convex_ours}
    \vspace{-3mm}
\end{figure}

We provide a comparison of the convergence for different methods. Fig. \ref{fig: convex_5methods} shows the curves of validation loss and validation SNR during the training process for the other five comparative methods on the SEG C3 dataset with multiple missing traces. It should be noted that they implement a fixed epoch length decay strategy for training learning rates except for cWGAN-GP, resulting in noticeable stepwise changes. We can observe that both the validation loss and interpolation performance on the validation set converge to a relatively stable state. Since the training of the diffusion model does not involve a validation set, Fig. \ref{fig: convex_ours} illustrates the training loss curves of our SeisDDIMCR method on the SEG C3 dataset and the MAVO dataset. To clearly observe each loss component, we have plotted $L_{\text {simple}}$ and $L_{\text {hybrid}}$ against the epochs. It is evident that the convergence of the model is ensured on both datasets.

\subsection{Computational Complexity}
\begin{figure}[!htbp]
\vspace{-3mm}
\centering
\subfloat[]{
    \includegraphics[width=2.2in]{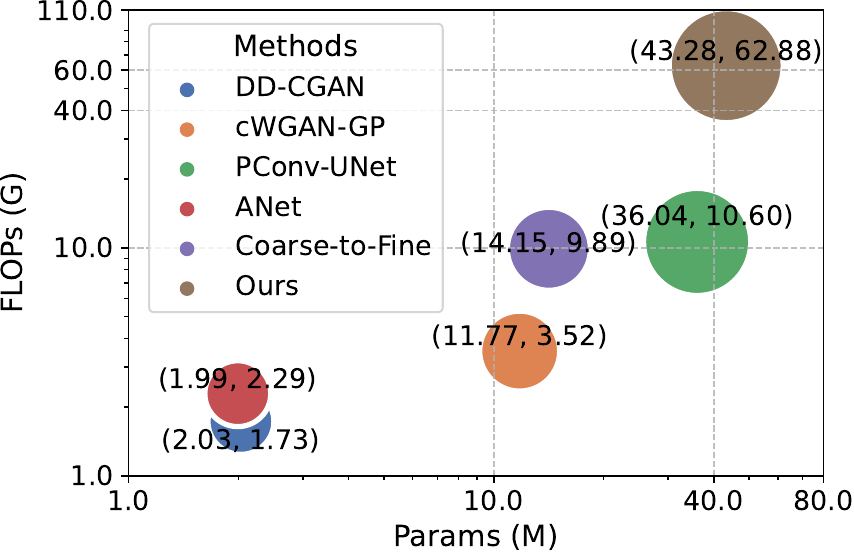}
    \label{fig:complexity}
}
\subfloat[]{
    \includegraphics[width=1.2in]{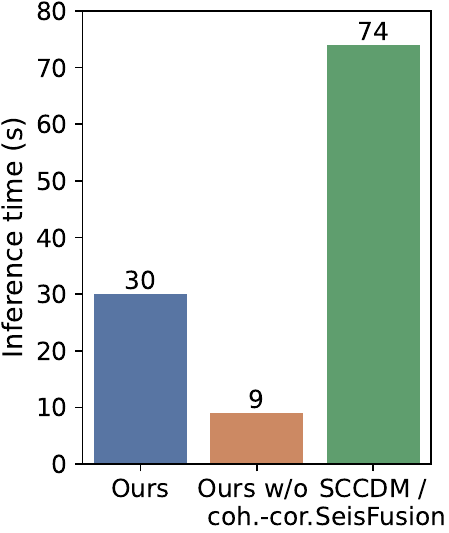}
    \label{fig:complexity_diffusion}
}
\vspace{-2mm}
\caption{(a) The computational complexity comparison of different methods. The axes are displayed on a logarithmic scale. (b) The inference time comparison. The yellow bar shows the inference time of our SeisDDIMCR model without coherence correction.}
\vspace{-5mm}
\end{figure}

We evaluate the computational complexity of our model and other five methods on the SEG C3 dataset. Fig. \ref{fig:complexity} visualizes the differences in FLOPs (Floating Point Operations Per Second) and Params (Parameters) across the models. Owing to the adoption of a lightweight U-Net architecture, the Params of our diffusion model maintain comparable to that of PConv-UNet. However, the inherent structure of the diffusion model significantly increases the FLOPs. In addition, we compare the inference times of our model with two recently proposed diffusion-based interpolation models, SCCDM\cite{deng2024seismic} and SeisFusion \cite{wang2024seisfusion}, as shown in Fig. \ref{fig:complexity_diffusion}. Although coherence correction increases the inference time compared to the version without it, our model still significantly reduces the inference time compared to these two methods. Moreover, CCSeis-DDPM \cite{wang2024reconstructing} adopts 4,570 as the inference timestep, yet our model just requires 296 steps as shown in Fig. \ref{fig: ablation resampling}.

\section{Conclusion}\label{secConclusion}
In this paper, we propose the SeisDDIMCR method, which tackles the seismic data interpolation problem with better model generalization on various missing data scenarios. SeisDDIMCR consists of two processes, including the training of seismic DDPM and implicit conditional interpolation with coherence correction and resampling. Seismic DDPM embeds seismic data into a denoising probability model framework. It achieves full-stage parameter sharing using the noise matching network based on the U-Net structure. The cosine noise schedule is introduced to speed up the transition during the high noise stage of seismic data. Implicit conditional interpolation with coherence-corrected resampling, serving as the inference process, achieves flexible interpolation for different missing data scenarios and missing rates by utilizing the existing traces of the seismic data as a condition. Coherence correction optimizes overall consistency by penalizing interpolation errors in the revealed traces. Resampling unifies the sampling distribution between adjacent steps through iteration.  
Interpolation experiments on synthetic and field seismic data with various patterns of missing data demonstrate that our SeisDDIMCR provides superior quality than existing methods and it also has advantages in robustness and generalization. 
In future studies, we will focus on extending our method to 3D or higher-dimensional seismic data interpolation. 
\section*{Acknowledgment}

The authors would like to thank the Sandia National Laboratory, Mobil Oil Company, and Geofizyka Torun S.A, Poland for providing open data sets. We are also grateful to the editors and reviewers for their valuable comments and suggestions.

\ifCLASSOPTIONcaptionsoff
  \newpage
\fi

\small
\bibliographystyle{IEEEtran}
\end{document}